\documentclass [aps,pra,amssymb,amsmath,showpacs,eqsecnum,twocolumn]{revtex4-1}
\usepackage[latin9]{inputenc}

\pdfoutput=1

\usepackage{times}
\usepackage{amsfonts}
\usepackage{amssymb}
\usepackage{amsmath}

\usepackage{lipsum}
\usepackage{graphicx}

\usepackage[caption=false]{subfig}
\usepackage{bm}
\usepackage{verbatim}
\usepackage[cspex,bbgreekl]{mathbbol}
\allowdisplaybreaks

\usepackage{hyperref}
\hypersetup{
    colorlinks,
    citecolor=blue,
    filecolor=blue,
    linkcolor=blue,
    urlcolor=blue
}

\usepackage{bm} 
\usepackage{color}
\usepackage{stackrel}
\usepackage{accents}

\usepackage{latexsym}

\usepackage{mathtools}

\usepackage{ulem}

\newcommand{\beg}{\begin{equation}}
\newcommand{\en}{\end{equation}}
\newcommand{\begs}{\begin{subequations}}
\newcommand{\ens}{\end{subequations}}
\newcommand \bea {\begin{eqnarray}}
\newcommand \eea {\end{eqnarray}}
\newcommand{\bem}{\begin{bmatrix}}
\newcommand{\enm}{\end{bmatrix}}
\newcommand{\bpm}{\begin{pmatrix}}
\newcommand{\epm}{\end{pmatrix}}
\newcommand{\bvm}{\begin{vmatrix}}
\newcommand{\evm}{\end{vmatrix}}
\newcommand{\ba}{\begin{array}}
\newcommand{\ea}{\end{array}}

\def\a{\alpha}

\def\t{\tau}

\def\ra{\rightarrow}

\def\lra{\longrightarrow}

\def\inv#1{\frac{1}{#1}}

\def\mean#1{\langle #1 \rangle}

\newcommand{\wt}[1]{\widetilde{#1}}

\newcommand{\re}[1]{(\ref{#1})}

\newcommand{\eref}[1]{Eq.~(\ref{#1})}
\newcommand{\fref}[1]{Fig.~\ref{#1}}
\newcommand{\fsref}[1]{Figs.~\ref{#1}}

\newcommand{\Rsref}[1]{Refs.~\onlinecite{#1}}
\newcommand{\sref}[1]{Sect.~\ref{#1}}

\newcommand{\esref}[1]{Eqs.~(\ref{#1})}

\newcommand{\aref}[1]{Appendix~\ref{#1}}
\newcommand{\tref}[1]{Table~\ref{#1}}

\newcommand{\Tr}{\mathrm{Tr}\,}
\def\Z2{\mathbb{Z}_{2}}
\def\S{\mathbb{\Sigma}}
\def\R{\mathbb{R}}

\def\fd#1{\frac{d #1}{d t}}

\def\cn{\textrm{cn}}

\newcommand{\eps}{\epsilon}
\newcommand{\lam}{\lambda}

\makeatletter
\renewcommand*\env@matrix[1][\arraystretch]{%
  \edef\arraystretch{#1}%
  \hskip -\arraycolsep
  \let\@ifnextchar\new@ifnextchar
  \array{*\c@MaxMatrixCols c}}
\makeatother

\begin{document}

\title{Driven-Dissipative Dynamics of  Atomic Ensembles in a Resonant Cavity I: Nonequilibrium Phase Diagram and Periodically Modulated Superradiance}

\author{Aniket Patra$^1$, Boris L. Altshuler$^2$ and Emil A. Yuzbashyan$^1$}
\affiliation{$^1$Department of Physics and Astronomy, Rutgers University, Piscataway, NJ 08854, USA \\ 
$^2$Department of Physics, Columbia University, New York, NY 10027, USA}

\begin{abstract} 

We study the dynamics of two ensembles of atoms (or equivalently, atomic clocks) coupled to a bad cavity and pumped incoherently by a Raman laser. Our main result is the nonequilibrium phase diagram for this experimental setup in
terms of two parameters - detuning between the clocks and the repump rate. There are three main phases - trivial steady state (Phase I), where all atoms are maximally pumped, nontrivial steady state corresponding to monochromatic superradiance (Phase II), and amplitude-modulated superradiance (Phase III). Phases I and II are fixed points of the mean-field dynamics, while in most of Phase III stable attractors are limit cycles. Equations of motion possess an axial symmetry and a $\Z2$ symmetry with respect to the interchange of the two clocks. Either one or both of these symmetries are spontaneously broken in various phases. The trivial steady state loses stability via a supercritical Hopf bifurcation bringing about a $\Z2$-symmetric limit cycle. The nontrivial steady state goes through a subcritical Hopf bifurcation responsible for coexistence of monochromatic and amplitude-modulated superradiance. Using Floquet analysis, we show that the $\Z2$-symmetric limit cycle eventually becomes unstable and gives rise to two $\Z2$-asymmetric limit cycles via a supercritical pitchfork bifurcation. Each of the above attractors has its own unique fingerprint in the power spectrum of the light radiated from the cavity. In particular, limit cycles in Phase III emit frequency combs - series of equidistant peaks, where the symmetry of the frequency comb reflects the symmetry of the underlying limit cycle. For typical experimental parameters, the spacing between the peaks is several orders of magnitude smaller than the monochromatic superradiance frequency, making the lasing frequency highly tunable.

\end{abstract}

\maketitle

\tableofcontents

\newpage

\section{Introduction and Summary of Main Results}

Atomic condensates trapped in optical cavities host a range of intriguing collective behaviors, such as superradiance  \cite{SR_Exist_1, SR_Exist_2, SR_Exist_3, SR_Exist_4}, Bragg crystals \cite{Bragg_1, Bragg_2, Bragg_3} and collective atomic recoil lasers~\cite{CARL_1, CARL_2, CARL_3}. Here pumping and dissipation play a key role. For example, they facilitate superradiance --  a macroscopic population of photons in the cavity mode~\cite{SR_Debate_1,Keeling_Ref_1,Keeling_Ref_2,Keeling_Ref_3, Keeling_Ref_4,Keeling_Ref_5,Keeling_Ref_6, Keeling_Ref_7,Keeling_Ref_8,Keeling_Ref_9, Keeling_Ref_10,Keeling_Ref_11,Keeling_Ref_12, Keeling_Ref_13,Keeling_Ref_14, Keeling_1, Keeling_2}.  In addition to providing a platform for studying  far-from-equilibrium physics,   atom-cavity systems  have interesting technological applications, such as an ultrastable superradiant   laser~\cite{Holland_One_Ensemble_Expt_1, Holland_One_Ensemble_Expt_2,Holland_One_Ensemble_Theory_1, Holland_One_Ensemble_Theory_2}. In this setup a large number  of ultracold atoms exchange photons with an isolated mode of a ``bad cavity".  This type of cavity leaks photons to the environment very fast.  The phases of  individual atoms  synchronize  in the process resulting in superradiance.  In the bad cavity limit, the frequency of the emitted light is solely determined by the atomic transitions, thereby   circumventing  thermal noise plaguing   lasers operating in the good cavity limit.  Moreover, if $N$ is the number of atoms in   a superradiant laser, the intensity of the emitted light is proportional to $N^{2}$; unlike a usual laser, where  atoms do not radiate in a correlated fashion and the intensity is proportional to $N$. Recent work proposed to utilize such high intensity light for an atomic clock \cite{Holland_atomic_clock}.
Having this in mind, below we use the terms ``atomic ensemble" (coupled to a bad cavity mode) and ``atomic clock" interchangeably. 

\begin{figure}[tbh!]
\begin{center}
\includegraphics[scale = 0.5]{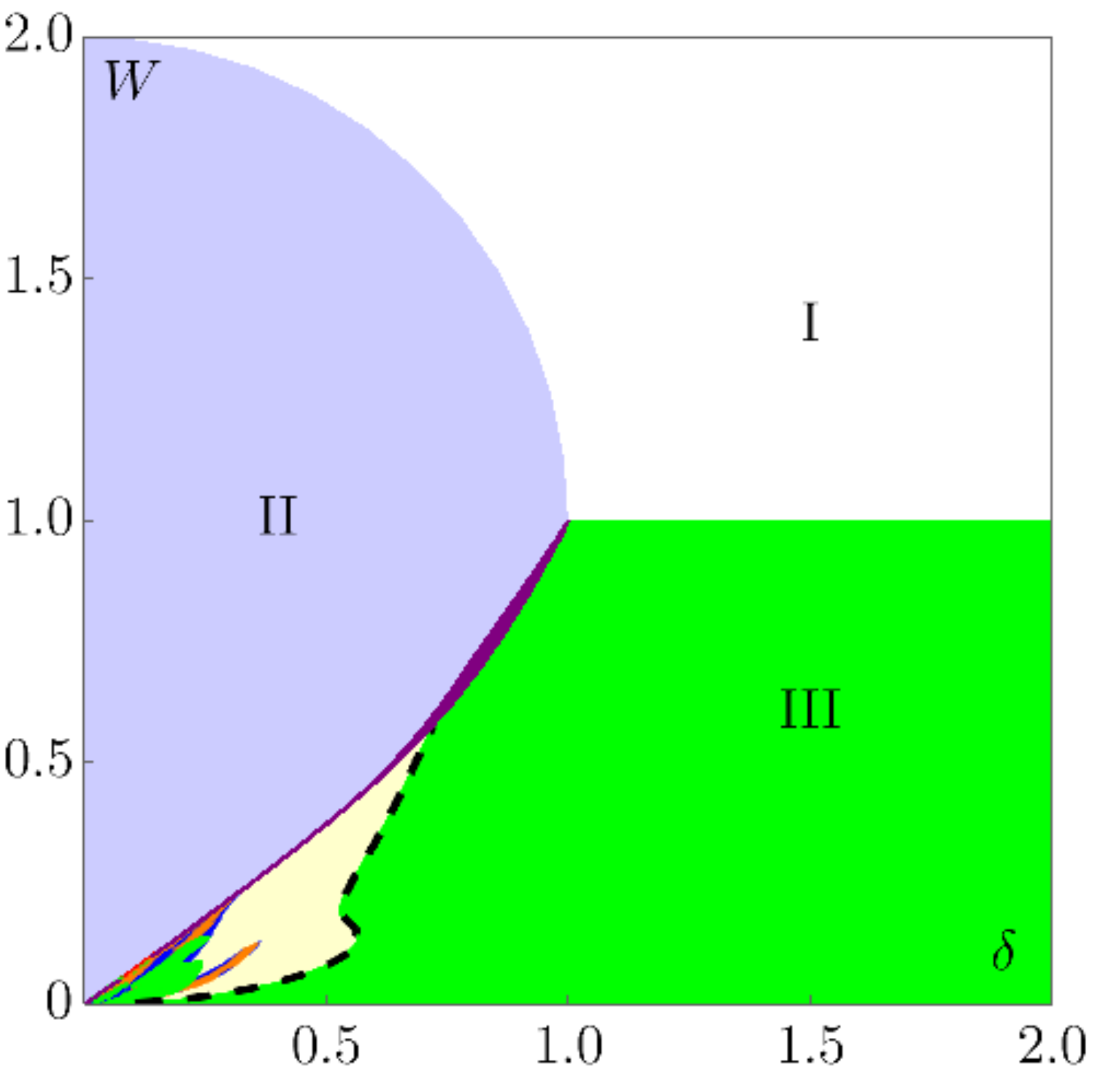}
\caption{Nonequilibrium phase diagram of two atomic ensembles resonantly coupled to a bad cavity, where $\delta$ and $W$  are the detuning  between the ensembles and incoherent repump rate, respectively, in units of the collective decay rate.  In Phase I (the normal phase) atoms interfere destructively and produce no light. In Phase II, we observe monochromatic superradiance, i.e., the ensembles synchronize   and radiate light with the mean frequency. Phase III features various types of amplitude-modulated superradiance -- frequency combs (limit cycles) in green and  yellow subregions as well as quasiperiodic and chaotic behaviors near the origin (blue and orange).  Inside the green   part of region III the limit cycles possess a $\Z2$ symmetry with respect to the interchange of the two ensembles (see the text), which is  spontaneously broken  to the left of the dashed line. We depict the region near the boundary of II and III, where monochromatic and amplitude-modulated superradiance coexist, in purple.}
\label{Phase_Diagram}
\end{center}
\end{figure}

External driving and dissipation are also major factors  in exciton-polariton condensates confined inside  semiconductor microcavities \cite{Wouters_1, Wouters_2, Keeling,Wertz, Littlewood, Balili, Lai}.  The condensate often fragments into several interacting droplets due to intrinsic inhomogeneities. Already two such   droplets   reveal several complicated synchronized phases. Besides lasing with a fixed frequency, which is referred to as ``weak lasing" \cite{Kavokin_1, Boris_Steady_State} and is similar to  the usual, monochromatic  superradiance, the droplets can synchronize to produce a frequency comb, i.e., light with periodically modulated amplitude \cite{Boris_Freq_Comb,Kavokin_3}.

\begin{figure}[tbp!]
\footnotesize
\centering
\subfloat[\large (a)]{\label{Power_Spectrum_Symmetric_LC_Intro}\includegraphics[scale=0.45]{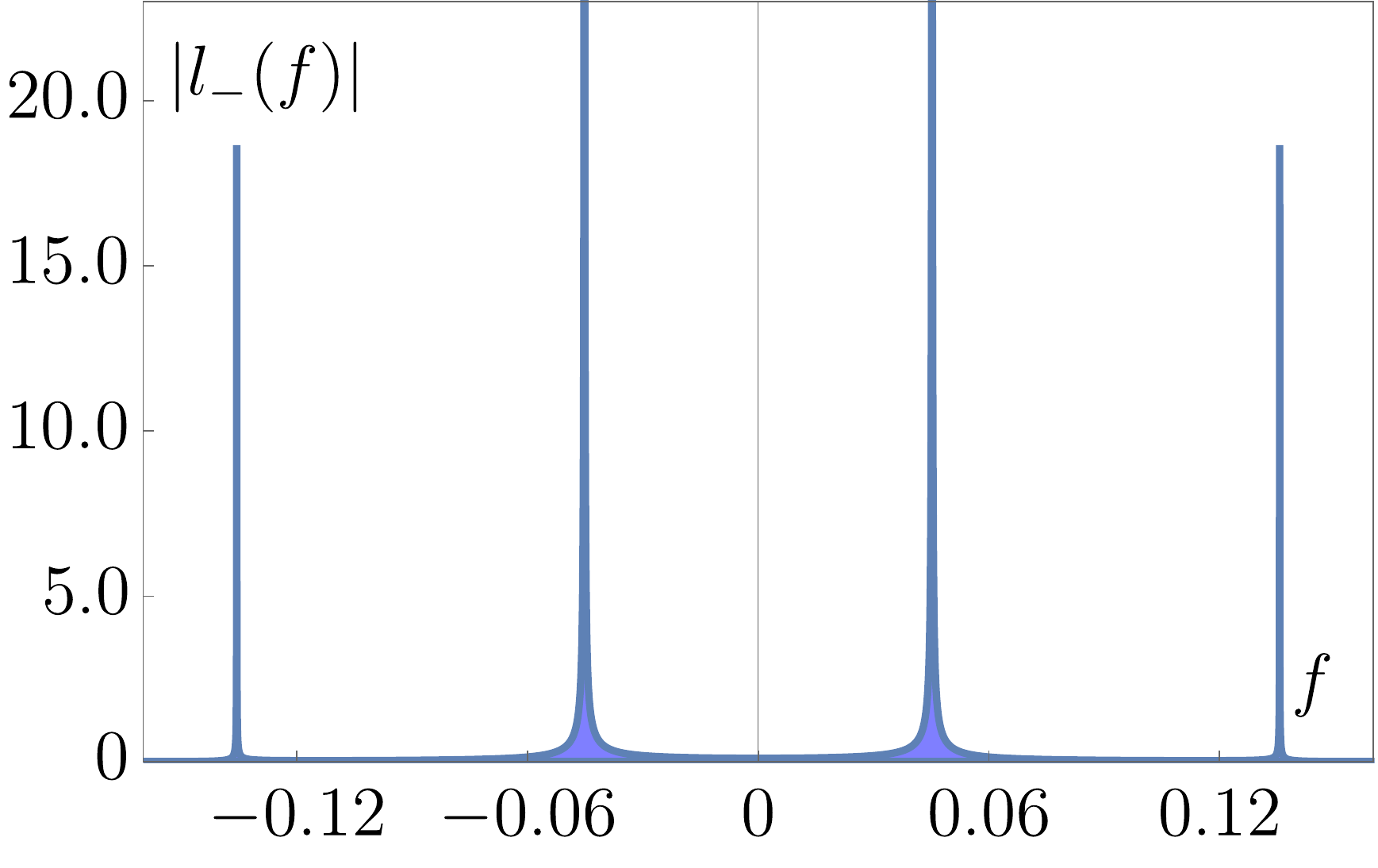}}\\
\subfloat[\large (b)]{\label{Power_Spectrum_Asymmetric_LC_Intro}\includegraphics[scale=0.45]{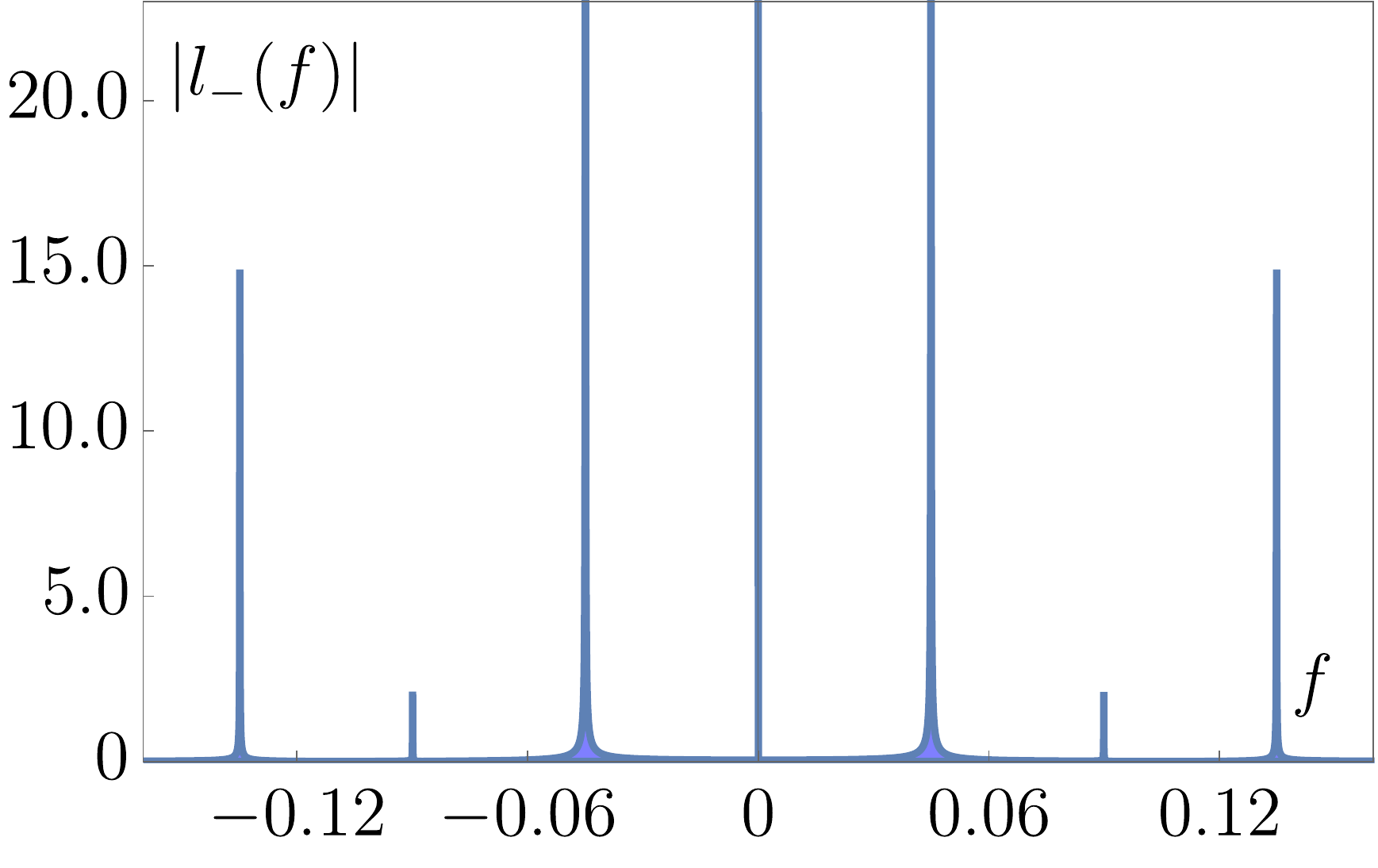}}\\
\subfloat[\large (c)]{\label{Power_Spectrum_No_Reflection_Intro}\includegraphics[scale=0.45]{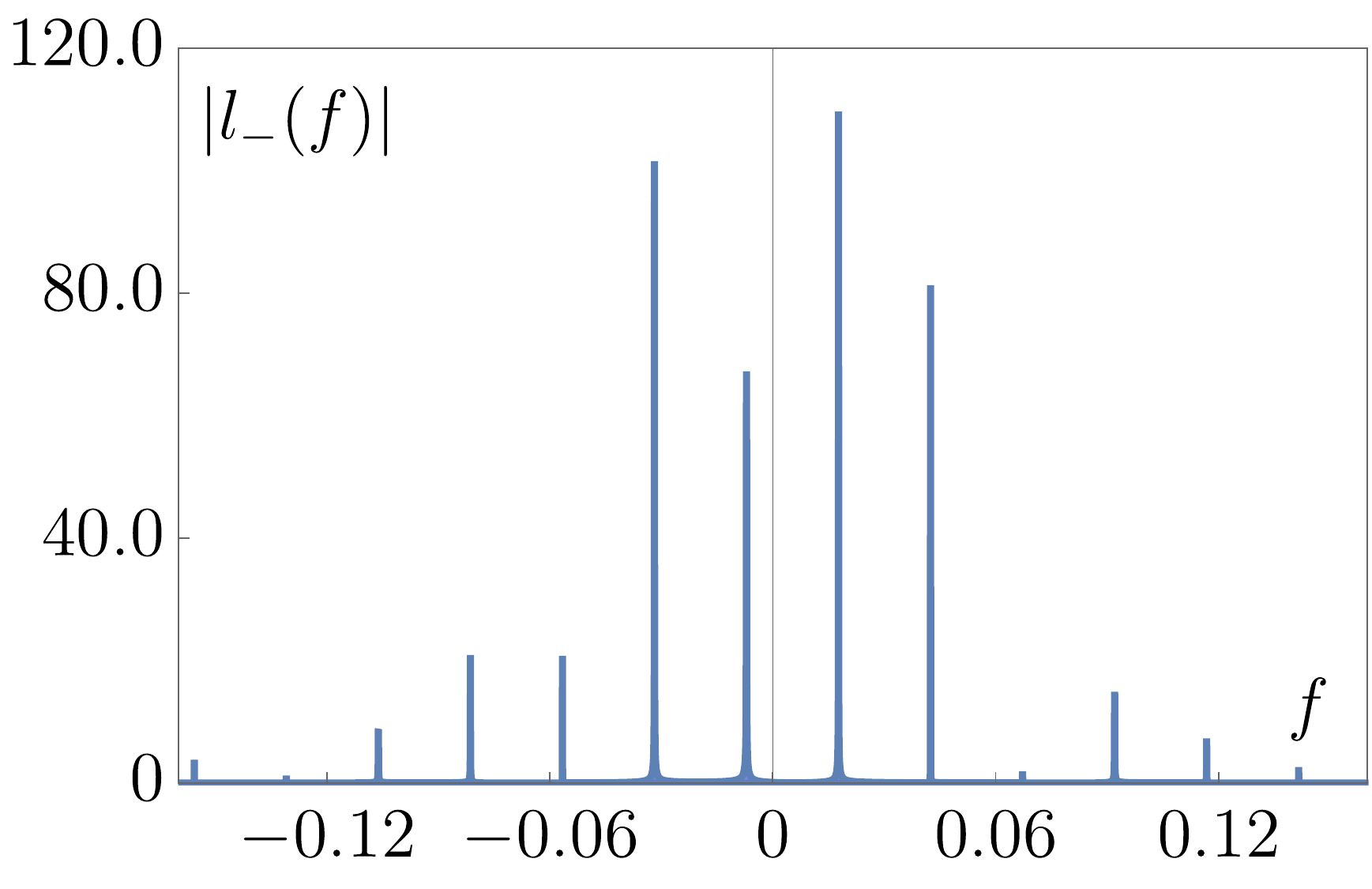}}
\caption{Power spectra of  periodically modulated superradiance in a rotating frame with the monochromatic superradiance frequency set as the origin. The unit of frequency $(f)$ is  the collective decay rate $N\Gamma_{c}$. Top to bottom, $\Z2$-symmetric limit cycle ($\delta = 0.50$, $W = 0.0802$) and two $\Z2$-symmetry-broken limit cycles [($\delta, W) = (0.49, 0.0802)$ and $(\delta, W) = (0.225, 0.05)$].     Both in \textbf{(a)} and \textbf{(b)} the fundamental frequency is  $f_{0} \approx 0.044$. Unlike  \textbf{(a)}, where only  odd harmonics are present, in \textbf{(b)}  even harmonics appear   (most prominently at zero). In spite of both not having the $\Z2$-symmetry,   \textbf{(c)}   visibly breaks the reflection symmetry about the vertical axis and features an overall frequency shift unlike   \textbf{(b)}. }
\label{Power_Spectrum_Intro}
\end{figure}

Recent research points out that two atomic clocks Rabi coupled to the same optical cavity mode synchronize and radiate with a common frequency \cite{Holland_Two_Ensemble_Theory, Holland_Two_Ensemble_Theory_Followup, Holland_Two_Ensemble_Expt, Shammah}. This is an analog of the weak lasing phenomenon and is reminiscent of the original synchronization experiment performed by C. Huygens in the 18th century.  He studied the long-time dynamics of the pendulums  of two clocks suspended from a common support and observed that after some time their phases and frequencies  synchronize  \cite{Pikovsky}.   In this paper, we map out the nonequilibrium phase diagram of two atomic ensembles  in a bad cavity, see \fref{Phase_Diagram}. In addition to  monochromatic  superradiance, we discover a plethora of fascinating dynamical behaviors -- periodic, quasiperiodic and chaotic modulations  of superradiance amplitude. Here we  focus on monochromatic and  periodic regimes, leaving more complicated behaviors for later~\cite{Patra_2, Patra_3}. An interesting feature of the periodically modulated superradiance, where the power spectra are frequency combs (see \fref{Power_Spectrum_Intro}), is that it makes the frequency of the ultrastable superradiant laser tunable.

\begin{figure*}[tbp!]
\centering
\subfloat[\large (a)]{\label{Set_up}\includegraphics[scale=0.5]{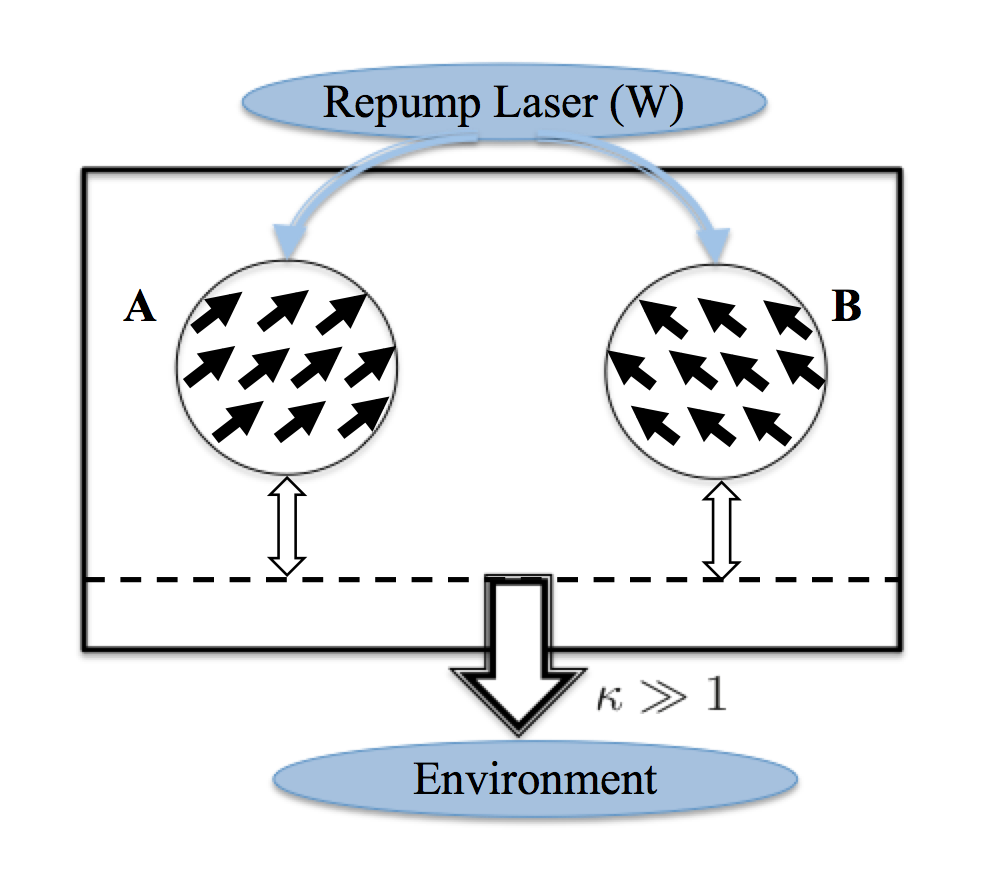}}\qquad\qquad
\subfloat[\large (b)]{\raisebox{1.7cm}{\includegraphics[scale=0.5]{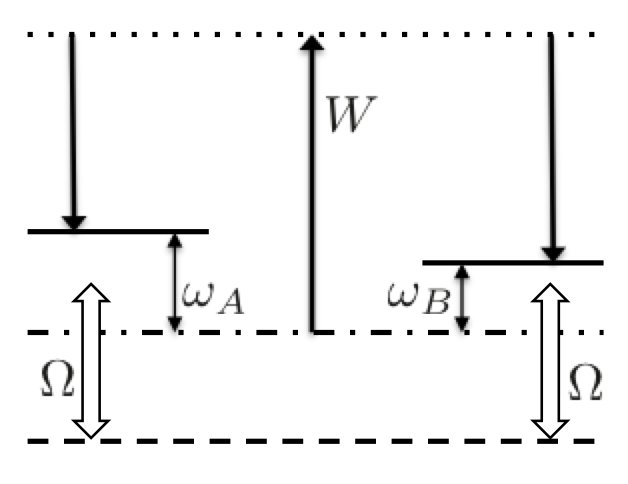}}\label{Energy_Level}}
\caption{Schematics of the setup \textbf{(a)} and energy level diagram \textbf{(b)}.  Two atomic ensembles A and B  couple with Rabi coupling $\Omega$ (double-headed block arrows) to a heavily damped cavity mode (dashed line).   Cavity intensity decays with a rate  $\kappa$.  The atoms (shown with solid arrows) in the two ensembles  are effective two-level systems    with level splittings   $\omega_{A}$ and $\omega_{B}$. The dot-dashed line in \textbf{(b)} shows their shared ground state.   The atoms are pumped incoherently  from the ground state   to their excited states via a third metastable state at a rate $W$.}
\label{Intro_Cartoon}
\end{figure*}

In the reminder of this section we describe our setup and summarize main results.
We model the two atomic ensembles $(\tau = A, B)$ coupled to each other through a bad cavity mode, in the presence of dissipation and pumping (see \fref{Intro_Cartoon}), by the following master equation for the density matrix  $\rho$:
\beg
\begin{split}
\dot{\rho} = -\imath \big[\hat{H}, \rho\big] + \kappa\mathcal{L}[a]\rho + W\sum_{\tau = A,B}\sum_{j = 1}^{N}\mathcal{L}[\hat{\sigma}^{\tau}_{j+}]\rho, \\  
\hat{H} = \omega_{0}\hat{a}^{\dagger}\hat{a} + \!\!\! \sum_{\tau = A,B}\big[\omega_{\tau}\hat{S}_{z}^{\tau} + \frac{\Omega}{2}\big(\hat{a}^{\dagger}\hat{S}_{-}^{\tau}+\hat{a}\hat{S}_{+}^{\tau}\big)\big],   
\end{split}
\label{Full_Master}%
\en        
where the Hamiltonian for the system without  energy nonconserving processes is $\hat{H}$ and the creation (annihilation) operator for the cavity mode $\omega_{0}$ is $\hat{a}^{\dagger} (\hat{a})$. Each  ensemble contains $N (\approx 10^{6})$ atoms of the same type (e.g.,  $\prescript{87}{}{\textrm{Sr}}^{}_{}$ or $\prescript{87}{}{\textrm{Rb}}^{}_{}$). We regard the atoms as two-level systems and only focus on the two atomic energy levels  most strongly coupled to the cavity mode. As a result, it is sufficient to represent the two ensembles with two collective spin operators 
\beg
\hat{S}^{A,B}_{z} = \frac{1}{2}\sum_{j = 1}^{N}\hat{\sigma}^{A,B}_{jz}, \quad \hat{S}^{A,B}_{\pm} = \sum_{j = 1}^{N}\hat{\sigma}^{A,B}_{j\pm},
\en
 where the Pauli operators $\hat{\bm \sigma}_j$ stand for individual atoms.  Level spacings $\omega_{A}$ and $\omega_{B}$ of the two-level  atoms in ensembles A and B, respectively, are controlled by separate Raman dressing lasers   \cite{Holland_Two_Ensemble_Expt}.

Besides the atom-cavity coherent coupling, we consider   two energy nonconserving processes: (1) decay of the cavity mode with a rate $\kappa$ and (2) incoherent pumping of the atoms with a transverse laser at an effective repump rate $W$.   We model these processes via  Lindblad superoperators acting on the density matrix
\beg 
\mathcal{L}[\hat{O}]\rho = \frac{1}{2}\big(2\hat{O}\rho \hat{O}^{\dagger} - \hat{O}^{\dagger}\hat{O}\rho -\rho\hat{O}^{\dagger}\hat{O}\big).
\label{Lindblad_def}
\en%
In the bad cavity  regime $\kappa \gg 1$, we neglect other sources of dissipation, such as spontaneous emission and background dephasing.

Our final goal is to analyze the light emitted by the cavity. To this end,   we  eliminate the cavity mode using the adiabatic approximation,  which is exact in the $\kappa \ra \infty$ limit, and then derive the following mean-field equations  of motion   in \sref{sec: Mean-Field Equations} that describe the dynamics of the system in terms of two classical spins $\bm s^A$ and $\bm s^B$:
 \begs
\bea
\dot{s}^{\t}_{\pm} &=& \biggl(\pm\imath \omega_{\t}- \frac{W}{2}\biggr)s^{\tau}_{\pm} + \frac{1}{2}s^{\t}_{z}l_{\pm}, \\ \label{REOMPM}
\dot{s}^{\t}_{z} &=&  W\big(1 - s^{\t}_{z}\big) - \frac{1}{4}s^{\t}_{+}l_{-} - \frac{1}{4}s^{\tau}_{-}l_{+}, \label{REOMz} 
\eea 
\label{Mean-Field_1}%
\ens%
where  
\beg
s^{\t}_{\pm} = s_{x}^{\tau} \pm \imath s_{y}^{\tau} = \frac{2}{N} \mean{\hat{S}_{\pm}^{\tau}},\quad s^{\t}_{z} = \frac{2}{N}\mean{\hat{S}_{z}^{\tau}},
\label{classicalspindef}
\en
 are  components of the classical spins and
 \beg
 \bm{l} = \sum_\tau \bm s^\tau=\bm s^A+\bm s^B,
 \en
  is the total classical spin. 
 In the coordinate frame rotating  with the angular frequency $\frac{1}{2}\sum_{\tau}\omega_{\tau}$ around the $z$-axis,  the level spacings are
\beg 
\omega_{A} =  -\omega_{B} = \frac{\delta}{2},\quad\delta = \omega_{A} - \omega_{B},
\label{Detuning}
\en 
where $\delta$ is the ``detuning" between the two level-spacings. 
We note that  \eref{Mean-Field_1} is valid for an arbitrary number $n$ of spatially separated ensembles of $N$ atoms each, identically coupled to the cavity.  In \sref{sec: Fokker-Planck Equation}, we further derive the Fokker-Planck equation  governing quantum fluctuations over the mean-field dynamics for $n$  ensembles.
In the  equations of motion  and from now on we set the units of time and energy so that
 \beg
 N\Gamma_c=1,
 \label{unitsdef}
 \en
 where $\Gamma_c=\Omega^2/\kappa$ is the collective decay rate. Thus, the mean-field dynamics and  nonequilibrium phase diagram for two atomic ensembles depend on only two dimensionless parameters $\delta$ and $W$. In a typical experiment $N\Gamma_{c}$ is approximately $1$ kHz, whereas $\delta$ and $W$ can be varied between zero and $4\pi$ MHz~ \cite{Holland_One_Ensemble_Theory_1, Holland_One_Ensemble_Expt_1, Holland_One_Ensemble_Expt_2, Holland_Two_Ensemble_Expt}.    
 
Equations of motion (\ref{Mean-Field_1})  are axially symmetric, i.e., they are invariant with respect to  $\bm{s}^{\tau}\to\R(\phi)\cdot \bm{s}^{\tau}$, where $\R(\phi)$ is  a rotation by $\phi$ about the $z$-axis,  
\beg 
\R(\phi): (s^{\tau}_{\pm}, s^{\tau}_{z}) \longrightarrow (s^{\tau}_{\pm}e^{\pm \imath \phi}, s^{\tau}_{z}).
\label{Rot_Def}  
\en 
 Using axial symmetry and introducing a set of new variables,
\beg 
 s^{\tau}_{\pm} = s^{\tau}_\perp e^{\pm\imath\phi_{\tau}}, \quad \phi_{A} = \Phi + \varphi, \quad \phi_{B} = \Phi - \varphi,
\label{New_Var_Rot}
\en%
 where $\varphi$ is defined modulo $\pi$, we factor out the evolution of the overall phase $\Phi$ from \eref{Mean-Field_1},
\beg 
\dot{\Phi} = \frac{1}{2}\big(\omega_{A} + \omega_{B}\big) + \frac{\sin{2\varphi}}{4}\bigg(\frac{s^{A}_{z}s^{B}_\perp }{s^{A}_\perp} - \frac{s^{B}_{z}s^{A}_\perp }{s^{B}_\perp }\bigg).
\label{Mean-Field_Group1}
\en%
Note,  $\dot{\Phi}$  as well as the equations of motion for the remaining   five variables,
 \beg
\begin{split}
\dot{s}^{A}_\perp = - \frac{W}{2}s^{A}_\perp + \frac{s^{A}_{z}}{2}\big(s^{A}_\perp + s^{B}_\perp\cos{2\varphi}\big), \\ 
\dot{s}^{B}_\perp = - \frac{W}{2}s^{B}_\perp + \frac{s^{B}_{z}}{2}\big(s^{A}_\perp\cos{2\varphi} + s^{B}_\perp\big), \\ 
\dot{s}^{A}_{z} =  W\big(1 - s^{A}_{z}\big) - \frac{s^{A}_{\perp}}{2}\big(s^{A}_\perp + s^{B}_\perp\cos{2\varphi}\big), \\ 
\dot{s}^{B}_{z} =  W\big(1 - s^{B}_{z}\big) - \frac{s^{B}_{\perp}}{2}\big(s^{A}_\perp\cos{2\varphi} + s^{B}_\perp\big), \\ 
\dot{\varphi}= \frac{1}{2}\big(\omega_{A} - \omega_{B}\big) - \frac{\sin{2\varphi}}{4}\bigg(\frac{s^{A}_{z}s^{B}_\perp}{s^{A}_{\perp}} + \frac{s^{B}_{z}s^{A}_\perp}{s^{B}_{\perp}}\bigg),
\label{Mean-Field_Group4}
\end{split}
\en
do not contain $\Phi$ as a consequence of the  axial symmetry. This ensures that the values of $s^{\tau}_\perp, s_{z}^{\tau}$ and $\varphi$ at subsequent times do not depend on the initial value of $\Phi$.

For two ensembles, \eref{Mean-Field_1} and \eref{Mean-Field_Group4} also possess a $\Z2$ symmetry. \eref{Mean-Field_1}  remains unchanged upon the replacement $\bm{s}^{\tau}\to \S \circ \R(\phi_{0})\cdot\bm{s}^{\tau}$, i.e., a rotation of the spins by a fixed angle $\phi_{0}$ about the $z$-axis, followed by an interchange of the two atomic clocks with a simultaneous change of the sign of the $y$-component  
\beg  
\mathbb{\Sigma}:\big( s_{\pm}^{A}, s_{z}^{A}, s_{\pm}^{B}, s_{z}^{B} \big) \longrightarrow \big( s_{\mp}^{B}, s_{z}^{B}, s_{\mp}^{A}, s_{z}^{A} \big).
\label{Z2}
\en%
 $\Z2$-symmetric asymptotic solutions (attractors)   obey $\bm{s}^{\tau} = \mathbb{\Sigma}\circ \R(\phi_{0}) \cdot \bm{s}^{\tau}$, where $\phi_{0}$ depends on the initial condition.  This condition defines a confining 4D submanifold of the full 6D phase space defined (independently of the initial conditions) by  the following two  constraints:
\beg 
\begin{split} 
 s_\perp^A =  s_\perp^B,\quad s_{z}^{A} = s_{z}^{B}.
\end{split} 
\label{Gen_Z2_Symm_MF}%
\en%
 In a reference frame rotated by $\phi_{0}$ around the $z$-axis,  the $\Z2$-symmetric solutions satisfy
\beg 
s^{A}_{x} = s^{B}_{x},\quad s^{A}_{y} = -s^{B}_{y},\quad s^{A}_{z} = s^{B}_{z}.
\label{Z2_Expl}
\en%
These three constraints define an invariant 3D submanifold, which is obtained by considering a fixed value of $\phi_{0}$ along with \eref{Gen_Z2_Symm_MF}. An initial condition on this submanifold restricts the future dynamics on the same. The geometric meaning of the $\Z2$ transformation is a reflection of the spin configuration through the plane containing the total spin $\bm l$ and the $z$-axis. In \eref{Mean-Field_Group4} it amounts to an interchange of $s_\perp^A$ with  $s_\perp^B$ and of $s_{z}^{A}$ with $s_{z}^{B}$. In \eref{Mean-Field_Group1}, one additionally needs to replace $\Phi\to -\Phi$ and set $\omega_A+\omega_B=0$.

For  $\Z2$-symmetric attractors,  \eref{Z2_Expl} implies that in a suitable coordinate frame $l_{x}  = 2s^{A}_{x}$, $l_{y}  = 0$, and  \eref{Mean-Field_1} yields   closed equations of motion for  $\bm{s}^{A}$,
\begs
\bea
\dot{s}_{x} &=& - \frac{\delta}{2}s_{y}- \frac{W}{2}s_{x} + s_{z}s_{x},\label{ReducedX} \\ 
\dot{s}_{y} &=& \frac{\delta}{2}s_{x} - \frac{W}{2}s_{y}, \label{ReducedY}\\ 
\dot{s}_{z} &=&  W\big(1 - s_{z}\big) - \big(s_{x}\big)^{2}, \label{ReducedZ} 
\eea 
\label{Symm_One_Spin_Eqn}
\ens%
where we  dropped the superscript.  Therefore, for a $\Z2$ symmetric attractor it is sufficient to study \eref{Symm_One_Spin_Eqn}. Even though
\eref{Symm_One_Spin_Eqn} describes a motion of  one spin, it is    much more complex than   \eref{Mean-Field_1} for a single atomic ensemble. Indeed, as we show in Appendix~\ref{singleclock}, the phase diagram for the latter case is effectively 1D and contains only two phases (monochromatic superradiance and the normal phase).

 Mean-field dynamics of several ensembles in a bad cavity have two types of fixed points, which we derive by equating the time derivatives to zero in \eref{Mean-Field_1}. The  first one   is
 \beg 
s^{\tau}_{\pm} = 0, \quad s^{\tau}_{z} = 1.
\label{TSS}%
\en
This is a normal state, where  atomic clocks are not synchronized and no light emanates from the cavity $(\mean{\hat{a}^{\dagger}\hat{a}}\propto|l_{-}|^{2} = 0)$. In this phase (region I in \fref{Phase_Diagram}),  the atoms are maximally pumped $(s^{\tau}_{z} = 1)$ and the cavity mode is not populated. We call this fixed point   the trivial steady state (TSS).  It is the only attractor with both the axial and  $\Z2$ symmetry.

 The second  type (non-trivial steady state  or NTSS) corresponds to  monochromatic superradiance  (region II of \fref{Phase_Diagram}). Here the damping is nontrivially balanced by the external pumping leading to a macroscopic population of the cavity mode.   For two ensembles the NTSS reads
\beg 
\begin{aligned}
s^{A}_{-} &= e^{-\imath(\Phi + \varphi)}\frac{l_{\perp}}{2}\sqrt{1 + \frac{\delta^{2}}{W^{2}}}, \\
s^{B}_{-} &= e^{-\imath(\Phi - \varphi)}\frac{l_{\perp}}{2}\sqrt{1 + \frac{\delta^{2}}{W^{2}}}, \\
s^{A}_{z} &= s^{B}_{z} = \frac{\delta^{2} + W^{2}}{2W},
\end{aligned}
\label{NTSS}
\en%
where 
\beg 
l_{\perp} = \sqrt{2\big(1- (W-1)^{2}- \delta^{2}\big)}, \quad
\varphi = \arctan{\frac{\delta}{W}},
\label{NTSS_1}
\en%
and $\Phi$ is an arbitrary angle.  The NTSS comes to pass when the TSS loses  stability   on the quarter-circular arc  depicting the boundary of Phases I and II.   It spontaneously breaks   the axial symmetry, while retaining the  $\Z2$ symmetry. Specifically, it is invariant under $\S \circ \R(-\Phi)$, where $\Phi$ is the   overall phase   in \eref{NTSS}.
  NTSS is not a single point, but a  one-parameter ($\Phi$) family of fixed points related to each other by  rotations around the $z$-axis.   \esref{Detuning}, \re{Mean-Field_Group1} and (\ref{NTSS}) imply $\dot{\Phi} = 0$.  
When $\omega_{A} +\omega_{B}\ne0$, the  axial symmetry of the NTSS is restored, and it becomes a  trivial limit cycle with $\dot{\Phi} \ne0$.  Note also that for \eref{Mean-Field_Group4} the NTSS is a single fixed point, since $s_\perp^A$ and $s_\perp^B$ are independent of $\Phi$, while
for \eref{Symm_One_Spin_Eqn} it reduces to two fixed points corresponding to $\Phi=0$ and $\pi$ in \eref{NTSS}. 

\begin{figure}[tbp!]
\begin{center}
\includegraphics[scale=0.6]{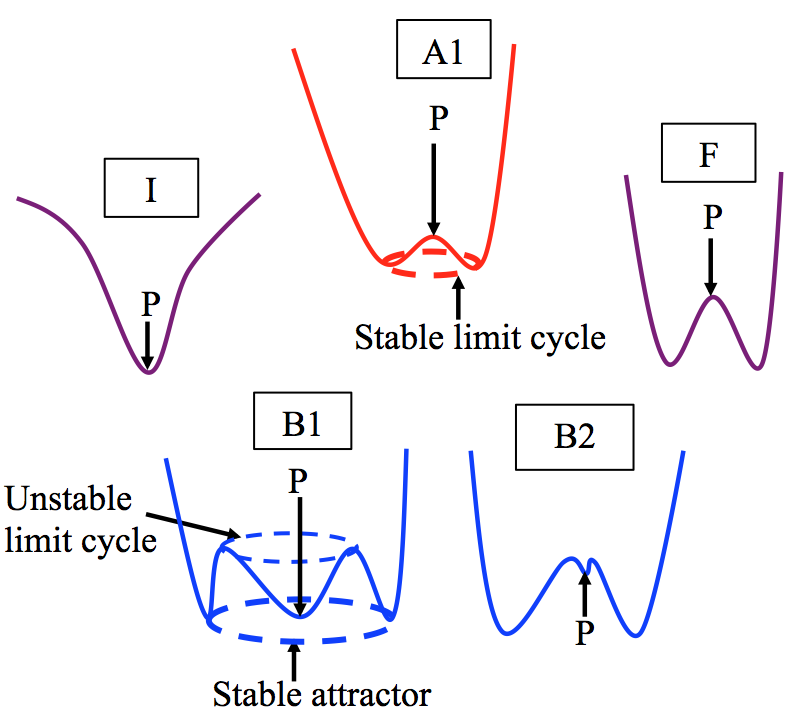}
\caption{Cartoon demonstrating the similarity between Hopf bifurcations and phase transitions. Supercritical (subcritical) Hopf bifurcations are analogous to second (first) order phase transitions. The curves  indicate 2D free energy plots  and the direction of the flow towards stable attractors  in the case of phase transitions and driven-dissipative dynamics, respectively. The leftmost curve depicts a stable equilibrium point  P at the minimum. In a supercritical Hopf bifurcation (I $\rightarrow$ A1 $\rightarrow$ F) the fixed point loses stability by giving rise to a stable   limit cycle. At the bifurcation the limit cycle is infinitesimally small just like the order parameter in a second order phase transition.    In  a subcritical Hopf bifurcation (I$\rightarrow$ B1 $\rightarrow$ B2 $\rightarrow$ F) an unstable limit cycle comes to exist \textit{before} the bifurcation. This  limit cycle acts as a separatrix  between the basins of attraction of the fixed point P and    another stable attractor.}\label{SuperSubCritical}
\end{center}
\end{figure}

  We determine the regions of stability of the TSS and  NTSS in \sref{Linear Stability Analysis of the Fixed Points}.  In \sref{Beyond_Lin_Stability}, we go beyond  linear stability  analysis (introducing the Poincar\'{e}-Birkhoff normal form) to explain the coexistence of NTSS with other attractors. 
 We find that, in addition to the I-II boundary,  TSS  losses stability   via a supercritical Hopf bifurcation (see \fref{SuperSubCritical}) on the $\delta\ge1$ part of the $W=1$ line separating Phases I and III  in \fref{Phase_Diagram}. \textit{After} the bifurcation it gives rise to an infinitesimal limit cycle (frequency comb)   as shown in \fref{supercritical_Evolution}. The NTSS, on the other hand, loses  stability via a subcritical Hopf bifurcation (see \fref{SuperSubCritical} and \fref{subcritical_Evolution}), bringing about an unstable limit cycle \textit{before} the bifurcation \cite{Intro_bifurc, Kuznetsov, Hilborn}. This limit cycle is the separatrix -- the boundary separating the basin of attraction of the NTSS from that of another attractor. As one nears criticality, the size of the separatrix shrinks, and after the bifurcation it disappears altogether. Therefore, any perturbation will push the dynamics far away from the fixed points right after the bifurcation, making it a catastrophic bifurcation. This not only explains the absence of any infinitesimal limit cycle after the bifurcation, but also justifies the coexistence (in the purple region of \fref{Phase_Diagram}) of the NTSS with other  attractors before the bifurcation.  As explained in \fref{SuperSubCritical},  the loss of stability of the NTSS is analogous to the first  (rather than second \cite{Holland_Two_Ensemble_Theory}) order phase transition. 

\begin{figure}[tbp!]
\begin{center}
\includegraphics[scale=0.6]{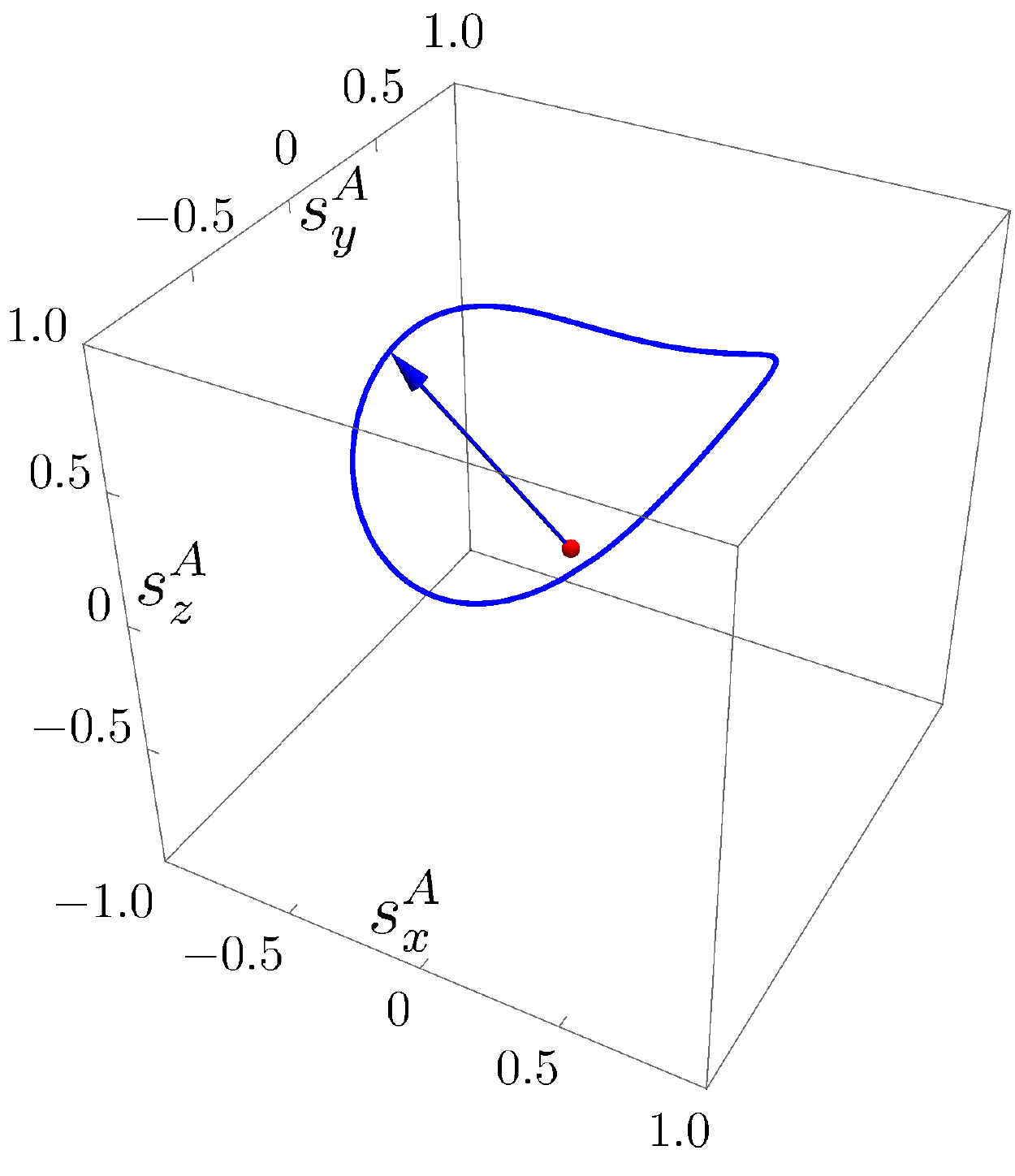}
\caption{ $\Z2$-symmetric limit cycles from the green parts of region III in \fref{Phase_Diagram} resemble  potato chips. Here $(\delta,W) = (1.5, 0.5)$.  We only show  spin $\bm s^{A}$ (blue arrow) representing the dynamics of atomic ensemble A, since    spin $\bm s^B$ is related to $\bm s^A$  by the $\Z2$ symmetry.   In this regime the cavity radiates a frequency comb similar to the one shown in \fref{Power_Spectrum_Symmetric_LC_Intro}. Individual spin trajectories for     $\Z2$-symmetry-broken limit cycles   are similarly potato chip-like shaped, but  the two spins  are no longer tied to each other in a simple way.}
\label{Limit_Cycle}
\end{center}
\end{figure}     

For the sake of completeness, let us mention that  in the absence of pumping (on the $\delta$ axis, sans the origin), the system goes to  a nonradiative fixed point  distinct from the TSS. For $W = 0$, the mean-field equations of motion  \re{Mean-Field_1} reduce to a variant of the Landau-Lifshitz-Gilbert equation. The asymptotic solution only retains the  axial symmetry, where both  spins point along the negative $z$-axis. Finally, at the origin ($\delta = W = 0$), the equations of motion are integrable.  The attractor is a nonradiative fixed point that breaks both  symmetries. We include a detailed discussion of these fixed points in \aref{Other Non-radiative Fixed Points}.

In region III of \fref{Phase_Diagram},  all stable asymptotic solutions of  \eref{Mean-Field_Group4}   are  time-dependent. In particular, there are   limit cycles that lead to  periodically modulated  superradiance or frequency combs. None of them retain the  axial symmetry. The limit cycle in the green part of region III possesses  $\Z2$ symmetry, while the ones in the light yellow subregion break it. A typical $\Z2$-symmetric limit cycle is shown in \fref{Limit_Cycle}. Using    \eref{Symm_One_Spin_Eqn}, we  are able to analytically determine this  limit cycle  in various limits in \sref{Harmonic_Soln}. For example, when either $ W\ll\delta$  or $1-W\ll 1$ (but $\delta$ is not too close to 1), in a suitably rotated frame,
\begs
\bea 
s^{A}_{x} & = s^{B}_{x} \approx & \sqrt{2W(1-W)}\cos{(\omega t - \alpha)}, \\ 
s^{A}_{y}& =-s^{B}_{y} \approx & \sqrt{2W(1-W)}\sin{(\omega t)}, \\
s^{A}_{z}& =s^{B}_{z} \approx & W,
\eea
\label{del>>W_Limit_Intro}%
\ens%
where 
\beg 
\omega = \frac{1}{2}\sqrt{\delta^{2} - W^{2}}, \qquad \tan{\alpha} = \frac{W}{2\omega}.
\en%
Then, the potato chip in  \fref{Limit_Cycle} is flat and normal to the $z$-axis.
 For a comparison of \eref{del>>W_Limit_Intro} with the numerical result, see \fref{Comparison_With_Perturbation}. We also note that the $z$-component of the limit cycle in the   $\delta\gg W$ limit   agrees with the result obtained with the help of quantum regression theorem in Ref.~\onlinecite{Holland_Two_Ensemble_Theory}. Especially interesting is the case when both $W$ and $\delta$ are close to 1, i.e., the vicinity of the tricritical point in \fref{Phase_Diagram}. In this case, the harmonic approximation  \re{del>>W_Limit_Intro} breaks down and the solution for the limit cycle is now in terms of the Jacobi elliptic function cn, see
 \sref{ellipticsect}.

 We analyze the stability of  $\Z2$-symmetric limit cycles  with the help of the Floquet technique  in \sref{Floquet} and find that it becomes unstable as we cross the dotted line in \fref{Phase_Diagram}. As a result, two new, symmetry-broken limit cycles   related to each other by the $\Z2$ symmetry operation emerge, see \sref{SBLC}.

Limit cycles in region III of the phase diagram  are periodic solutions of \eref{Mean-Field_Group4} for given $W$ and $\delta= (\omega_A-\omega)/2$. They are closed curves in the 5D space  with coordinates
 $s_\perp^\tau$, $s_z^\tau$, and $\varphi$ (mod $\pi$). \eref{Mean-Field_Group1} may introduce the second fundamental frequency $\omega_q$ depending on the reference frame and the limit cycle. Indeed,   this equation implies 
 \beg
 \Phi(t)=\Theta t+F(t),\quad \Theta=\frac{1}{2}(\omega_A+\omega_B)+ \omega_q,
 \label{Theta}
 \en
  where $F(t)$ is periodic with the same period as the limit cycle and $\omega_q$ is the zeroth harmonic term in the Fourier series of the second term on the right hand side of  \eref{Mean-Field_Group1}.  When $\Theta\ne0$, the limit cycle precesses with constant angular frequency $\Theta$ in the full 6D space of components
 of $\bm s^A$ and $\bm s^B$, i.e., the corresponding attractor of  \eref{Mean-Field_1} is an axially symmetric 2-torus. If $\Theta=0$, then $\Phi=\mathrm{const}$ and instead of a 2-torus we have a one parameter ($\Phi$) family of limit cycles related to each other via an overall rotation. Each of them  breaks the axial symmetry. Regardless of the value of $\Theta$, we  refer to all above attractors as a limit cycle   at a point $(\delta, W)$ throughout this paper, keeping in mind that it is always a single limit cycle for \eref{Mean-Field_Group4}. We
 are using a rotating frame such that $\omega_A+\omega_B=0$. In addition, $\omega_q=0$ for $\Z2$-symmetric limit cycles, since the second term on the right hand side of  \eref{Mean-Field_Group1} vanishes by  $\Z2$ symmetry. Therefore, $\Theta=0$ in this case.

 Each of the above nonequilibrium phases of  two atomic ensembles coupled to a heavily damped cavity mode has its unique signature in the power spectrum  of the light radiated by the cavity. Experimentally, one measures the autocorrelation function of the radiated complex electric field. The power spectrum is  the Fourier transform of this  function, i.e., the modulus squared  of the Fourier transform of the electric field. In terms of the classical spin variables, we identify the power spectrum to be proportional to $|l_{-}(f)|^{2}$ within mean-field approximation, where
\beg 
l_{-}(f) = \int_{-\infty}^{+\infty}dt\;l_{-}(t)\;e^{2\pi \imath f t},  \quad l_{-} = \sum_{\tau} s_{-}^{\tau}.
\label{l_Minus_f}
\en%
We derive this relationship between $l_-$ and the power spectrum in \aref{Expt} starting from the master equation.  

 The power spectrum of  monochromatic superradiance  (NTSS) consists of a single peak at $f_\mathrm{mc}=(\omega_A+\omega_B)/4\pi$, see \sref{sec: Time Independent superradiance}.  For example, for $\prescript{87}{}{\textrm{Sr}}^{}_{}$ in a bad cavity, the peak appears approximately at $4.3\times 10^{5}$ GHz \cite{3P0-1S0_87Sr}. Subsequently, we show all spectra in a rotating frame and set the above superradiant frequency to be the origin.

For the $\Z2$-symmetric limit cycle the power spectrum is a frequency comb that contains only odd harmonics  (see \fref{Power_Spectrum_Symmetric_LC_Intro}).   Moreover, because of the $\Z2$ symmetry the spectrum possesses a reflection symmetry about the vertical axis. As one moves into the yellow subregion (to the left of the dashed line in \fref{Phase_Diagram}), the $\Z2$ symmetry breaks spontaneously. The power spectra of these limit cycles  display both odd and even harmonics (see \fref{Power_Spectrum_Asymmetric_LC_Intro}). In particular, unlike the spectrum  of the $\Z2$-symmetric limit cycle,  they have a pronounced peak at the origin. Despite the loss of the $\Z2$ symmetry, the breaking of the reflection symmetry about the vertical axis is not so pronounced here. It is however possible to find examples of   symmetry-broken limit cycles,  where the reflection symmetry is visibly broken as in \fref{Power_Spectrum_No_Reflection_Intro}. Another interesting feature of the spectrum in \fref{Power_Spectrum_No_Reflection_Intro} is an overall shift of all frequencies by $ \omega_q/2\pi$, see the discussion around \eref{Theta}.   In \sref{sec: Frequency Combs}, we show  more examples of power spectra (frequency combs)  for different limit cycles.

Frequency combs arising from these limit cycles provide   a range of frequencies (harmonics)   around the main peak -- the  carrier frequency corresponding to the monochromatic superradiance.  We will see that the spacing between  consecutive peaks varies continuously in region III of the phase diagram  and can take arbitrary   values depending on $\delta$ and $W$.  When $\delta$ and $W$ are of order 1,  this spacing is many orders of magnitude smaller (tens of Hertz  for $\prescript{87}{}{\textrm{Sr}}^{}_{}$) than the carrier frequency.  For the ultrastable superradiant laser
mentioned above this implies that its frequency is in principle tunable to within this amount.

\section{Semiclassical Dynamics and Quantum Corrections}\label{Model_&_Approximations}

In this paper we primarily explore the semiclassical dynamics of the system depicted in \eref{Full_Master}, see also \fref{Intro_Cartoon}. As mentioned above \eref{Mean-Field_1}, one obtains the necessary evolution equations after adiabatically eliminating the cavity mode and employing the mean-field approximation  $\mean{\hat{O}_{1}\hat{O}_{2}}\approx\mean{\hat{O}_{1}}\mean{\hat{O}_{2}}$. To derive the Fokker-Planck equation  governing quantum fluctuations, we use the system size expansion (expansion in $ N^{-\frac{1}{2}}$) \cite{Carmichael_1, Carmichael_2, Carmichael_3}.  This also confirms the veracity of the mean-field equations   as we obtain the same equations from the system size expansion.

\subsection{Mean-Field Equations}     
\label{sec: Mean-Field Equations}

We write the mean-field equations in terms of  classical spins $\bm{s}^{\tau}$ introduced in \eref{classicalspindef}, where the average of an operator $\hat O$ is $\Tr [\hat O\rho]$. To obtain the evolution equations for these variables, we first adiabatically eliminate the cavity mode  \cite{Bad_cavity} and then apply the mean-field approximation to the expression  
\beg 
\dot{\bm{s}}^{\tau} = \frac{2}{N}\Tr\big[\hat{\bm{S}}^{\tau}\dot{\rho}\big]= \frac{2}{N}\Tr\big[\hat{\bm{S}}^{\tau}\dot \rho_\mathrm{at} \big],
\label{Evol_Eqn_MF}
\en 
 where $\rho_\mathrm{at} =\Tr_{\! F} (\rho)$ (traced over the cavity mode) is the atomic density matrix.  
 
We start by rewriting the master equation \re{Full_Master} in the interaction representation,
\beg
\begin{split}
\dot{\rho}_{I} = -\imath \big[\hat{H}_{I}, \rho_{I}\big] + \kappa\mathcal{L}[a]\rho_{I} + W\sum_{\tau, j}\mathcal{L}[\hat{\sigma}^{\tau}_{j+}]\rho_{I}, \\  
\hat{H}_{I} = \frac{\delta}{2}\big(\hat{S}_{z}^{A} - \hat{S}_{z}^{B}\big) + \frac{\Omega}{2}\big(\hat{a}^{\dagger}\hat{J}_{-}+\hat{a}\hat{J}_{+}^{+}\big). 
\end{split}
\label{Full_Master_Int}%
\en
Here
\beg
\begin{aligned}
\rho_{I} &= e^{\imath (\hat H_\mathrm{at}  +\hat  H_{F})t}\rho e^{-\imath (\hat H_\mathrm{at}  +\hat  H_{F})t}, \\
\hat  H_\mathrm{at}  &= \frac{(\omega_{A} + \omega_{B})}{2}\big(\hat{S}_{z}^{A}+\hat{S}_{z}^{B}\big),\quad \hat H_{F} = \omega_{0}a^{\dagger}a\\
\hat J_\pm & =\hat S^A_\pm + \hat S^A_\pm.\\
\end{aligned}
\label{Full_Master_Conv}%
\en
Next, we trace out the cavity mode in the above master equation and write the right hand side as a power series in $\Omega^{2} N/\kappa^{2}$. In the bad cavity regime ($\kappa \gg \Omega\sqrt{N}$), retaining only the zeroth order term and neglecting any memory effects, we derive   
\beg
\begin{split}
\dot{\rho}_\mathrm{at}  = -\imath \big[ \hat h, \rho_\mathrm{at} \big] + \Gamma_{c}\mathcal{L}[\hat{J}_{-}]\rho_\mathrm{at}  +W \sum_{\tau, j}^{N}\mathcal{L}[\hat{\sigma}^{\tau}_{j+}]\rho_\mathrm{at},\\ 
\hat h = \frac{\delta}{2}\big(\hat{S}_{z}^{A} - \hat{S}_{z}^{B}\big), \\ 
\end{split}
\label{Effective_Master}%
\en 
 where we introduced the collective decay rate
\beg 
\Gamma_{c} = \frac{\Omega^{2}}{\kappa}.
\label{Collective_Decay_Rate}
\en
This procedure is exact in the limit $\kappa\to\infty$. 

 A heuristic explanation of the adiabatic approximation is as follows. In the interaction representation, the classical equation of motion for the cavity mode is  
\beg 
\frac{d \mean{\hat{a}(t)}}{dt} = -\frac{1}{2}\big(\kappa \mean{\hat{a}(t)} + \imath\Omega \mean{\hat{J}_{-}(t)}\big),
\label{Cavity_Decay}
\en%
 In the bad cavity regime, the cavity mode decays very quickly. As a result,  the time derivative on the left  hand side of \eref{Cavity_Decay}   is negligible and we obtain
\beg 
\mean{\hat{a}} \approx -\frac{\imath \Omega}{\kappa}\mean{\hat{J}_{-}}= -\frac{2\imath \Omega}{N\kappa}l_-.
\label{Ad_Elimination}
\en%
If we extend this equality to the operator level and  replace $\hat{a}$ with $\frac{\imath \Omega}{\kappa}\hat{J}_{-}$ in \eref{Full_Master}, we immediately arrive at \eref{Effective_Master}. From \eref{Ad_Elimination} it is also apparent why,  within the mean-field approach the intensity of emitted light ($\mean{\hat{a}^{\dagger}\hat{a}}$) is proportional to $|l_{-}|^{2}$.

Finally, using \esref{Evol_Eqn_MF}, (\ref{Effective_Master}), and with the help of the mean-field approximation, we derive \eref{Mean-Field_1}. In terms of the $x$, $y$ and $z$ components  \eref{Mean-Field_1} reads
\begs
\bea
\dot{s}^{\tau}_{x} &=& - \omega_{\tau}s^{\t}_{y}- \frac{W}{2}s^{\tau}_{x} + \frac{1}{2}s^{\tau}_{z}l_{x}, \\ \label{REOMX}
\dot{s}^{\tau}_{y} &=& \omega_{\tau}s^{\tau}_{x} - \frac{W}{2}s^{\tau}_{y} + \frac{1}{2}s^{\t}_{z}l_{y}, \\ \label{REOMY}
\dot{s}^{\t}_{z} &=&  W\big(1 - s^{\t}_{z}\big) - \frac{1}{2}s^{\t}_{x}l_{x} - \frac{1}{2}s^{\tau}_{y}l_{y}. \label{REOMZ} 
\eea 
\label{Mean-Field_2}%
\ens%
 Note also that  our choice of units \re{unitsdef} of time and energy  implies that $\Gamma_c$ scales as $N^{-1}$ with the number of atoms $N$.  
This ensures that  the pumping and  decay terms in \eref{Effective_Master} are comparable in magnitude. Moreover, in \sref{sec: Fokker-Planck Equation} we show that assuming $\Gamma_{c} \propto N^{-1}$ helps  achieve proper scaling factors in front of the semiclassical and the Fokker-Planck terms in the system size expansion.

\subsection{Fokker-Planck Equation}     
\label{sec: Fokker-Planck Equation} 

In this subsection, we derive  the Fokker-Planck equations for  quantum fluctuations for $n$ atomic ensembles inside a bad cavity.    The master equation is  \eref{Effective_Master}, but with  
\beg 
\hat h= \sum_{\tau = 1}^{n} \omega_{\tau}\hat{S}_{z}^{\tau}, 
\label{Effective_H_Gen}
\en 
 where $\omega_\tau$ are arbitrary.   In the process, we also rederive the mean-field equations \re{Mean-Field_2}. We  summarize the main steps and  final results, referring the reader to a similar derivation in Refs.~\onlinecite{Carmichael_1,Carmichael_3} for further details.
 
 Define the characteristic function $\chi$,  
 \beg
\chi\big(\bm{\xi}, \bm{\xi^{*}}, \bm{\eta}, t\big)  = \Tr\left[\prod_{\tau}\rho_\mathrm{at} e^{\imath \xi^{*}_{\tau} \hat{S}_{+}^{\tau}}e^{\imath \eta_{\tau} \hat{S}_{z}^{\tau}}e^{\imath \xi_{\tau} \hat{S}_{-}^{\tau}}\right]. 
\label{P_Distribution}
\en
Taking the time derivative of $\chi$ and using \eref{Effective_Master}, we obtain, after some algebra, a partial differential equation for $\chi$. Then, we trade $\chi$ for the Glauber-Sudarshan $\cal P$-distribution function~\cite{Glauber,Sudarshan}, which is the Fourier transform of $\chi$. In other words, we substitute
\beg
\chi=\int d^{2}\bm{v}\int d\bm{m} {\cal P}\big(\bm{v}, \bm{v^{*}}, \bm{m}, t\big) e^{\imath \bm{\xi^{*}}\cdot \bm{v^{*}}}e^{\imath \bm{\xi}\cdot\bm{v}}e^{\imath \bm{\eta}\cdot\bm{m}}\!,
\en
into the partial differential equation for $\chi$. This allows us to make the following replacements:
\begin{equation}
\begin{aligned}
\partial_{\xi_{\tau}} &\ra \imath v_{\tau} ,  & \partial_{\xi^{*}_{\tau}} &\ra \imath v^{*}_{\tau},  &   \partial_{\eta_{\tau}} &\ra \imath m_{\tau} \\
\xi_{\tau} &\ra -\imath\partial_{v_{\tau}},  &  \xi^{*}_{\tau} &\ra -\imath\partial_{v^{*}_{\tau}},  &  e^{\pm \imath \eta_{\tau}} &\ra e^{\pm \partial_{m_{\tau}}}.
\end{aligned}
\label{n_Clock_Chi_to_P}
\end{equation}
After some additional manipulations (integration by parts to shift the differential operators onto ${\cal P}$), we arrive at a partial differential equation for
the distribution function  ${\cal P}\big(\bm{v}, \bm{v^{*}}, \bm{m}, t\big)$ known as the Krammers-Moyal expansion, 
 \beg 
\frac{\partial {\cal P}}{\partial t} = \bigg(\sum_{\tau}\mathbb{L}_{\tau}\bigg){\cal P},
\label{n_Clock_Krammers-Moyal}
\en 
where 
\begin{multline}
\mathbb{L}_{\tau} = \imath\omega_{\tau}\partial_{v_{\tau}}v_{\tau} + \frac{\Gamma_{c}}{2}\mathbb{A}_{\tau}\sum_{\tau^{\prime}}v_{\tau^{\prime}} + \\ + \frac{W}{2}\bigg[\frac{N}{2}\big(e^{\partial_{m_{\tau}}}\mathbb{B}^{2}_{\tau} - 1\big) + \mathbb{C}_{\tau}m_{\tau} + \\ + \partial_{v_{\tau}}\big(2\mathbb{B}_{\tau} - 1\big)v_{\tau} \bigg]  + \textrm{c.c},
\label{n_Clock_L}
\end{multline}
\begin{equation}
\begin{aligned}
\mathbb{A}_{\tau} &= v_{\tau}^{*} - e^{-\partial_{m_{\tau}}}v^{*}_{\tau} + 2\partial_{v_{\tau}}m_{\tau} - \partial^{2}_{v_{\tau}}v_{\tau}, \\
\mathbb{B}_{\tau} &= e^{-\partial_{m_{\tau}}} + \partial_{v_{\tau}}\partial_{v^{*}_{\tau}}, \\
\mathbb{C}_{\tau} &= 1 - e^{-\partial_{m_{\tau}}} + \partial^{2}_{v_{\tau}}\partial^{2}_{v^{*}_{\tau}},
\end{aligned}
\label{n_Clock_Krammers-Moyal_ABC}
\end{equation} 
 and `c.c' stands for complex conjugate. The complex conjugate of $\partial_z$ is $\big(\partial_{z}\big)^{*} \equiv \partial_{z^{*}}$.
 The right-hand side of \eref{n_Clock_Krammers-Moyal} contains derivatives of all orders. 

To separate out the classical motion from the quantum fluctuations, we partition  variables $\big(\bm{v}, \bm{v^{*}}, \bm{m}\big)$   as   
\beg
\begin{aligned} 
v_{\tau} = N\left( \frac{s^{\tau}_{-}}{2} + N^{-\inv{2}}\nu_{\tau} \right),\\
v_{\tau}^{*} = N \left( \frac{s^{\tau}_{+}}{2} + N^{-\inv{2}}\nu_{\tau}^{*} \right), \\ 
m_{\tau} = N\left( \frac{s^{\tau}_{z}}{2} + N^{-\inv{2}}\mu_{\tau} \right),
\end{aligned}
\label{n_Clock_Class_Var_Quant_Fluc}
\en%
where $s^{\tau}_{\pm,z}$ are the classical spins defined in \eref{classicalspindef}, and $\big(\bm{\nu}, \bm{\nu^{*}}, \bm{\mu} \big)$ correspond to  quantum fluctuations over the classical motion.  Let $\bar{{\cal P}}\big(\bm{\nu}, \bm{\nu^{*}}, \bm{\mu}, t\big)$ be the probability density function of these  fluctuations.  By definition  
\begin{multline}
\int d^{2}\bm{v}\int d\bm{m} {\cal P}\big(\bm{v}, \bm{v^{*}}, \bm{m}, t\big) = \\ = \int d^{2}\bm{\nu}\int d\bm{\mu} \bar{{\cal P}}\big(\bm{\nu}, \bm{\nu^{*}}, \bm{\mu}, t\big) \equiv 1.
\label{Dim_Match_Int_1}
\end{multline}%
Using \esref{n_Clock_Class_Var_Quant_Fluc} and (\ref{Dim_Match_Int_1}), we conclude
\beg 
\bar{{\cal P}}\big(\bm{\nu}, \bm{\nu^{*}}, \bm{\mu}, t\big) = N^{3n/2}{\cal P}\big(\bm{v}, \bm{v^{*}}, \bm{m}, t\big).
\label{Dim_Match_Int_2}
\en%
  Substituting \esref{n_Clock_Class_Var_Quant_Fluc} and \re{Dim_Match_Int_2}  into \eref{n_Clock_Krammers-Moyal} and recalling that $N\Gamma_{c} = 1$, we obtain an expansion in powers of $N^{-\frac{1}{2} }$
\begin{multline}
\frac{\partial }{\partial t}\bar{{\cal P}}\big(\bm{\nu}, \bm{\nu}, \bm{\mu}, t\big) = N^{\frac{1}{2}}\big(\textrm{semiclassical part}\big) + \\ + N^{0}\big(\textrm{Fokker-Planck equation}\big) + \mathcal{O}\left(N^{-\frac{1}{2}}\right),
\label{n-clock_System_Size}
\end{multline}%
 where the semi-classical part is
\beg 
  \sum_{\tau}\bigg[\big(\# \big)\partial_{\nu_{\tau}} + \big(\# \big)\partial_{\nu^{*}_{\tau}} + \big(\# \big)\partial_{\mu_{\tau}}\bigg]\bar{{\cal P}}
\label{n-clock_Semi-classical}
\en%
We note that the coefficient at $N^{\frac{1}{2}}$ contains only the first order derivatives  with respect to $\nu_{\tau}$, $\nu^{*}_{\tau}$ and $\mu_{\tau}$; the one at $N^0$ -- the first and second order derivatives  with respect to the same variables; etc. 
In  $N \ra \infty$ limit, the  semiclassical part  needs to vanish. Moreover, the coefficients at each partial derivative in \eref{n-clock_Semi-classical} must vanish separately, because otherwise we would end up with a time-independent constraint on the probability density function $\bar {\cal P}$ of quantum fluctuations. The resulting three conditions are precisely the mean-field equations of motion \re{Mean-Field_1}. 

Finally, neglecting terms of the order $N^{-\frac{1}{2}}$ in \eref{n-clock_System_Size}, we  obtain the   Fokker-Planck equation
\beg 
\frac{\partial \bar {\cal P}}{\partial t} = \bigg(\sum_{\tau}\mathbb{L}^{(2)}_{\tau}\bigg)\bar {\cal P},
\label{n-Clock_Fokker_Planck}
\en 
where 
\begin{multline}
\mathbb{L}^{(2)}_{\tau} = \bigg[\imath\omega_{\tau} + \frac{1}{2}\big(W - s_{z}^{\tau}\big)\bigg]\partial_{\nu_{\tau}}\nu_{\tau} + \frac{1}{2}\partial_{\mu_{\tau}}\mu_{\tau} + \\ + \frac{1}{4}s_{-}^{\tau}l_{-}\partial^{2}_{\nu_{\tau}} + \frac{1}{64}\big[2W(1 - 2s_{z}^{\tau}) + s_{-}^{\tau}l_{+}\big]\partial^{2}_{\mu_{\tau}} +\\ +\frac{W}{2}\partial_{\nu_{\tau}}\partial_{\nu^{*}_{\tau}} + \frac{W}{4}s_{-}^{\tau}\partial_{\nu_{\tau}}\partial_{\mu_{\tau}} + \textrm{c.c}.
\label{n_Clock_L2}
\end{multline}
This operator  does not contain derivatives of orders higher than the second.

\section{Stability Analysis of  TSS and  NTSS}
\label{Fixed_Pt_MF}

In this section, we determine  regions of stability for the TSS  (normal nonradiative phase) and  NTSS (monochromatic superradiance). These fixed points are described by  \esref{TSS} and (\ref{NTSS}), respectively. We show that linear stability analysis of the reduced spin equations  \re{Symm_One_Spin_Eqn} obtains the same regions of stability as that of full equations of motion~\re{Mean-Field_1}. This means that  perturbations that respect the $\Z2$-symmetry destabilize these steady states before or at the same time as the ones that do not. Then, going beyond the linear analysis, we establish that the TSS and NTSS undergo different types of Hopf bifurcations.      

\subsection{Linear Stability Analysis of the Fixed Points}
\label{Linear Stability Analysis of the Fixed Points}

\subsubsection{Jacobian Matrix}

To analyze the stability of a fixed point $({\bm s}^{A}_{0}, {\bm s}^{B}_{0})$ we linearize the right hand side of \eref{Mean-Field_2} about $({\bm s}^{A}_{0}, {\bm s}^{B}_{0})$. This produces a $6\times 6$ Jacobian matrix,
\beg 
\mathbb{J}=\bem[1.5]
 \mathbb{D}^{A} &  \mathbb{X}^{A}\\  
 \mathbb{X}^{B} & \mathbb{D}^{B}
\enm,
\label{6_by_6_Jacobian}
\en
where 
\beg
\begin{aligned}      
 \mathbb{D}^{\tau}&=\bem[1.5]
\frac{1}{2}\big(s^{\tau}_{z0} - W\big) & -\omega_{\tau} & \frac{1}{2}l_{x0}\\  
\omega_{\tau} & \frac{1}{2}\big(s^{\tau}_{z0} - W\big) & \frac{1}{2}l_{y0}\\ 
-\frac{1}{2}\big(l_{x0} + s^{\tau}_{x0}\big) & -\frac{1}{2}\big(l_{y0} + s^{\tau}_{y0}\big) & -W 
\enm, \\     
 \mathbb{X}^{\tau}&=\bem[1.5]
\frac{1}{2}s^{\tau}_{z0} & 0 & 0\\  
0 & \frac{1}{2}s^{\tau}_{z0} & 0\\ 
-\frac{1}{2}s^{\tau}_{x0} & -\frac{1}{2}s^{\tau}_{y0} & 0
\enm .
\label{6_by_6_Jacobian_MQ}
\end{aligned}
\en
Its eigenvalues   are the characteristic values of the fixed point $({\bm s}^{A}_{0}, {\bm s}^{B}_{0})$, while the eigenvectors  are the corresponding characteristic directions.   If a characteristic value has a positive real part at certain point $(\delta, W)$,  the fixed point is unstable. Accordingly, we define the region of stability of a fixed point  as the region in the $\delta-W$ plane, where all  characteristic values have negative real parts.    

\subsubsection{Stability of TSS}
\label{Stability_of_TSS_Sect}

We observe that the TSS exists  everywhere on the $\delta - W$ plane. Substituting $({\bm s}^{A}_{0}, {\bm s}^{B}_{0})$ from \eref{TSS} into \eref{6_by_6_Jacobian}, we determine the characteristic equation as
\begin{multline}
\big[ Q_3(\lambda)\big]^{2}\equiv\bigg[\big(W+\lambda\big) \times \\ \times \bigg(\lambda^{2}+\big(W - 1\big)\lambda + \frac{1}{4}\big(W^{2}+\delta^{2}-2W\big)\bigg)\bigg]^{2} = 0.
\label{Char_Eqn_TSS}
\end{multline}%
The roots of the above equation,  
\beg
\begin{split}
\lambda_{1,2} = -W,   \\
\lambda_{3,4} = \frac{1}{2}\big[\big(1-W\big)+\sqrt{1-\delta^{2}}\big], \\ 
\lambda_{5,6} = \frac{1}{2}\big[\big(1-W\big)-\sqrt{1-\delta^{2}}\big],  
\end{split}
\label{3values}
\en
are the characteristic values of the TSS. All eigenvalues are double degenerate. The symmetry responsible for the degeneracy is the $\Z2$   transformation \re{Z2}, i.e., $\mathbb{J}\S-\S \mathbb{J}=0$ for the TSS, where $\S$ is the $6\times 6$ matrix representation of the mapping \re{Z2}. This implies that if $|\lambda\rangle$ is a characteristic direction with the characteristic value $\lambda$, so is $\S|\lambda\rangle$.

For $\delta>1$   the characteristic values $\lambda_{3,4}$ and $\lambda_{5,6}$   are complex and conjugate to each other. Their real part changes sign from negative to positive as  $W$ goes from $1^+$ to $1^-$. This pair of degenerate characteristic values  simultaneously crosses the imaginary axis at $W=1$, i.e., a Hopf bifurcation takes place across the $W = 1, \delta\ge1$ half-line. 
In \sref{Beyond_Lin_Stability}, we will see that this   Hopf  bifurcation  is supercritical  giving rise to a stable limit cycle as illustrated in  \fref{SuperSubCritical}.  We will find in  Sect.~\ref{Beyond_Lin_Stability} below that the the real and imaginary parts of characteristic values  $\lambda_{3,4}$ and $\lambda_{5,6}$  determine the amplitude ($\propto\sqrt{1-W}$) and the period ($\approx 4\pi/\sqrt{\delta^{2} - 1}$) of the
limit cycle, respectively, at its inception.  The  analytical expression \re{del>>W_Limit_Intro}  for this $\Z2$-symmetric limit cycle  in various limits further corroborates these results.

Note also that  when $\omega_A+\omega_B\ne 0$ this limit cycle rotates around the $z$-axis with a constant angular frequency $\frac{\omega_A+\omega_B}{2}$ according to
\eref{Mean-Field_Group1} [the additional term on the right hand side vanishes by $\Z2$ symmetry]. It thus becomes a 2D torus and the bifurcation is fixed point $\to$ torus, rather than fixed point $\to$ limit cycle \cite{Resonant_Hopf}. This marks an important difference between this bifurcation and the
TSS $\to$ NTSS transition, which is fixed point $\to$ limit cycle  for any  $\omega_A+\omega_B\ne 0$.

For $\delta<1$ all roots are real. For these values of $\delta$, $\lambda_{3,4} < 0$ only when 
\beg 
(W-1)^{2}+\delta^{2}>1.
\label{Reality}
\en%
The above condition also ensures that $\lambda_{5,6} < 0.$ The quarter arc arising from this condition that separates Phases I and II in \fref{Phase_Diagram} depicts a supercritical Hopf bifurcation. Indeed, in a rotating frame where $\omega_A+\omega_B\ne0$, NTSS is a trivial limit cycle and the roots   $\lam_3$  and $\lam_4$ acquire imaginary parts $\pm\frac{\imath (\omega_A+\omega_B)}{2}$ and cross the imaginary axis in unison.  Outside the quarter arc,  TSS is a node, i.e., all characteristic values are negative. Inside it turns into a saddle point -- a few characteristic values become positive. At the same time a stable NTSS comes into existence. On the bifurcation line the NTSS is indistinguishable from the TSS, but deviates from it significantly as we move  deeper  into  Phase~II.   Later in this section we provide an alternative interpretation of the TSS $\to$ NTSS transition as a supercritical pitchfork bifurcation  using the reduced spin equations \re{Symm_One_Spin_Eqn}.

\subsubsection{Stability of NTSS}

From \eref{NTSS}, we observe that for the NTSS to exist, $l_{\perp}$ needs to be real. So, it only exists inside the semicircle $\delta^{2}+(W-1)^{2} = 1$. However, presently we show that it is not stable everywhere inside this semicircle. 
As we have done for the TSS earlier, we derive the characteristic equation for the NTSS,
\beg 
\lambda P_{2}(\lambda) P_{3}(\lambda) = 0,
\label{Char_NTSS}
\en%
where   the polynomials $P_{2}$ and $P_{3}$ are
\begs
\bea 
P_{2}(\lambda) &\equiv & \lambda^{2} + c_{2}\lambda + c_{1},\\
P_{3}(\lambda) &\equiv & \lambda^{3} + c_{2}\lambda^{2} + c_{1}\lambda + c_{0},\label{P3}
\eea
\label{P2_P3}
\ens%
with the coefficients 
\beg 
\begin{aligned}
c_{0} &= W\bigg(W - \frac{W^{2}+\delta^{2}}{2}\bigg), \\
c_{1} &= 2W - \frac{W^{2}+3\delta^{2}}{2},\\
c_{2} &= \frac{3W^{2}-\delta^{2}}{2W}.
\end{aligned}
\label{c_i_Defn}
\en%
We observe that one characteristic value is always zero. This corresponds to the characteristic direction along the overall rotation around the $z$-axis [changing $\Phi$ in \eref{NTSS}]. Recall that NTSS is a collection of fixed points on a circle. Thus, when we perturb one such point along this direction, it neither goes to infinity nor does it come back to the original point. Instead, it just lands onto the neighboring NTSS point.

\begin{figure}[tbp!]
\begin{center}
\includegraphics[scale=0.5]{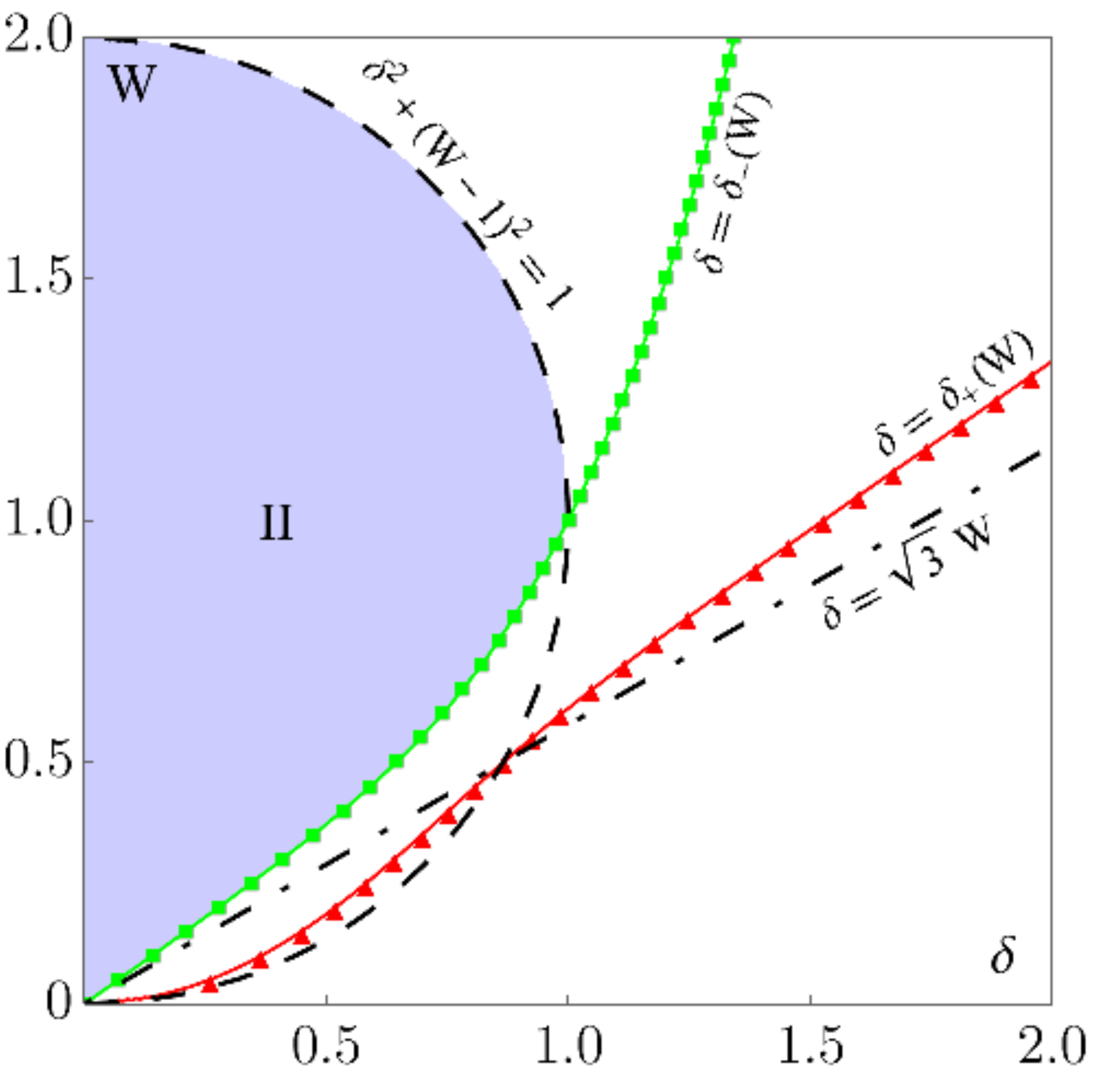}
\caption{The domain of stability of the NTSS (Phase II) is shown in blue.  NTSS exists inside the semicircle (dashed line) $\delta^{2} + (W-1)^{2} = 1$. The dot-dashed line is $\delta = \sqrt{3}W$ [\eref{P_2_c2}]. Red triangles and  green squares denote the lines $\delta = \delta_{+}(W)$ and $\delta = \delta_{-}(W)$, respectively, where $\delta_{\pm}(W)$ are defined in \eref{del_PM}.  This picture  shows that the minimal description of Phase II is $\big[\delta^{2} + (W-1)^{2} < 1\big]\wedge\big[\delta<\delta_{-}(W)\big]$. Since   the $\delta = \delta_{-}(W)$ line is inside the semicircle,  the NTSS loses  stability via a Hopf bifurcation (see the text).}
\label{NTSS_Stability_Final_Graph}
\end{center}
\end{figure}

We use the Routh-Hurwitz stability criterion \cite{Gradshteyn_Ryzhik} to determine   the boundary between  Phases II and III. According to this criterion, the NTSS is stable when $\big(c_{i}>0 \;\textrm{for all $i$}\big) \wedge \big(c_{2}c_{1}>c_{0}\big)$. Moreover, $\big(c_{2}c_{1}>c_0\big) \wedge   \big(c_{2}>0\big) \wedge \big(c_{0}>0\big)$ guarantees $\big(c_{1}>0\big)$. Therefore,   the NTSS  is stable if and only if
\begs
\bea
(W-1)^{2} + \delta^{2}&<&1, \label{P_2_c0} \\
W&>&\frac{\delta}{\sqrt{3}}, \label{P_2_c2} \\ 
3\delta^{4} - \big(6W^{2} + 4W\big)\delta^{2} + W^{3}\big(8 - W\big)&>&0,\label{Blue_Region}
\eea 
\label{region_of_Stability_NTSS}
\ens%
where the last inequality corresponds to $c_2c_1>c_0$.
Its left hand side   is a biquadratic polynomial in $\delta$. We write it as $\big[\delta^{2} - \delta_{+}^{2}(W)\big]\big[\delta^{2} - \delta_{-}^{2}(W)\big]$, where
\beg
\begin{aligned} 
&\delta_{\pm}^{2}(W) = \frac{2W}{3}\bigg(\frac{3W}{2} + 1 \pm \sqrt{3W^{2} - 3W + 1}\bigg), \\
&\delta_{+}(W)>\delta_{-}(W)>0, \textrm{ (for $\delta,W>0$).} 
\label{del_PM}
\end{aligned} 
\en 
Taking into account $\delta>0$ and $W>0$,   we see that   \eref{Blue_Region} requires
\beg 
\big[\delta > \delta_{+}(W)\big] \vee \big[\delta < \delta_{-}(W)\big].  
\en 
Plotting the above two conditions along with \esref{P_2_c0} and (\ref{P_2_c2})  in \fref{NTSS_Stability_Final_Graph}, we conclude that the equation for the lower ($W<1$)  part of the Phase  II boundary is 
\beg 
\delta = \delta_{-}(W).
\label{d-W}
\en
This means that the real part of at least one of the roots of $P_3(\lambda)$ changes sign from negative to positive as we cross the boundary.
 Since $P_3(\lambda)=0$ is a cubic equation with real coefficients, this implies one of the following two complimentary circumstances:
\begin{enumerate}

\item A real root is equal to zero at  criticality.  

\item Near the boundary, the cubic polynomial has one real and  two complex conjugate roots. None of them are zero on the boundary. The real part of the complex conjugate roots changes sign  at criticality. This second scenario entails a Hopf bifurcation.

\end{enumerate}
 $P_{3}(\lambda)$ has a zero root only when $c_{0} = 0$, i.e., on the semicircle $\delta^{2}+(W-1)^{2} = 1$. We observe from \fref{NTSS_Stability_Final_Graph}  that the condition $c_2c_1>c_0$ is violated first, while $c_i>0$ for all $i$ still holds.
 This proves that the NTSS must obey the condition 2 above. In other words,  it loses   stability via a Hopf bifurcation. Moreover, in \sref{Beyond_Lin_Stability} we prove that  it, in fact, undergoes a subcritical Hopf bifurcation  to bring about a coexistence region near the boundary between Phases II and III, see also \fsref{SuperSubCritical} and \ref{subcritical_Evolution}.

\subsubsection{Stability Analysis with Symmetric Spin Equations}
\label{symmstab}

 Since the TSS and NTSS both have  $\mathbb{Z}_{2}$ symmetry,  we also carry out  linear stability analysis  with the reduced spin equations  \re{Symm_One_Spin_Eqn} to  learn more about perturbations destabilizing these steady states.   The fixed points of  \eref{Symm_One_Spin_Eqn}    are 
 \beg
 s_-=0,\quad s_z=1,\quad
 \label{TSSsym}
 \en
 and
 \beg
 s_{-} = \pm\frac{l_{\perp}e^{-\imath \varphi }}{2}\sqrt{1 + \frac{\delta^{2}}{W^{2}}}, \quad
s_{z}   = \frac{\delta^{2} + W^{2}}{2W},
\label{NTSSsym}
\en
 where $\varphi$ and $l_\perp$ are given in \eref{NTSS_1}. \eref{TSSsym} is the TSS~\re{TSS},
  where we now only need the components of spin $\bm s^A$. Similarly, \eref{NTSSsym} is
  the NTSS \re{NTSS}, but now we also need to pick a specific frame where $l_y=0$ [see the text above \eref{Symm_One_Spin_Eqn}]. Note that with this choice of initial-condition-dependent frame, NTSS turns from a one parameter family of fixed points into two fixed points.
  
  The linearization of \eref{Symm_One_Spin_Eqn} about a fixed point $\bm s=\bm s_0$ yields the Jacobian matrix
\beg      
\mathbb{J}=\bem[1.5]
s_{z0}-\frac{1}{2}W &- \frac{1}{2}\delta & s_{x0}\\  
\frac{1}{2}\delta & -\frac{1}{2}W & 0\\ 
-2s_{x0} & 0 & -W
\enm .
\label{3_by_3_Jacobian}
\en%
 Substituting explicit solutions for the TSS and the NTSS, we obtain
\begs
\begin{align}
\textrm{TSS}: \qquad Q_3(\lambda) &= 0, \label{Char_TSS_One_Spin}\\
\textrm{NTSS}: \qquad P_{3}(\lambda) &= 0, \label{Char_NTSS_One_Spin}
\end{align}
\label{Char_One_Spin}
\ens
where    $Q_3(\lambda)$ and $P_{3}(\lambda)$ are defined in \esref{Char_Eqn_TSS} and (\ref{P3}). We saw above that it was the
behavior of the roots of $Q_3(\lambda)$ and $P_3(\lambda)$ that determined the regions of stability of the TSS and NTSS.
Thus, linear stability analysis with both \esref{Mean-Field_1} and (\ref{Symm_One_Spin_Eqn}) produces the same regions, but for  different reasons. For the TSS, factoring out the $\Z2$ symmetry from \eref{Mean-Field_1} reduces the degree of the degeneracy for each of the three distinct characteristic values from two to one. On the other hand, for the NTSS the deviations destabilizing the fixed point are $\Z2$-symmetric. Therefore,  for our problem it is sufficient to analyze the reduced spin equations \re{Symm_One_Spin_Eqn} to study the properties of the TSS and the NTSS near criticality.

From the point of view of the reduced spin equation, the TSS $\to$ NTSS transition is what is known as a supercritical pitchfork bifurcation \cite{Intro_bifurc}.  This is a situation when  there is a single stable fixed point before  and three fixed points -- one unstable and two stable -- after the bifurcation. At criticality all these fixed points coincide. 

\subsection{Different Types of Hopf Bifurcations: Beyond Linear Stability}
\label{Beyond_Lin_Stability}

\begin{figure*}[tbp!]
\centering
\subfloat[\large (a)]{\label{supercritical_Evolution_1_XY}\includegraphics[scale=0.38]{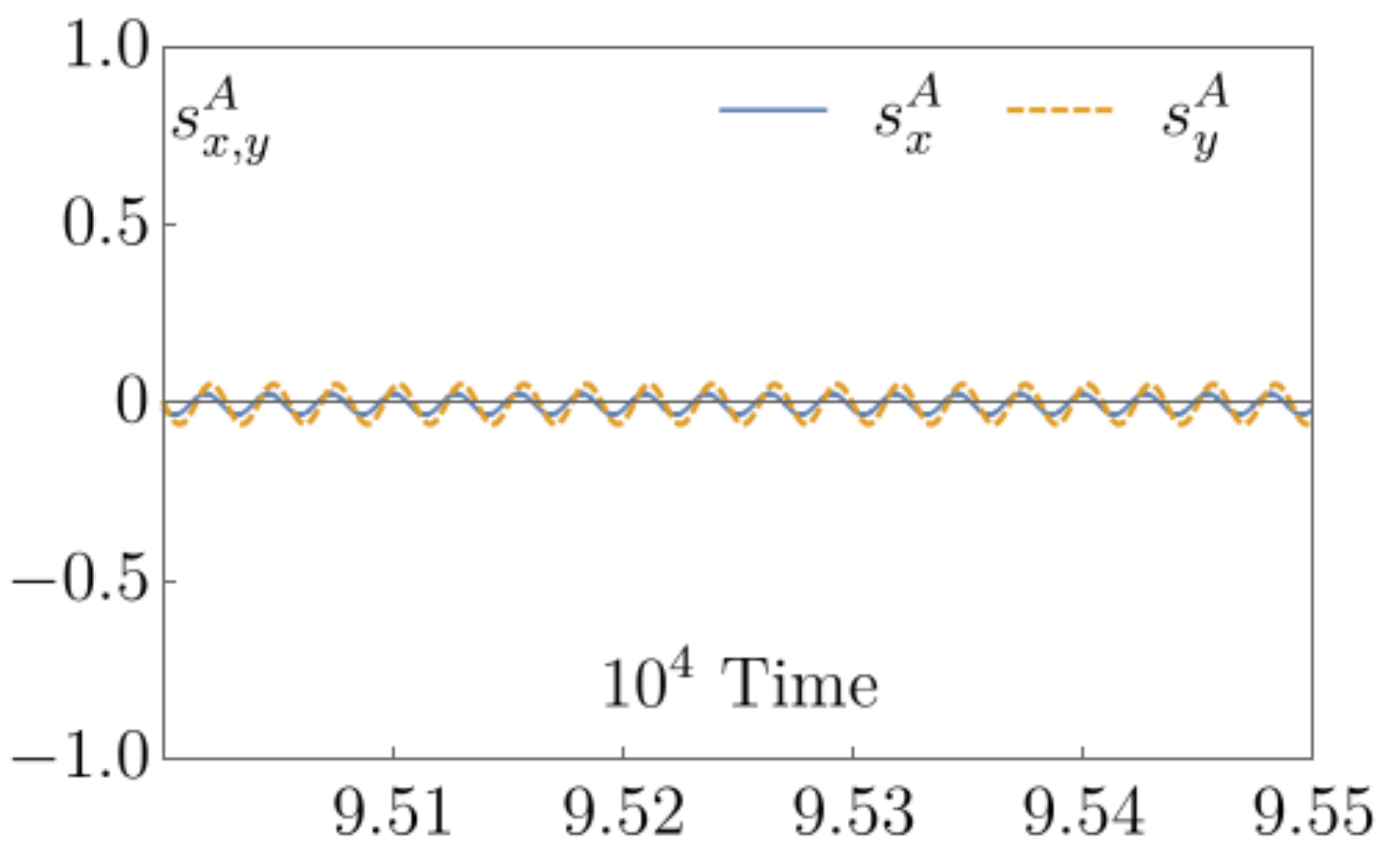}}\qquad\qquad
\subfloat[\large (b)]{\label{supercritical_Evolution_1_Z}\includegraphics[scale=0.38]{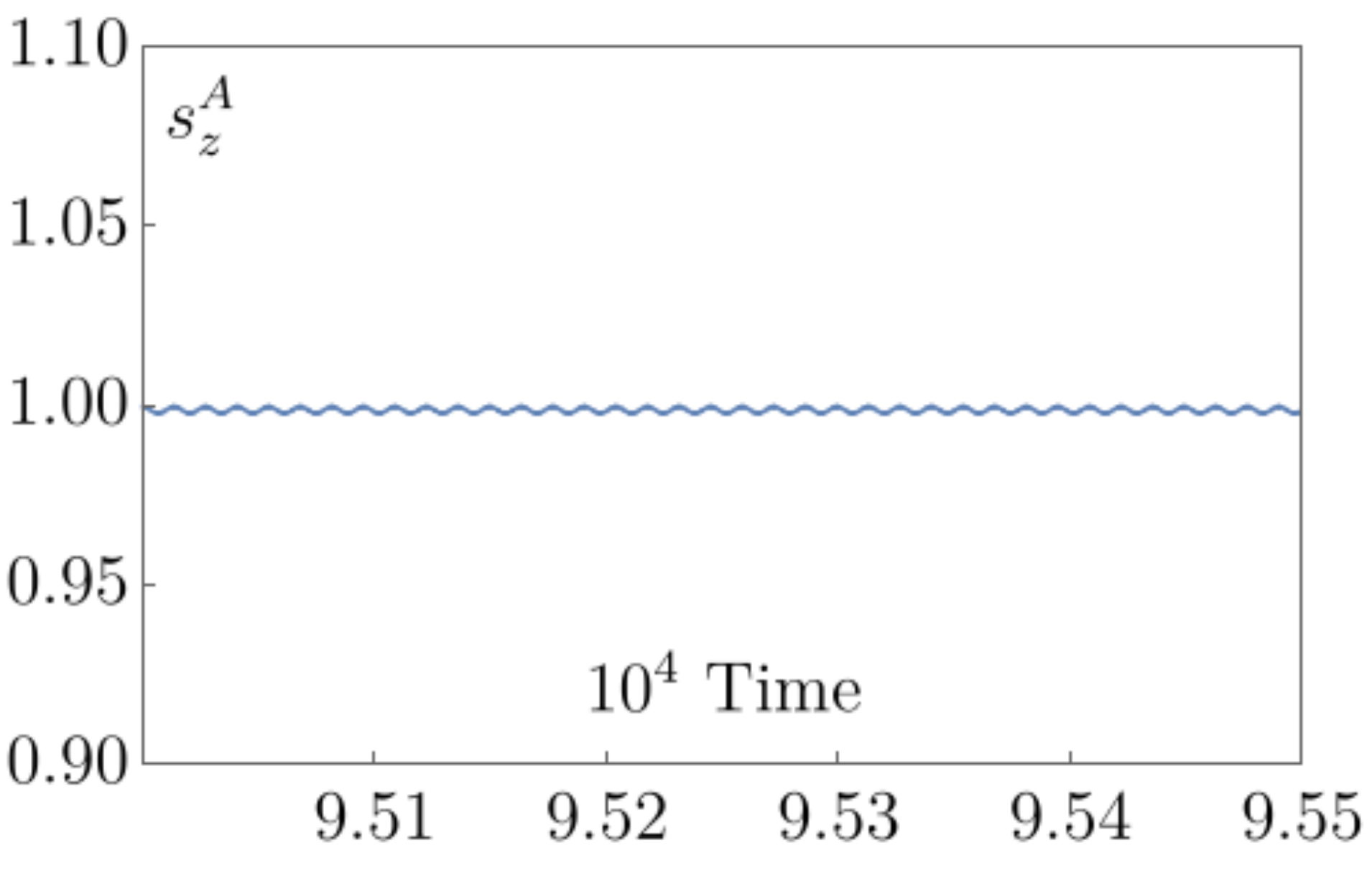}}\\
\subfloat[\large (c)]{\label{supercritical_Evolution_2_XY}\includegraphics[scale=0.38]{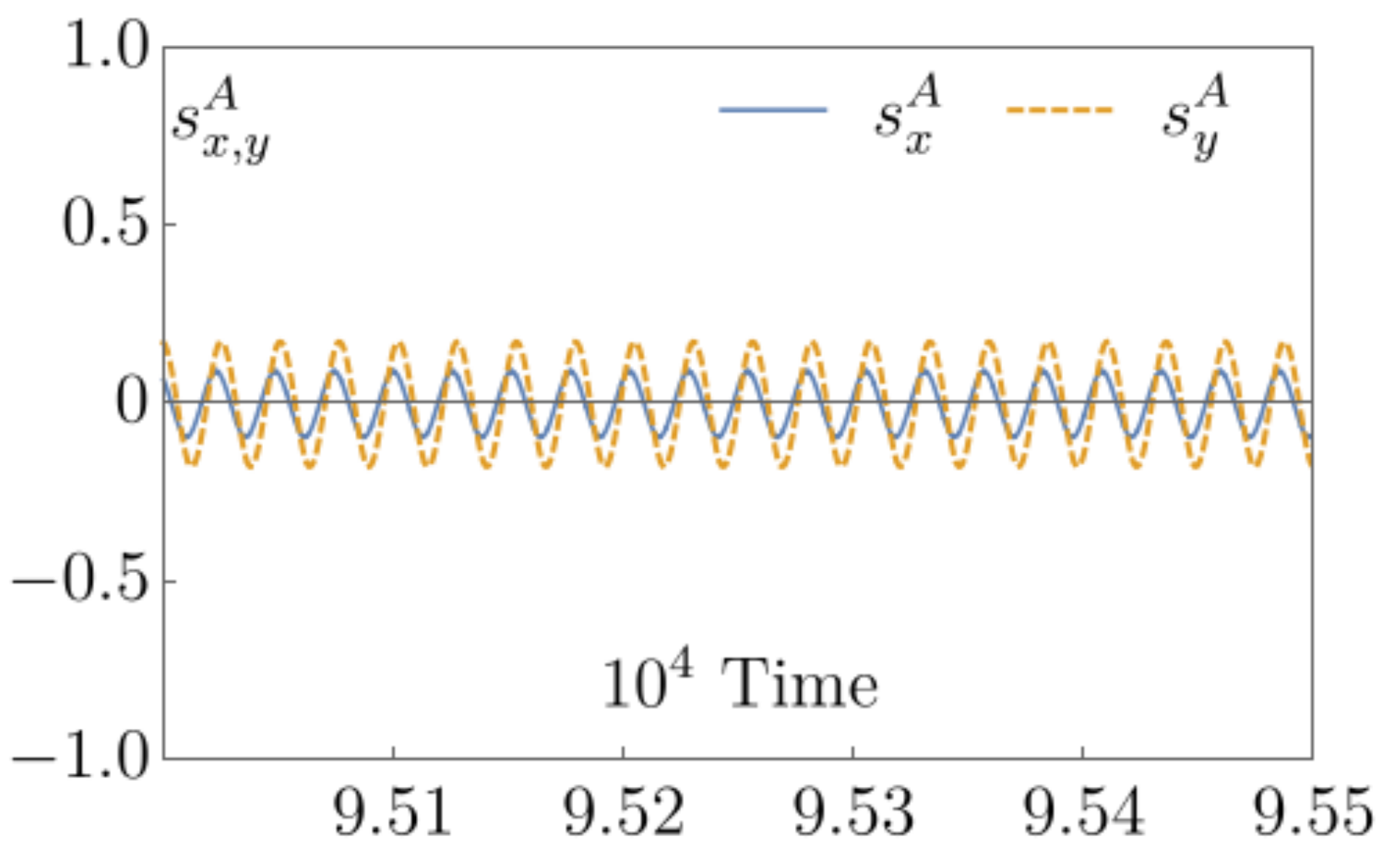}}\qquad\qquad
\subfloat[\large (d)]{\label{supercritical_Evolution_2_Z}\includegraphics[scale=0.38]{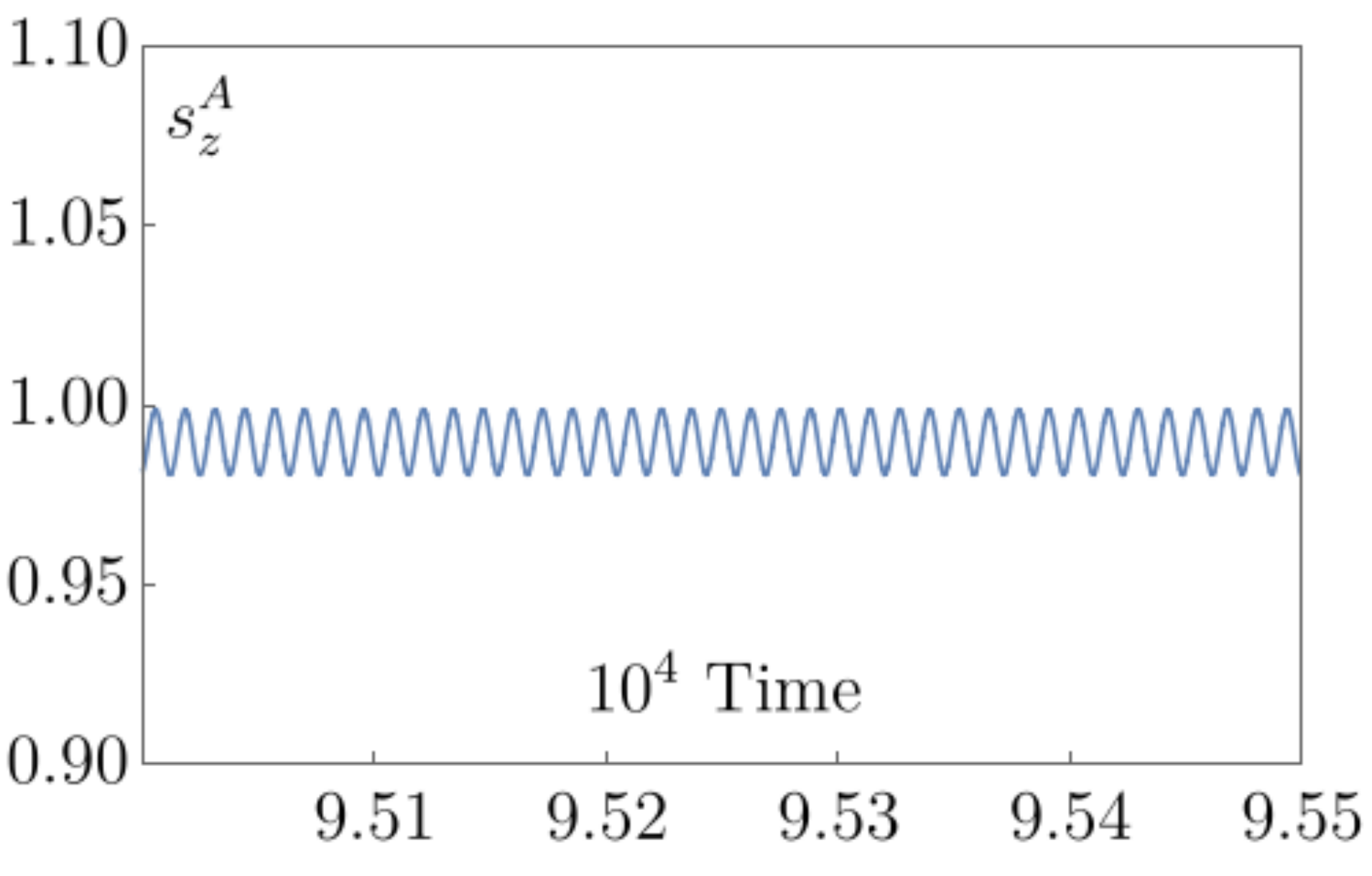}}\\
\subfloat[\large (e)]{\label{supercritical_Evolution_3_XY}\includegraphics[scale=0.38]{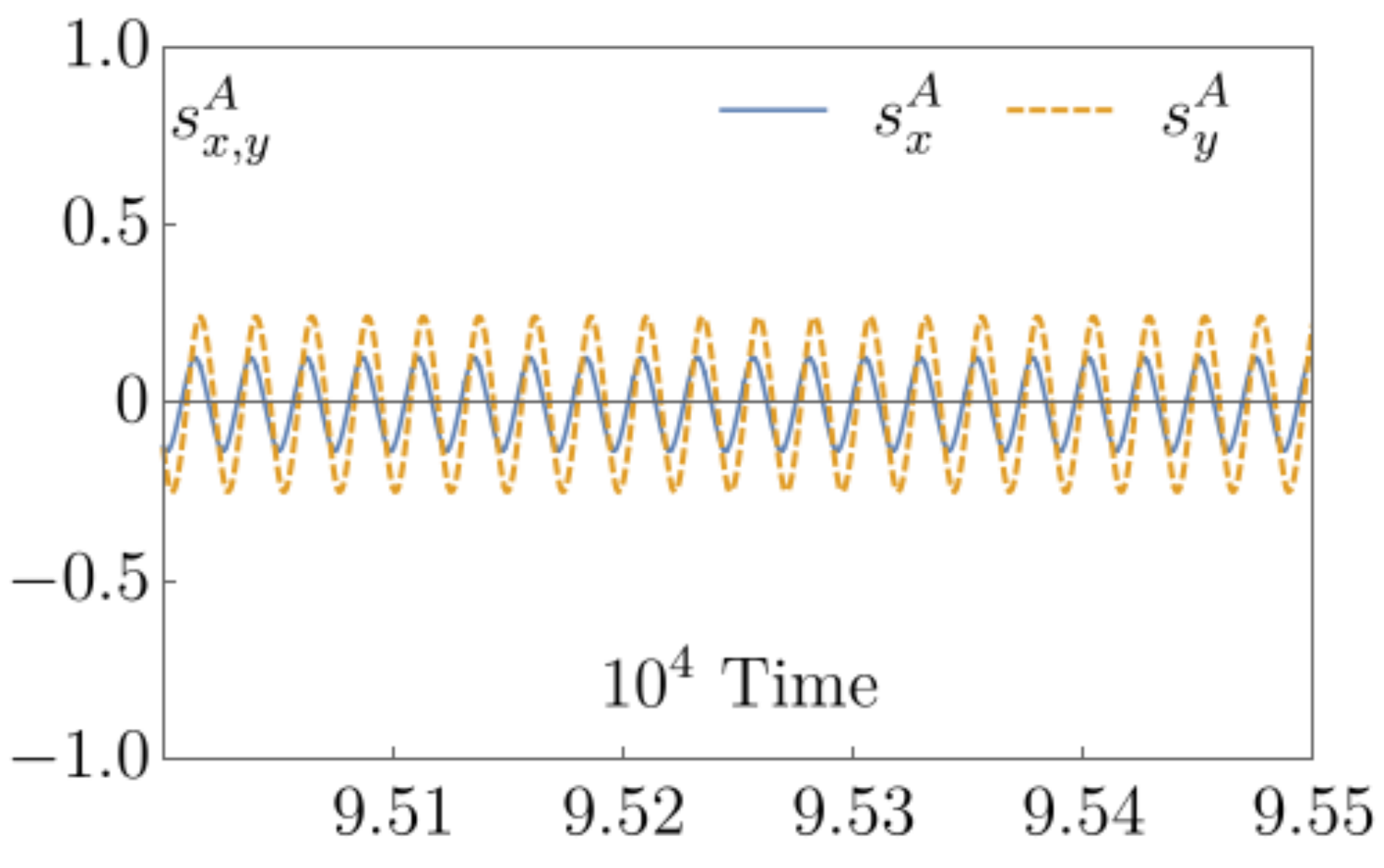}}\qquad\qquad
\subfloat[\large (f)]{\label{supercritical_Evolution_3_Z}\includegraphics[scale=0.38]{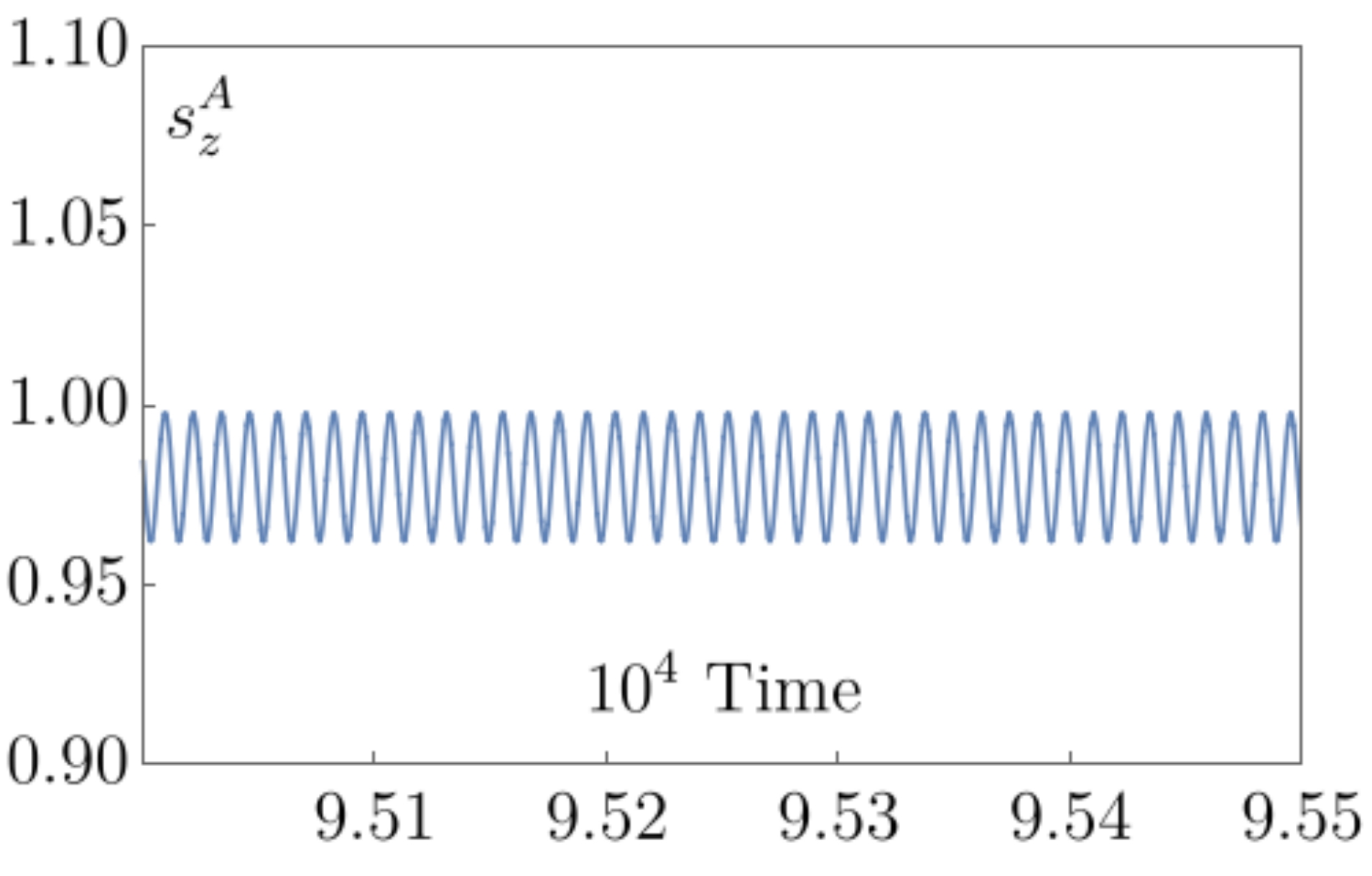}}
\caption{Evolution of the  limit cycle  born out of  the supercritical Hopf bifurcation at $(\delta, W) = (1.1, 1)$.  We fix $\delta = 1.1$ and gradually decrease $W$ moving vertically from Phase  I (TSS) into the green part of region  III ($\Z2$-symmetric limit cycle) in \fref{Phase_Diagram}.  Top to bottom,  $W = 0.999,  0.99$ and 0.98 in the first, second and  third row. We plot $s^{A}_{x}$ and $s^{A}_{y}$ vs. time in the first column and $s^{A}_{z}$ vs. time in the second. }
\label{supercritical_Evolution}
\end{figure*}

\begin{figure*}[tbp!]
\centering
\subfloat[\large (a)]{\label{subcritical_Evolution_1_XY}\includegraphics[scale=0.38]{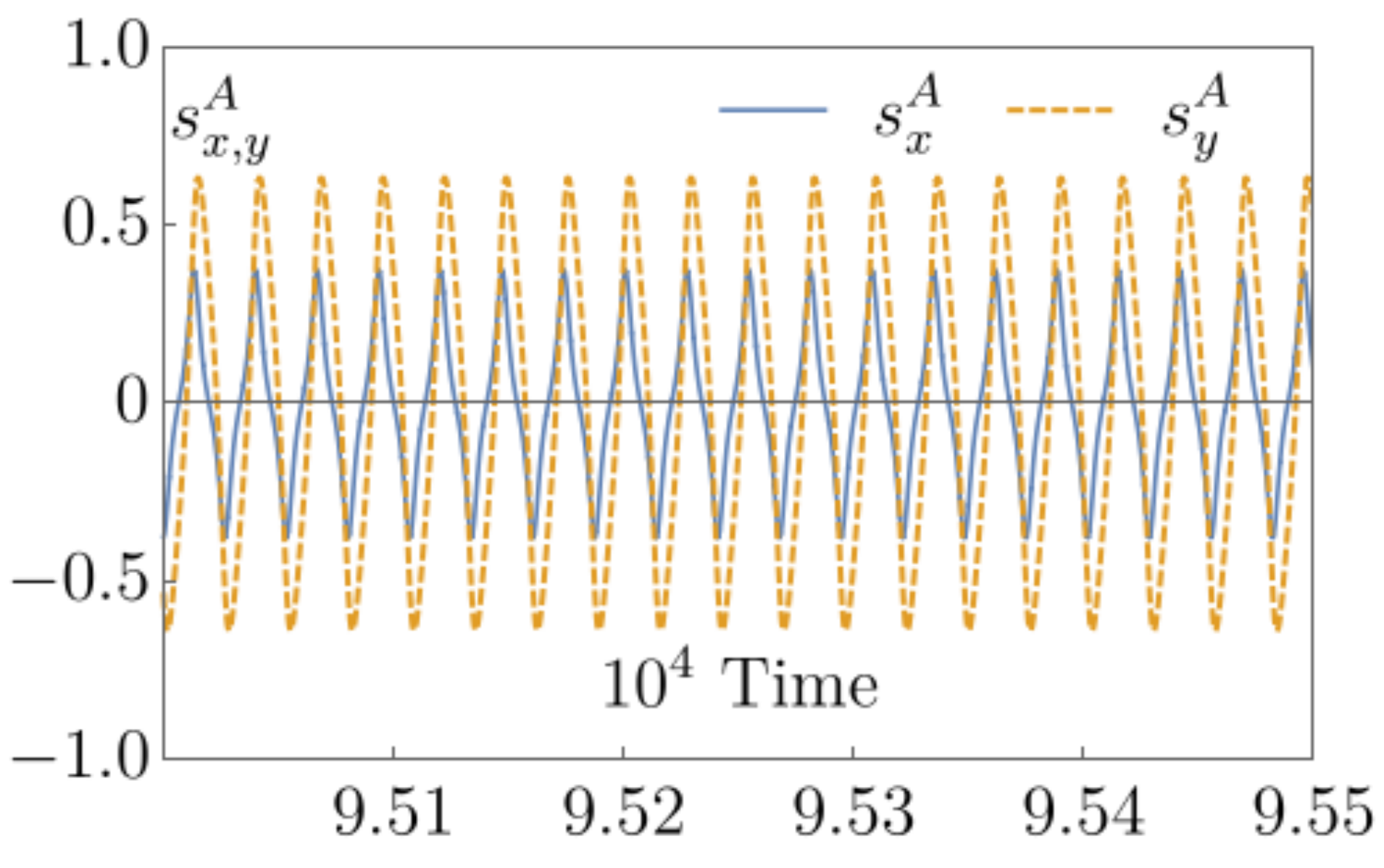}}\qquad\qquad
\subfloat[\large (b)]{\label{subcritical_Evolution_1_Z}\includegraphics[scale=0.37]{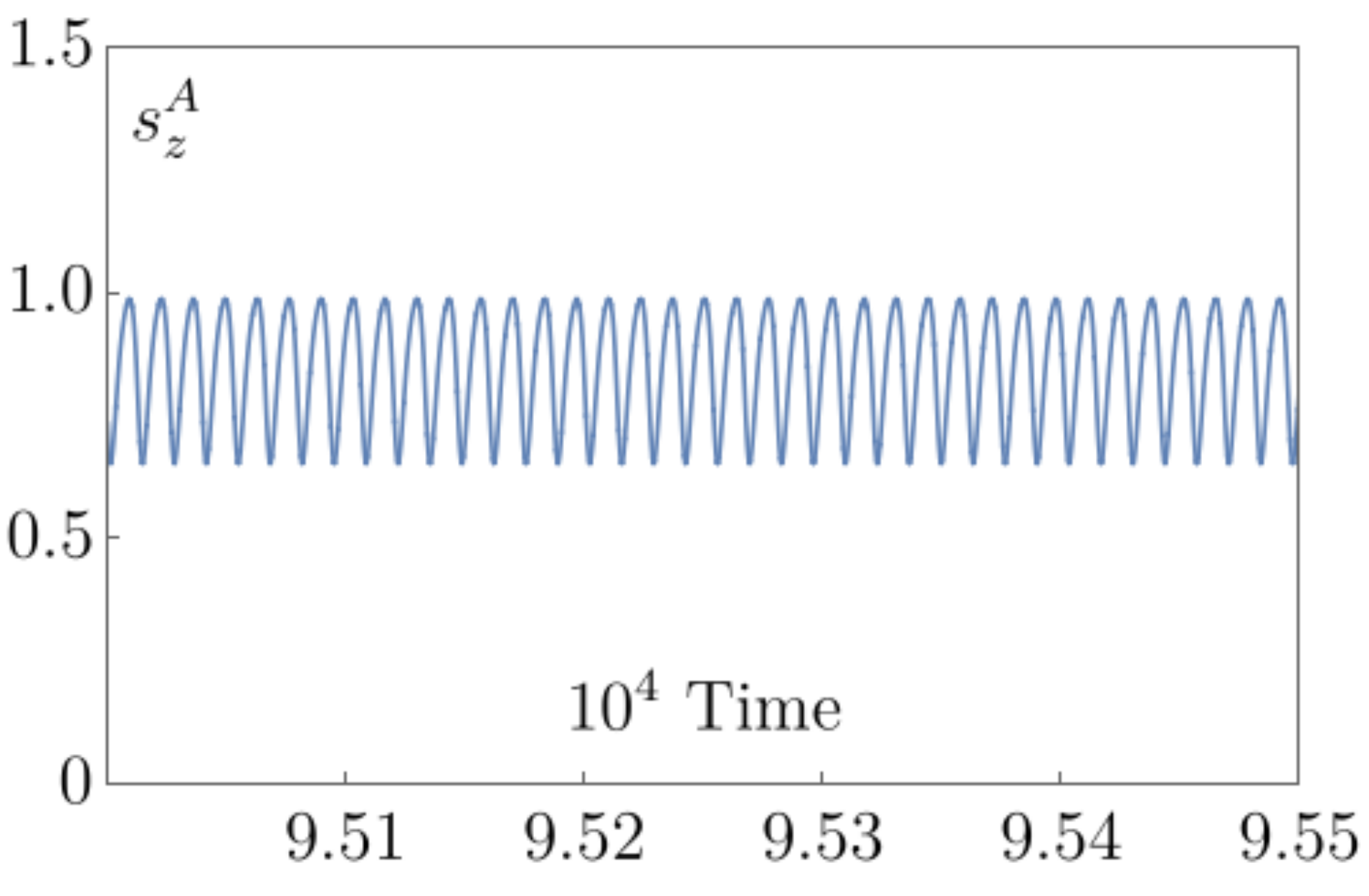}}
\caption{Limit cycle with finite amplitude   right after subcritical Hopf bifurcation of the NTSS at $(\delta, W) = (0.888,  0.800)$.   Here   $(\delta, W) =(0.890, 0.800)$. We show $s^{A}_{x}$ and $s^{A}_{y}$ vs. time in \textbf{(a)} and $s^{A}_{z}$ vs. time in \textbf{(b)}.}
\label{subcritical_Evolution}
\end{figure*}

To differentiate between  Hopf bifurcations, we need to go beyond  the linear stability analysis \cite{Intro_bifurc, Kuznetsov, Hilborn}. Since  only two characteristic directions become unstable in  a Hopf bifurcation, one can determine its essential features  by projecting the dynamics onto a  2D manifold called the ``center manifold". At criticality, this manifold is the   vector space spanned by the   two unstable characteristic directions.     The dynamics on the center manifold near criticality in terms of  polar coordinates $(s, \theta)$   take the  form
\begs
\bea
\dot{s}  &=& s \bigg(\gamma + \sum_{j = 1}^{\infty}a_{j} s^{2j}\bigg),\label{Normal_Form_Intro_1} \\ 
\dot{\theta}  &=& \omega + \sum_{j = 1}^{\infty}b_{j} s^{2j}, \label{Normal_Form_Intro_2}
\eea 
\label{Normal_Form_Intro}
\ens%
where the origin is at the fixed point and $\gamma \pm \imath\omega$ are the complex conjugate characteristic values responsible for the instability of the  fixed point; $\gamma=0$   at criticality.    \eref{Normal_Form_Intro} is a perturbative expansion near the fixed point. Its right hand side   is  known as the ``Poincar\'{e}-Birkhoff normal form".  We derive \eref{Normal_Form_Intro} for the TSS and NTSS starting from \eref{Symm_One_Spin_Eqn}, including the coefficient $a_1$ as a function of $\delta$ and $W,$ in  \aref{Derivation_PB_Normal_Form_appdx}.


\begin{figure*}[tbp!]
\centering
\subfloat[\large (a)]{\label{Supercritical_numerical}\includegraphics[scale=0.42]{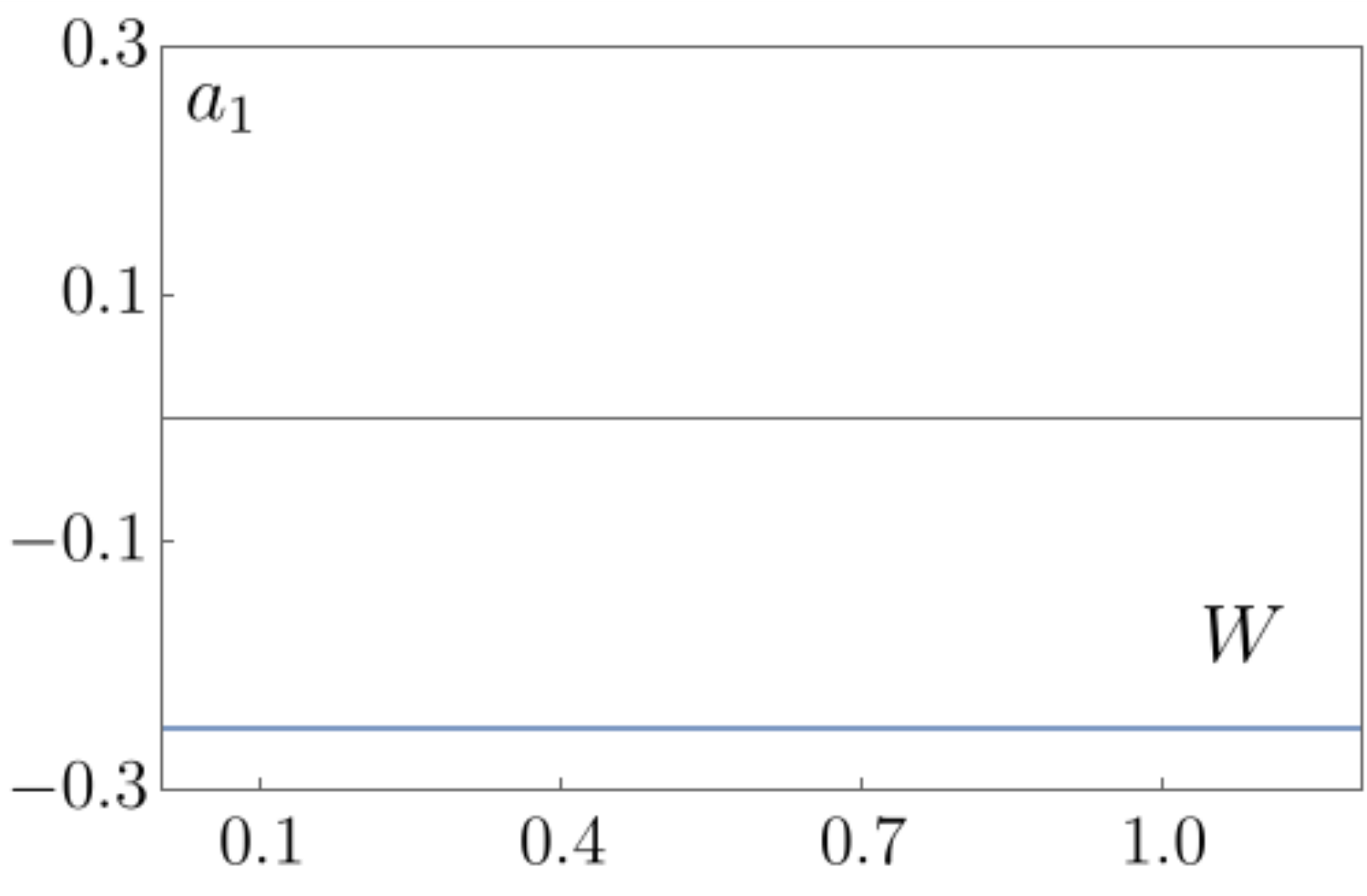}}\qquad\qquad
\subfloat[\large (b)]{\label{Subcritical_numerical}\includegraphics[scale=0.42]{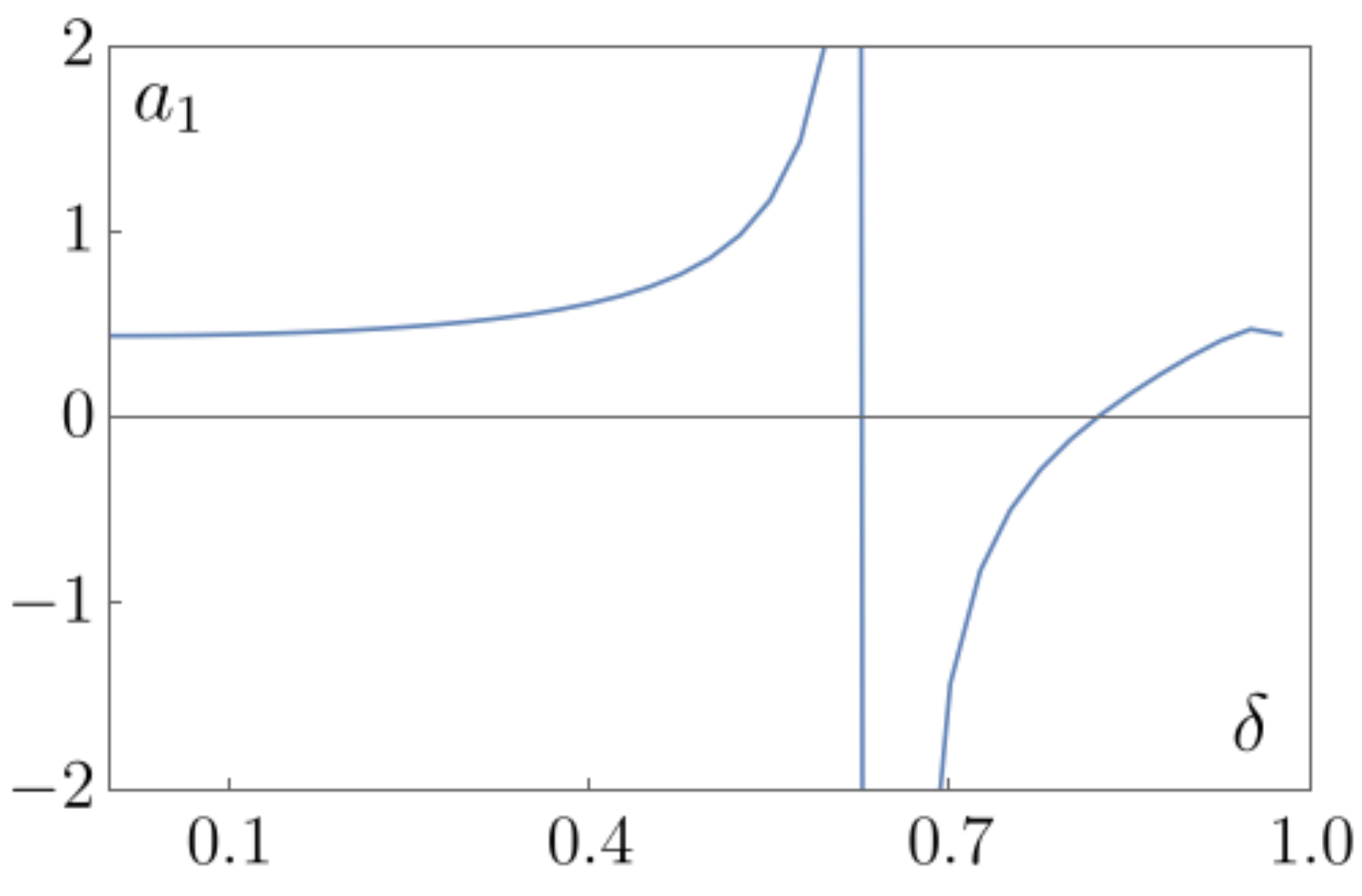}}
\caption{ Plots of the coefficient $a_{1}$ in the Poincar\'{e}-Birkhoff normal form~\re{Normal_Form_Intro}  near Hopf bifurcations of the TSS and NTSS. The bifurcation is supercritical (subcritical) if  $a_1<0$ ($a_1>0$). \textbf{(a)}   $a_{1}$ is negative for the TSS at $\delta = 1.5$ for a substantial range of $W$ around $W = 1$. It behaves in a similar manner elsewhere near the bifurcation along the $W=1,\delta\ge1$ boundary of Phases I and III in \fref{Phase_Diagram}. \textbf{(b)} Here $W = 0.95$. The Hopf bifurcation of the NTSS takes place at $\delta_{H}(0.95) = 0.97$, where $a_{1} = 0.46$. Note the finite neighborhood around $\delta_{H}$ where $a_{1}$ is positive. The qualitative variation of $a_{1}$ for NTSS remains  the same elsewhere near the bifurcation along the boundary of Phases II and III.}
\label{Super_Sub_critical_numerical}
\end{figure*}

 Suppose $\dot{s} = 0$.  Since $s$ is small near the bifurcation, \eref{Normal_Form_Intro_1} yields $s\big(\gamma + a_{1}s^{2}\big) \approx 0$. There are two solution, $s = 0$ and  $s = s_{H} = \sqrt{-\gamma/a_{1}}$. In order to have a limit cycle we need $s_{H}$ to be real. Thus, if $a_{1} < 0$, the limit cycle comes to exist only after $\gamma(\delta, W)$ becomes positive, i.e., after the  bifurcation,   signaling a supercritical Hopf bifurcation. 
 Linearizing the right hand side of  \eref{Normal_Form_Intro_1} near $s_{H}$, we find  $d\Delta s/dt = -2\gamma \Delta s$, where $\Delta s = s-s_{H}$. Since after the bifurcation $\gamma > 0$,   the limit cycle is stable  in the supercritical scenario.  \eref{Normal_Form_Intro_2} shows  that near the bifurcation the polar angle $\theta=\omega t$. Therefore, the period of the limit cycle is $2\pi / \omega$, where $\omega$ is  evaluated at the criticality.
On the other hand,   if $a_{1}>0$, the limit cycle must exist before the bifurcation has taken place. This is because in order for $s_{H}$ to be real we need $\gamma<0$, which is the case only before the bifurcation. This type of bifurcation is known as a subcritical Hopf bifurcation. From the stability analysis we observe that this limit cycle is unstable. As a consequence,  it serves as the separatrix of the basin of attraction for the fixed point, see \fref{SuperSubCritical}. 

 We plot representative examples of $a_{1}$  across the Hopf bifurcation for the TSS and  NTSS in  \fref{Super_Sub_critical_numerical} .   For the TSS, $a_{1} < 0$  near the half-line $W=1, \delta\ge1$  corroborating our earlier claim that  it loses stability via a supercritical Hopf bifurcation. 
 \fref{supercritical_Evolution} shows the (initially infinitesimal) limit cycle emerging right after the bifurcation of the TSS. 
In contrast, the NTSS undergoes   a subcritical Hopf bifurcation on the boundary of Phases II and III.  Indeed, there is a region in \fref{Subcritical_numerical} around $\delta_{H}(W)$ -- the value of $\delta$ on the boundary at a given $W$ -- where $a_{1}> 0$.

\subsubsection{Coexistence Due to Subcritical Hopf Bifurcation}

 In accordance with the above discussion, the unstable limit cycle existing before the bifurcation separates  basins of attraction of the NTSS and   another attractor, which continues into Phase   III after NTSS loses stability. An example of such an attractor is the limit cycle shown in \fref{subcritical_Evolution}. 
As we approach the bifurcation from inside Phase   II, $\gamma$    tends  to zero. Since the size of the separatrix (unstable limit cycle) is proportional to $\sqrt{|\gamma|}$, the basin of attraction of the NTSS shrinks  to zero. Thus, there is a region of coexistence  of NTSS with other attractors inside Phase   II (shown in purple in  \fref{Phase_Diagram}). Its right boundary coincides with the Phase   II-III boundary, while the left boundary is somewhere inside Phase   II. In Sect.~\ref{taper} we determine the shape of this region analytically in the vicinity of the tricritical point  $\delta=W=1$. 

Observe that the dashed line in \fref{Phase_Diagram} merges with the Phase   II-III boundary at a certain  point. Numerically, we find
 that the value of $W$ at this point is $W_c \approx 0.575$. For $W>W_c$, the  attractor to the right of Phase   II is the $\Z2$-symmetric limit cycle. Therefore,  NTSS coexists with this limit cycle in the purple sliver near the boundary for $W_c<W<1$.  For $W<W_c$,  it coexists with other time-dependent asymptotic solutions of \eref{Mean-Field_1}, such as chaotic superradiance \cite{Patra_2,Patra_3}. We  have also observed empirically that the coexistence region is an order of magnitude thinner in the latter case.
 
 We determine the left boundary of the coexistence region in \fref{Phase_Diagram}  using the following numerical method:

\begin{enumerate}

\item Determine the time-dependent asymptotic solution immediately to the right of $\delta_{H}(W)$ for a fixed $W$.
 
\item Record ten random $\big({\bm s}^{A}, {\bm s}^{B}\big)$ on this solution. We use these as initial conditions in the next step. 

\item Decrease $\delta$   and see whether any of these initial conditions   lead to a time-dependent asymptotic solution.  If yes, repeat steps 2 and 3. If not, record the value of $\delta$ as $\delta_{\textrm{End}}(W)$. Check that for $0<\delta\le \delta_{\textrm{End}}(W)$ the time evolution for all these initial conditions converges to the NTSS.

\item Repeat this procedure for  other values of $W$.  

\end{enumerate} 
Note that $a_{1}$  in \fref{Subcritical_numerical} decreases to zero as we move horizontally into Phase   II, i.e., decrease $\delta$ below $\delta_{H}(W)$ keeping $W$ fixed. 
 A  way to estimate the  left boundary of the coexistence  region is to obtain the value of $\delta=\delta_{a_{1}=0}(W)$ where $a_{1}$ becomes zero. However, we find that this estimate is rather inaccurate. Consider, for example,  four distinct values of $W$ in \tref{Hopf_Bifurc_Coexist}.
 Two of them are greater than $W_c$ and the other two are smaller. For each $W$, we report the values of $\delta_{H}(W), \delta_{a_{1} = 0}(W)$ and $\delta_{\textrm{End}}(W)$. In all these cases, an independent numerical analysis using the procedure outlined above reveals that the coexistence ends well before $\delta_{a_{1} = 0}(W)$. This indicates that  to accurately determine its left boundary, one needs to consider higher order terms in the  Poincar\'{e}-Birkhoff normal form. 
 \begin{table}[tbt!]
\centering 
\begin{tabular}{| c | c | c | c |}
\hline
$W$ & $\delta_{H}(W)$ & $\delta_{a_{1} = 0}(W)$ & $\delta_{\textrm{End}}(W)$ \\ [0.5ex]
\hline\hline
0.30 & 0.410 & 0.358 & 0.408 \\
\hline
0.45 & 0.592 & 0.520 & 0.588 \\
\hline
0.65 & 0.782 & 0.687 & 0.759 \\
\hline
0.95 & 0.974 & 0.820 & 0.967 \\ 
\hline
\end{tabular}
\caption{Comparison of $\delta_{H}(W),\delta_{a_{1} = 0}(W)$ and $\delta_{\textrm{End}}(W)$. For all  four values of $W$,   $\delta_{\textrm{End}}(W)>\delta_{a_{1} = 0}(W)$, i.e., the coexistence of NTSS with other attractors ends before a naive estimate based on
the first two orders of the perturbative expansion~\re{Normal_Form_Intro}.}
\label{Hopf_Bifurc_Coexist}
\end{table}    

\section{Limit Cycles}\label{Limit_Cycles}

In this section, we study  limit cycles   in region   III of the nonequilibrium phase diagram in \fref{Phase_Diagram}.  We will see in \sref{expsign} 
that   these attractors  translate into periodic modulations of the superradiance amplitude -- the  cavity radiates frequency combs in this regime. The radiation power spectrum   has various features depending on the symmetry of the limit cycle, such as  presence or absence  of  even harmonics of the limit cycle frequency, symmetry with respect to the  vertical axis, and a shift of the carrier frequency.

\begin{figure*}[tbp!]
\centering
\subfloat[\large (a)]{\includegraphics[scale=0.37]{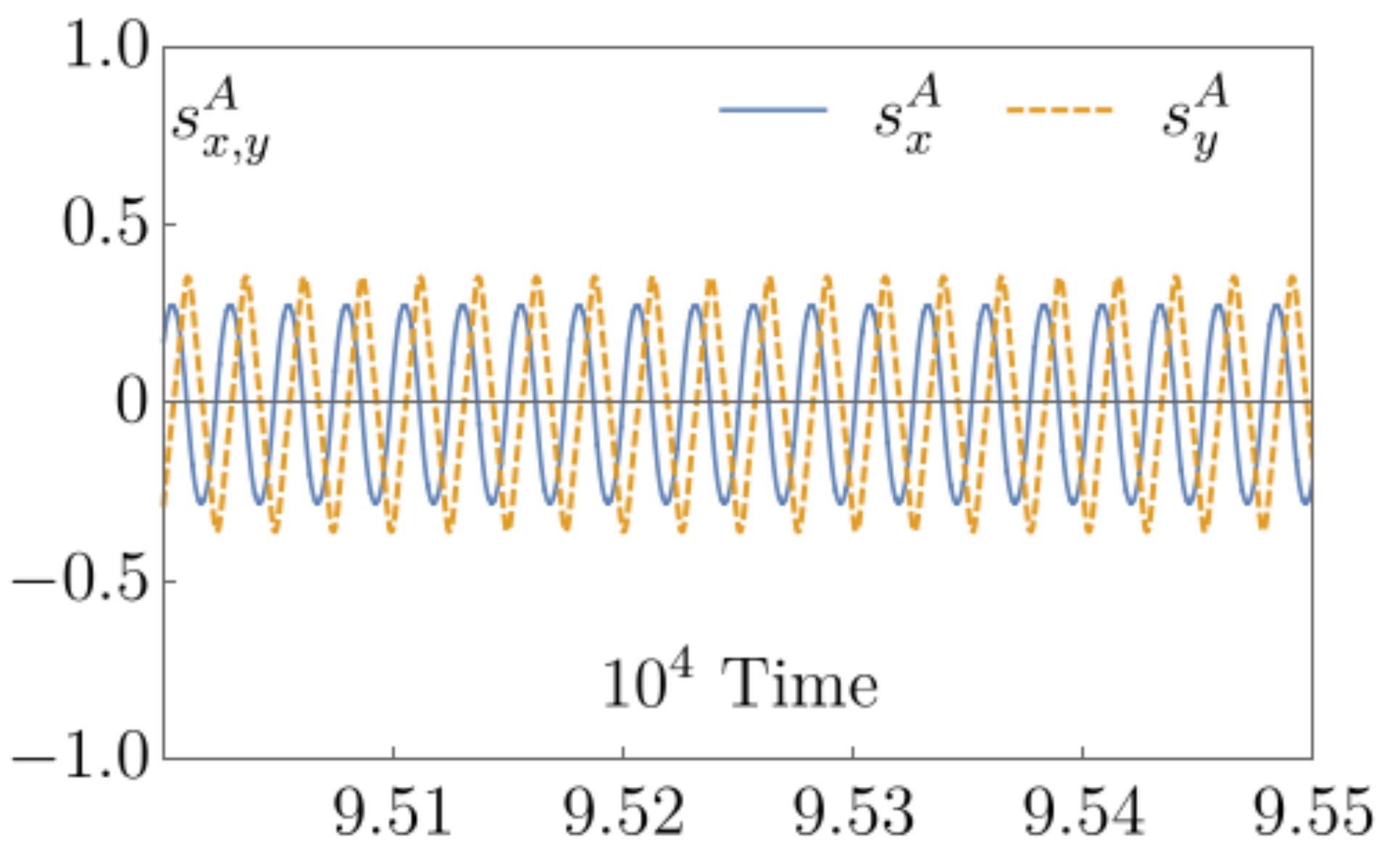}}\qquad\qquad
\subfloat[\large (b)]{\includegraphics[scale=0.37]{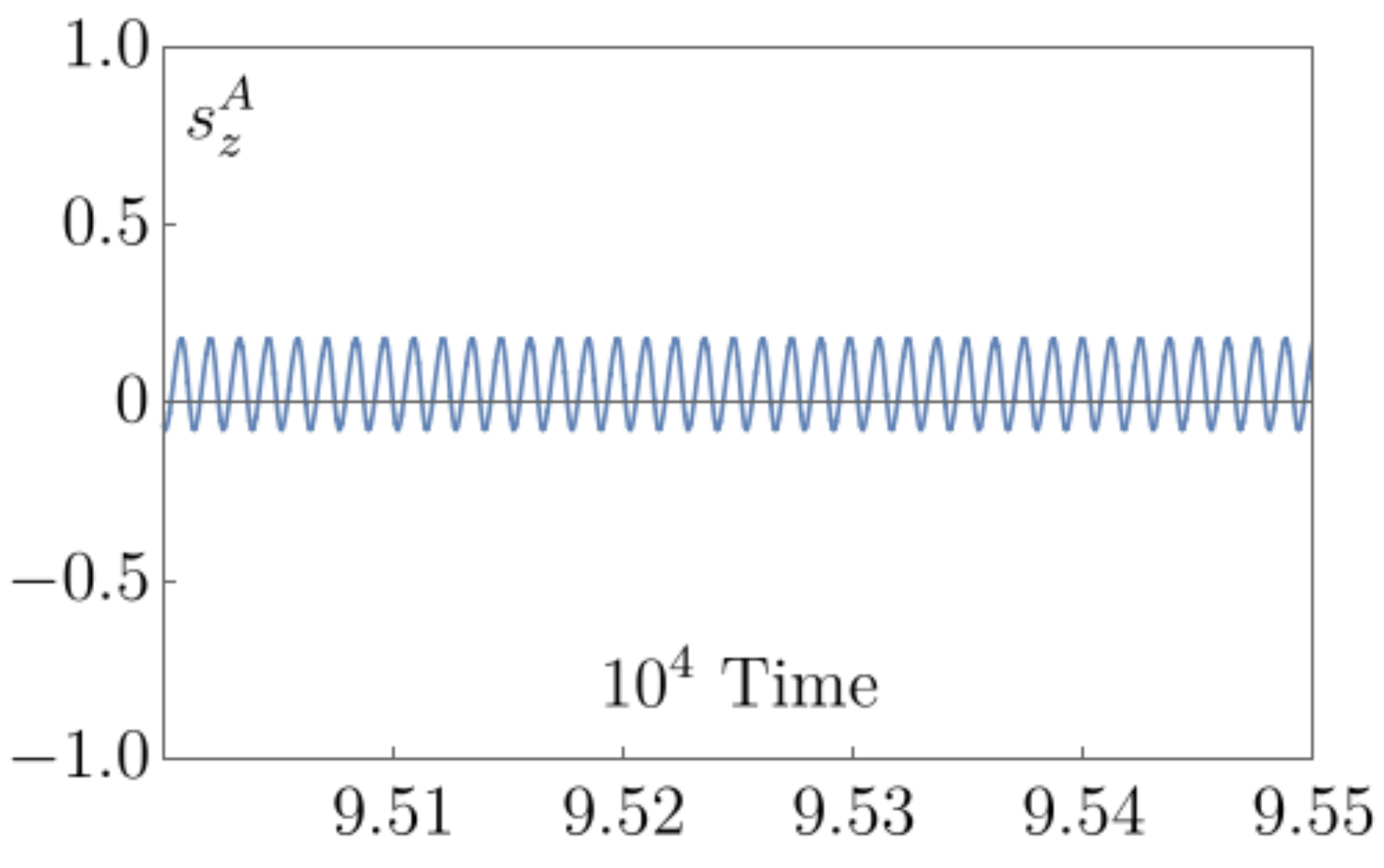}}\\
\subfloat[\large (c)]{\includegraphics[scale=0.37]{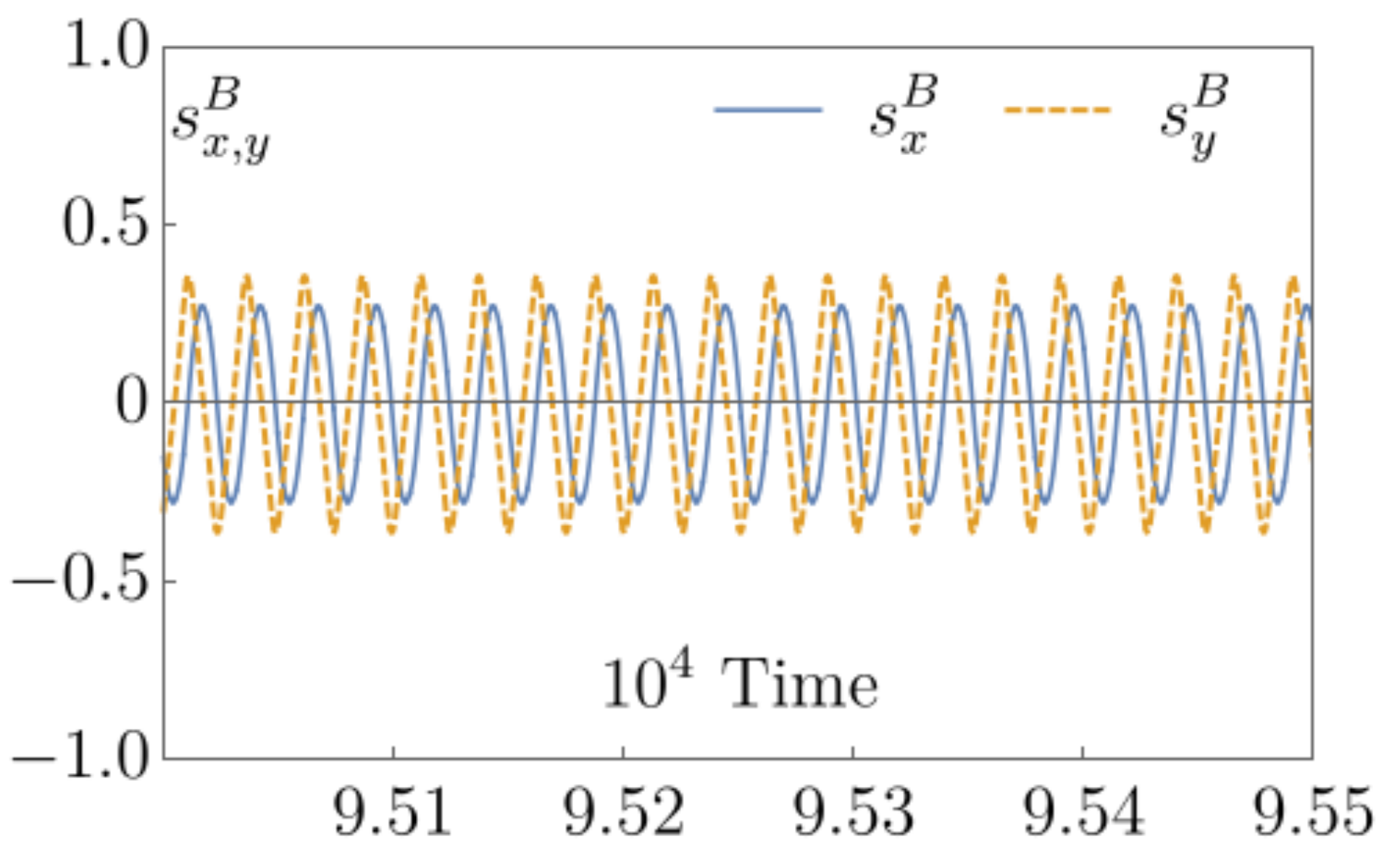}}\qquad\qquad
\subfloat[\large (d)]{\includegraphics[scale=0.37]{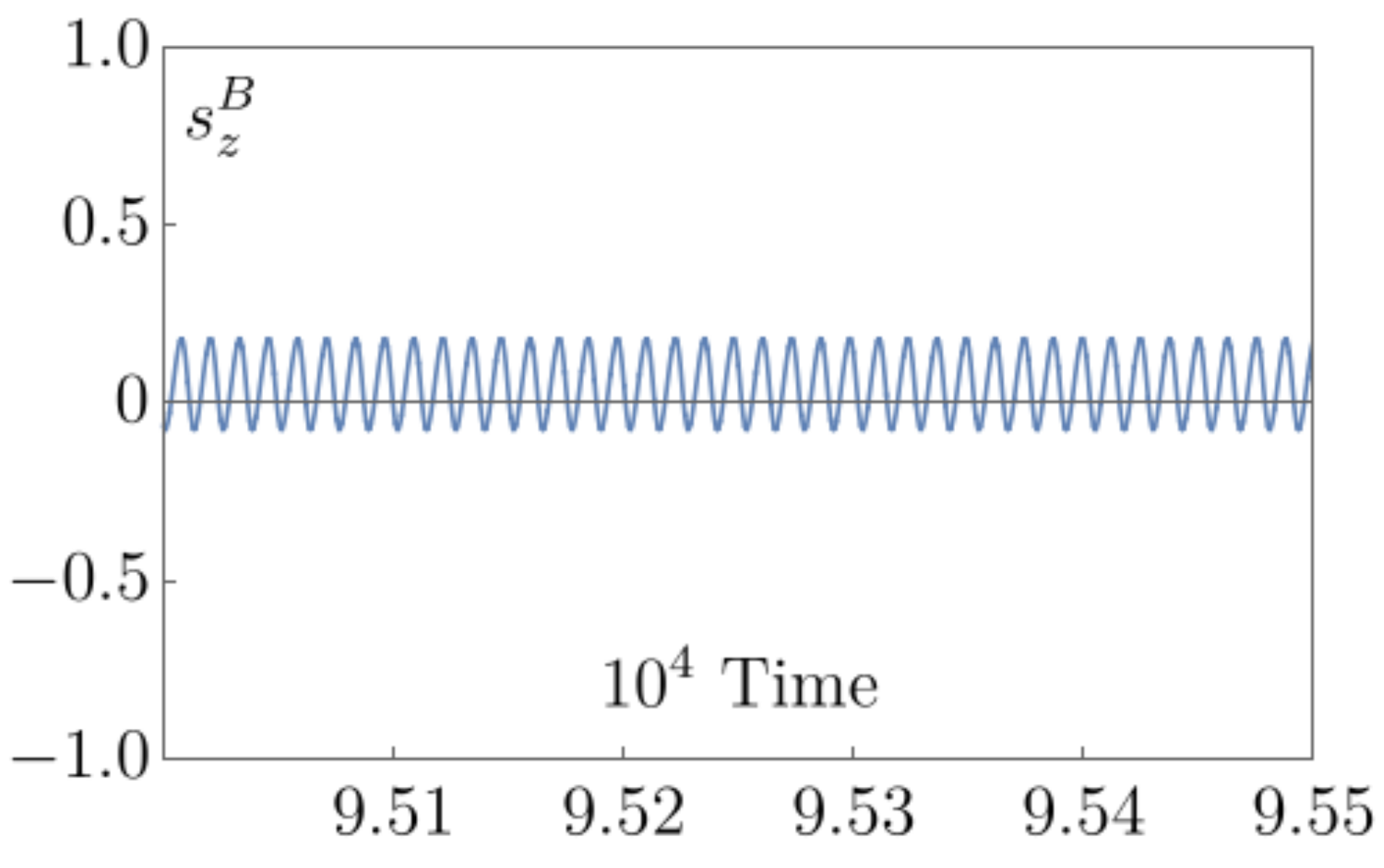}}\\
\subfloat[\large (e)]{\label{S_LC_SA-SB}\includegraphics[scale=0.37]{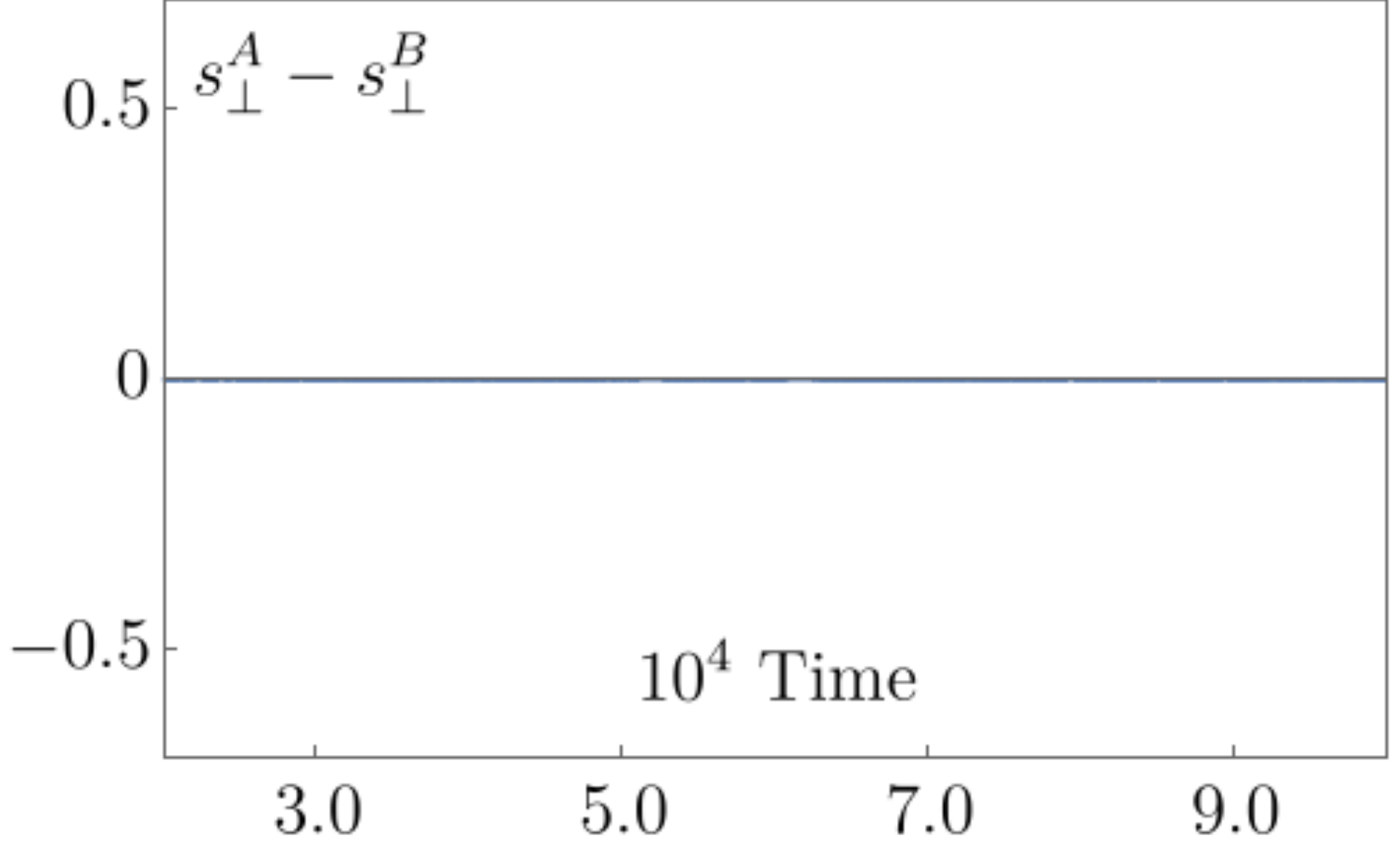}}\qquad\qquad
\subfloat[\large (f)]{\includegraphics[scale=0.37]{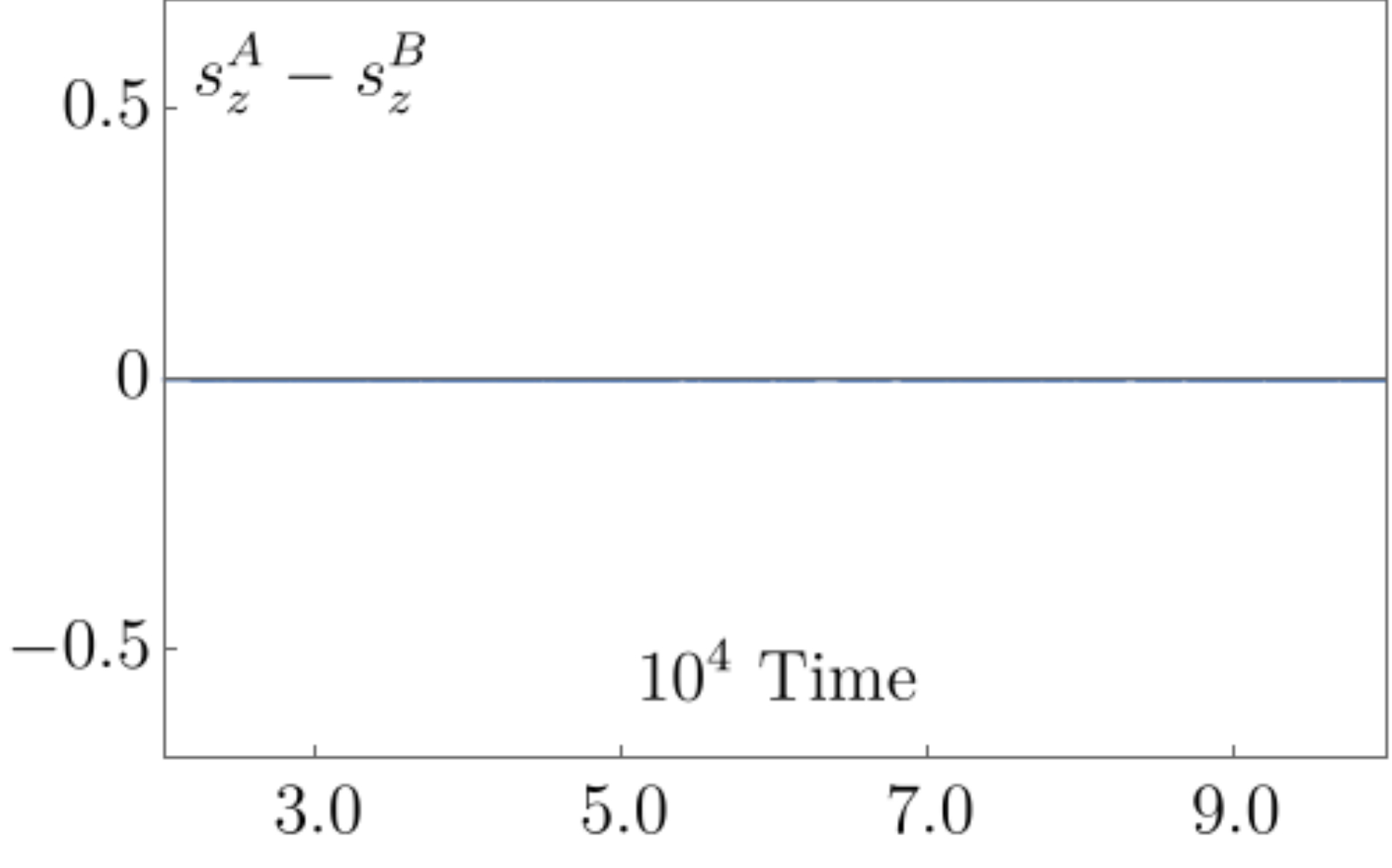}}
\caption{A $\Z2$-symmetric limit cycle at $(\delta, W) = (0.44, 0.056)$. Classical spins $\bm s^A$ and $\bm s^B$ describe the mean-field dynamics of two atomic ensembles in a bad cavity. Spin components fulfill  $\Z2$-symmetry conditions $s_\perp^A=s_\perp^B$ and $s_z^A=s_z^B$, see \eref{Gen_Z2_Symm_MF}. For these values of the detuning $\delta$ between level spacings of the two ensembles and pumping $W,$ the cavity radiates the frequency comb shown in \fref{Power_Spectrum_Symmetric_LC}.}
\label{Symmetric_LC}
\end{figure*}

\begin{figure*}[tbp!]
\centering
\subfloat[\large (a)]{\includegraphics[scale=0.34]{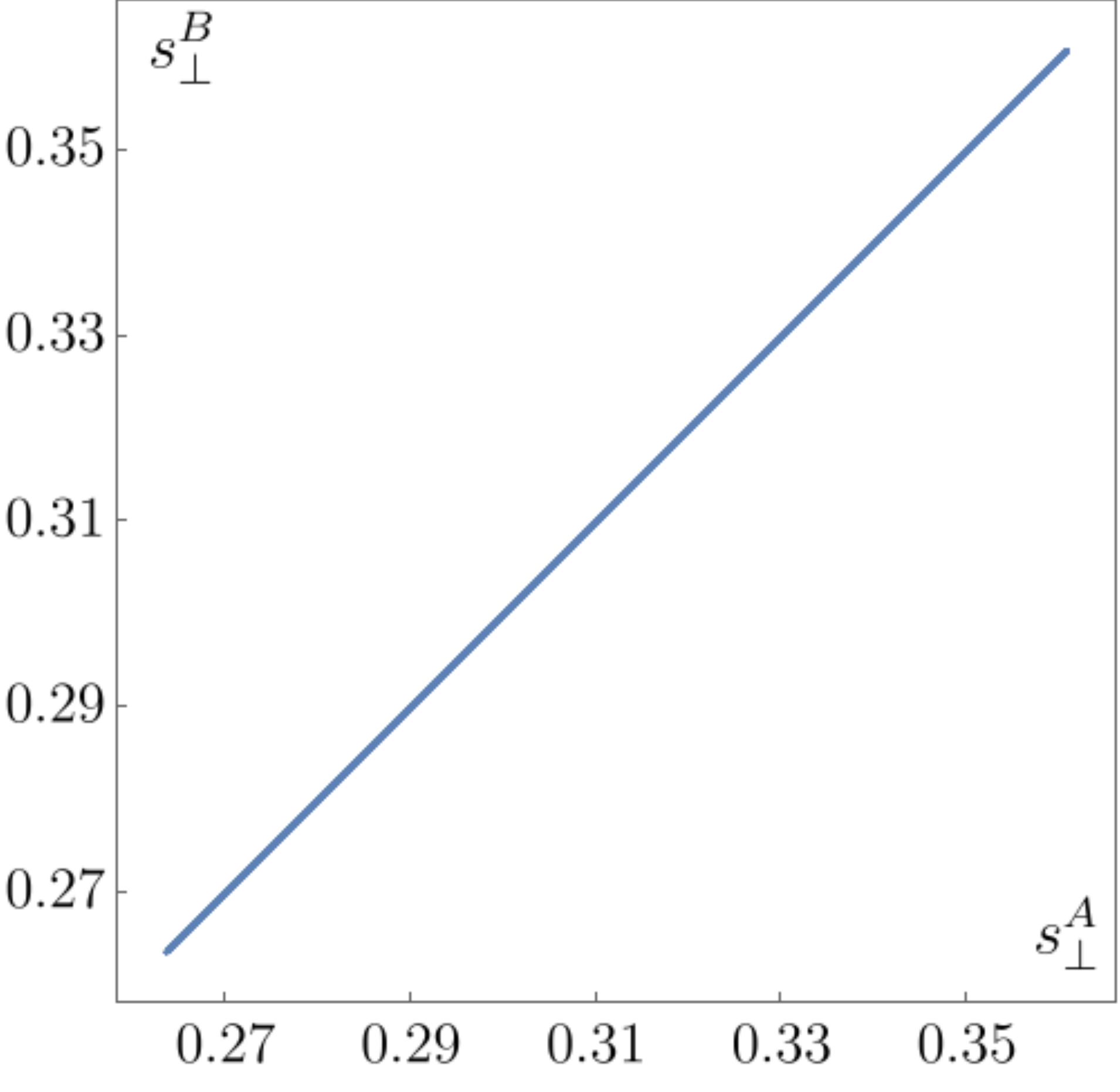}}\qquad\qquad\qquad\qquad
\subfloat[\large (b)]{\includegraphics[scale=0.35]{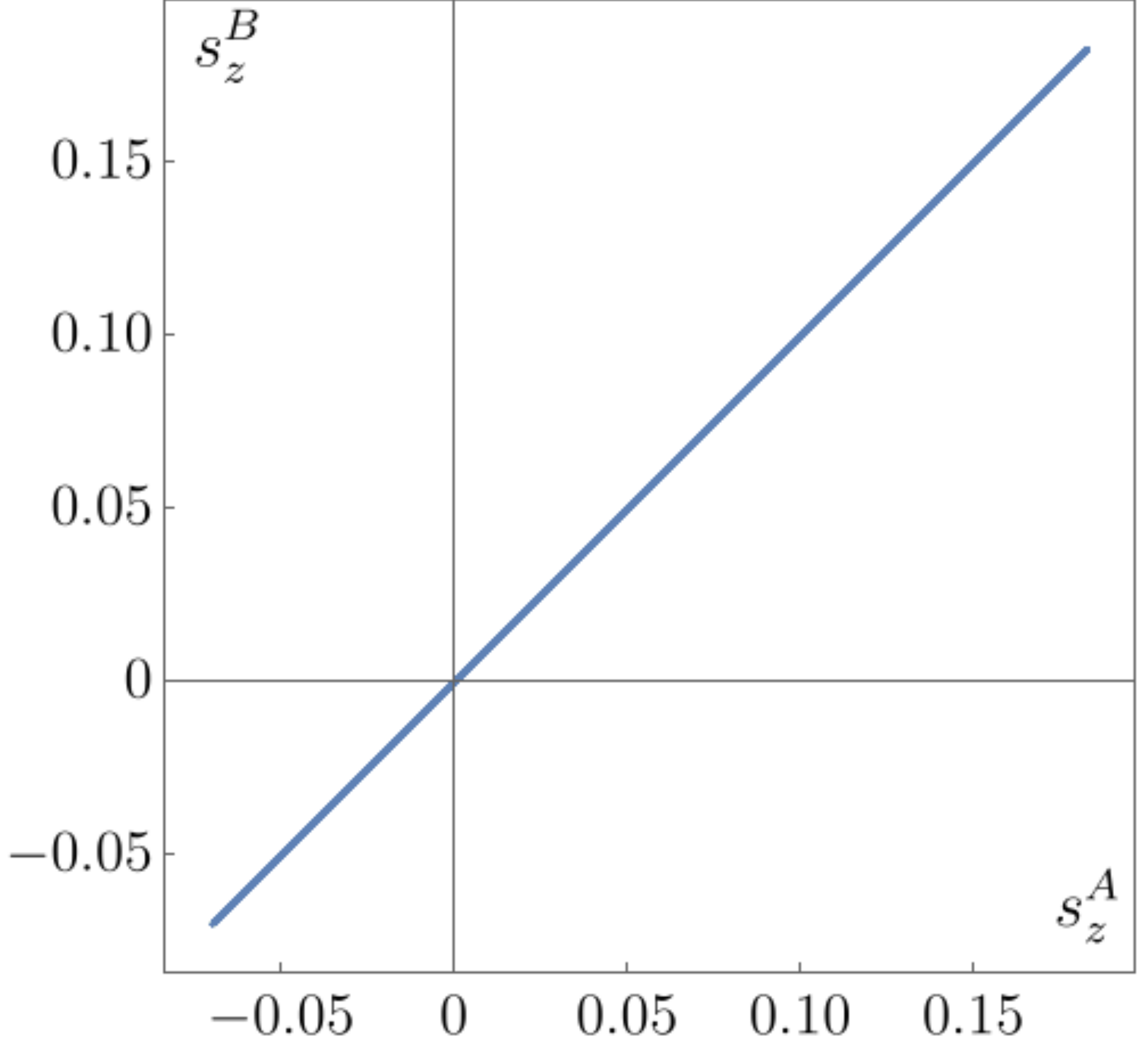}}
\caption{$s^{A}_\perp$ vs. $s^{B}_\perp$ and $s^{A}_z$ vs. $s^{B}_z$  projections of the $\Z2$-symmetric limit cycle at $(\delta, W) =( 0.44, 0.056)$, same as in \fref{Symmetric_LC}.  These are  straight lines passing through the origin with $45^{\circ}$ slope  due to the $\Z2$ symmetry of the attractor. Note the   difference with the $\Z2$-asymmetric limit cycle  at a nearby point $(\delta, W) =(0.42, 0.056)$ in \fref{Symmetry-broken_LC_X_Sect}.   }
\label{Symmetric_LC_X_Sect}
\end{figure*}

In most of  Phase   III, we observe $\Z2$-symmetric limit cycles, such as the ones   in  \fsref{Limit_Cycle}, \ref{Symmetric_LC}, and   \ref{Symmetric_LC_X_Sect}.   When $\delta$ is large or $W(1-W)$ is small, oscillations of spin components  become   harmonic and we  determine the approximate form of the limit cycle analytically. A special situation arises in the vicinity of the tricritical point 
$(\delta, W)=(1, 1)$ in \fref{Phase_Diagram}. Now the oscillations  are anharmonic, but we are still able to derive analytic expressions for the spins in terms of the Jacobi elliptic function cn.  This also helps us determine the shape of the coexistence region of the NTSS with  $\Z2$-symmetric limit cycles near the tricritical point.
Eventually,  limit cycles lose the $\Z2$-symmetry   across the dashed line in \fref{Phase_Diagram}. We determine this line  and independently explain the mechanism of the symmetry breaking with the help of Floquet stability analysis. We  conclude this section by discussing  properties of  limit cycles with broken $\Z2$ symmetry, such as the ones in \fsref{Symmetry-broken_LC}, \ref{Symmetry-broken_LC_X_Sect} and~\ref{Refection_Z2_Symmetry-broken_Pseudo_LC}.

\subsection{Solution for the $\mathbb{Z}_{2}$-Symmetric Limit Cycle in Various Limits}  
\label{Z2 Symmetric Limit Cycle}

As discussed below \eref{Theta}, a $\Z2$-symmetric limit cycle at a given $(\delta, W)$ is in fact a one parameter family of limit cycles that differ from each other only by   the constant value of the net phase $\Phi$. \eref{New_Var_Rot} shows that the projection $\bm l_\perp=(l_x, l_y)$ of the total spin   onto the $xy$-plane moves on a line  making a constant angle $\Phi$ with the $x$-axis. It is convenient to rotate the coordinate system, as we did
in \eref{Symm_One_Spin_Eqn}, so that $\bm l_\perp$ is along the $x$-axis.  \eref{Z2_Expl} then relates the  components of $\bm s^A$ and $\bm s^B$ and we see that it is sufficient to study reduced spin equations~\re{Symm_One_Spin_Eqn} to describe $\Z2$-symmetric limit cycles. 

\begin{figure*}[tbp!]
\centering
\subfloat[\large (a)]{\label{Perturb_W=0}\includegraphics[scale=0.44]{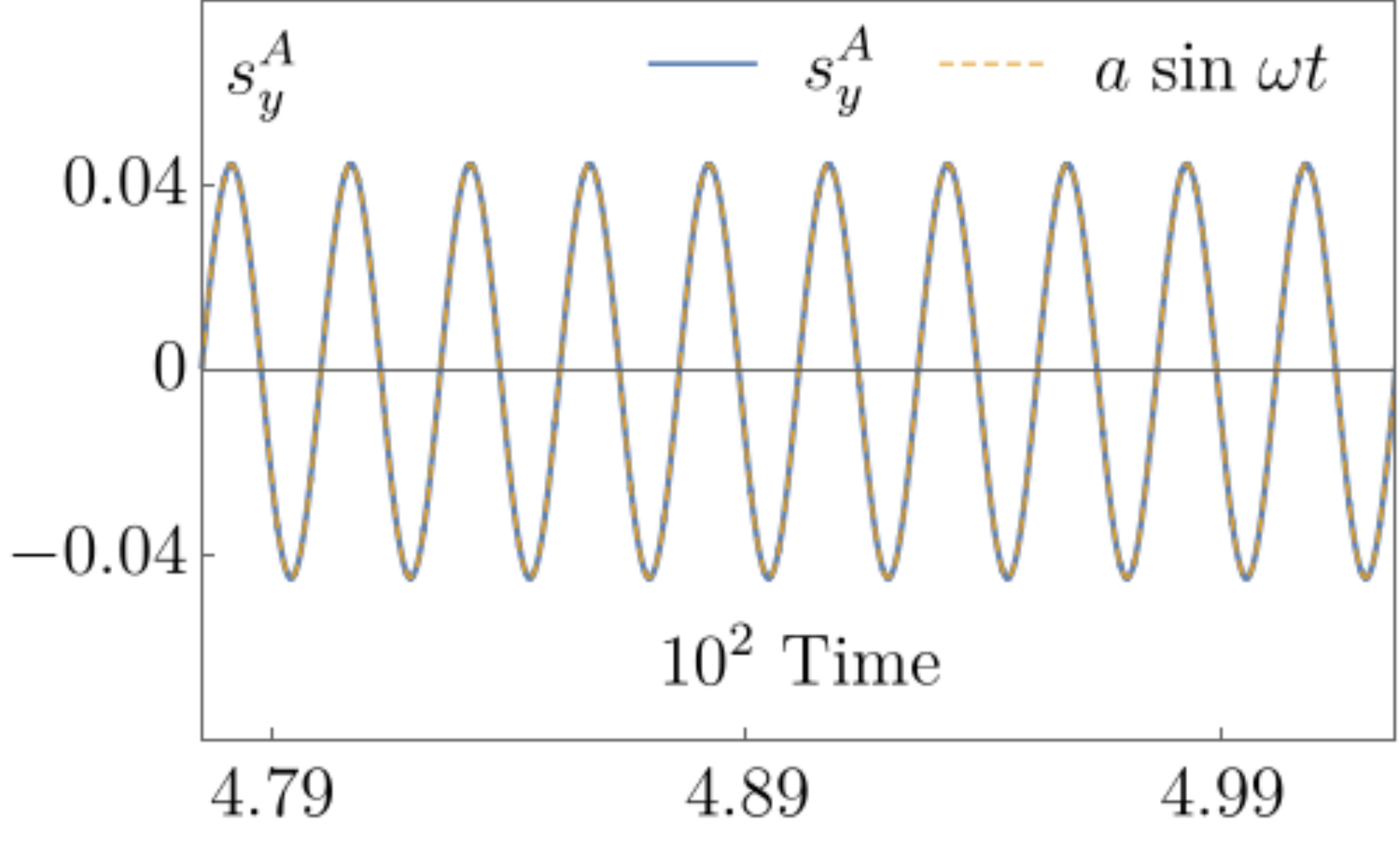}}\qquad\qquad
\subfloat[\large (b)]{\label{Perturb_W=1}\includegraphics[scale=0.44]{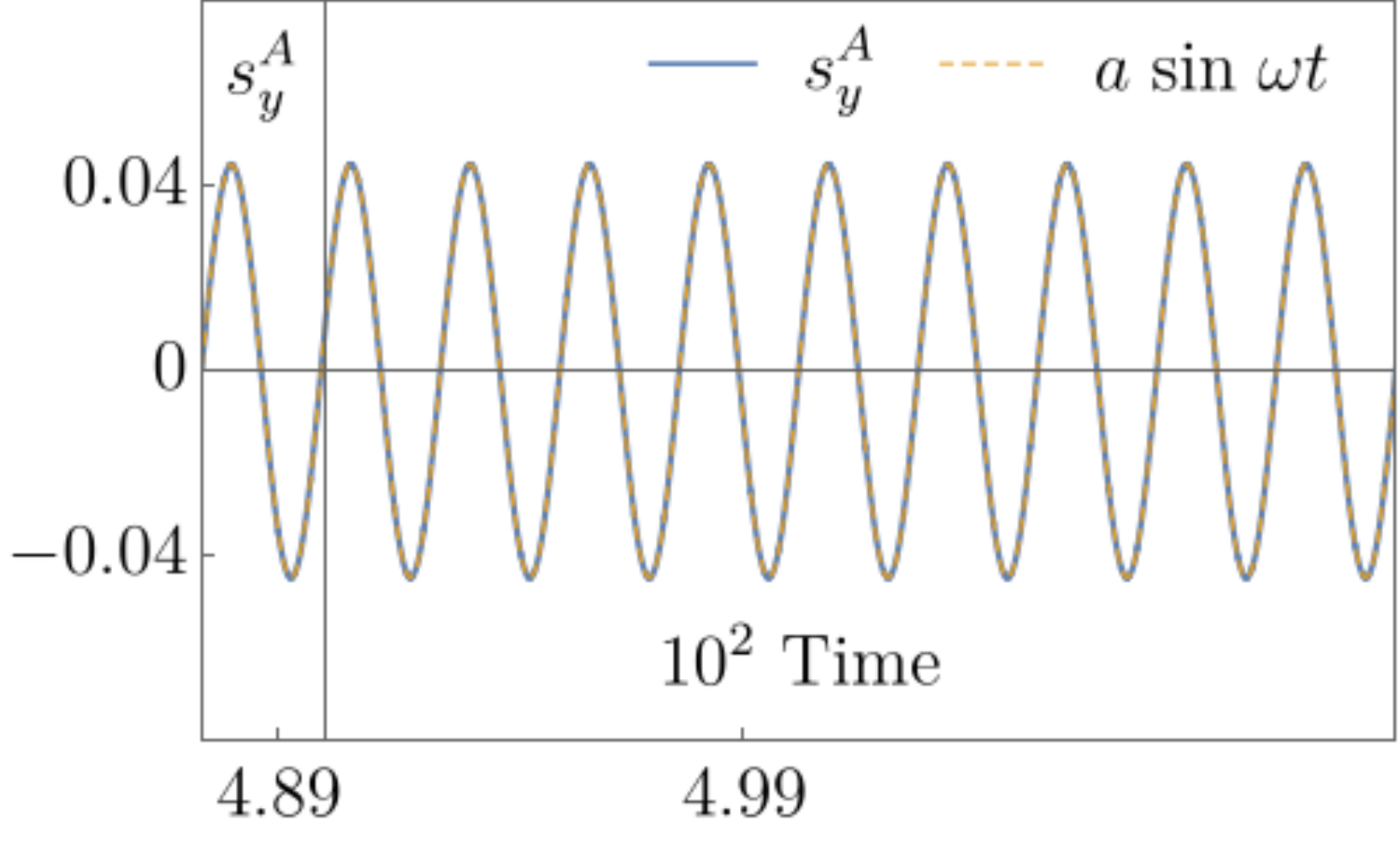}}\\
\subfloat[\large (c)]{\label{Perturb_W=0_5_del=500}\includegraphics[scale=0.44]{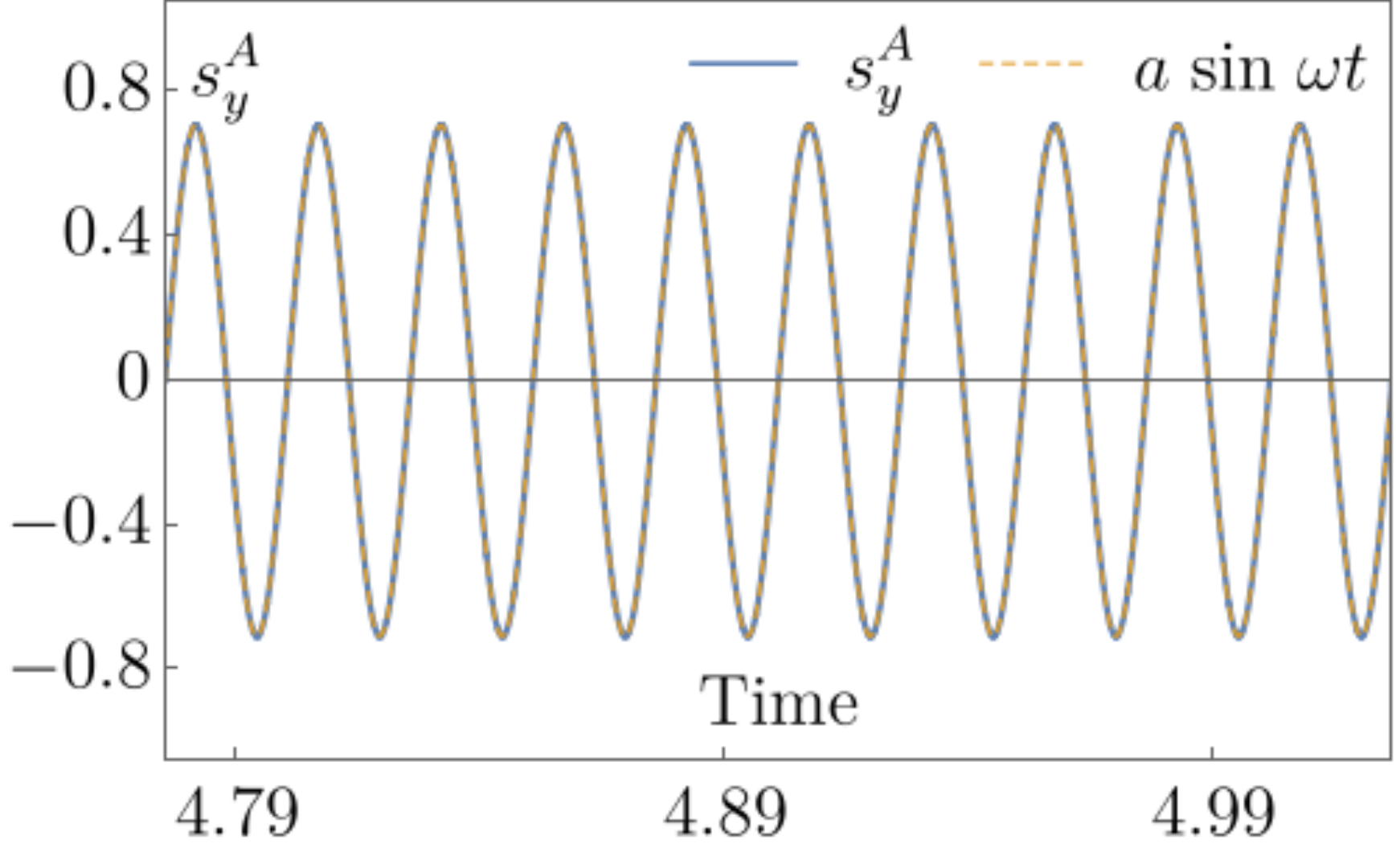}}\qquad\qquad
\subfloat[\large (d)]{\label{Perturb_W=del=1}\includegraphics[scale=0.44]{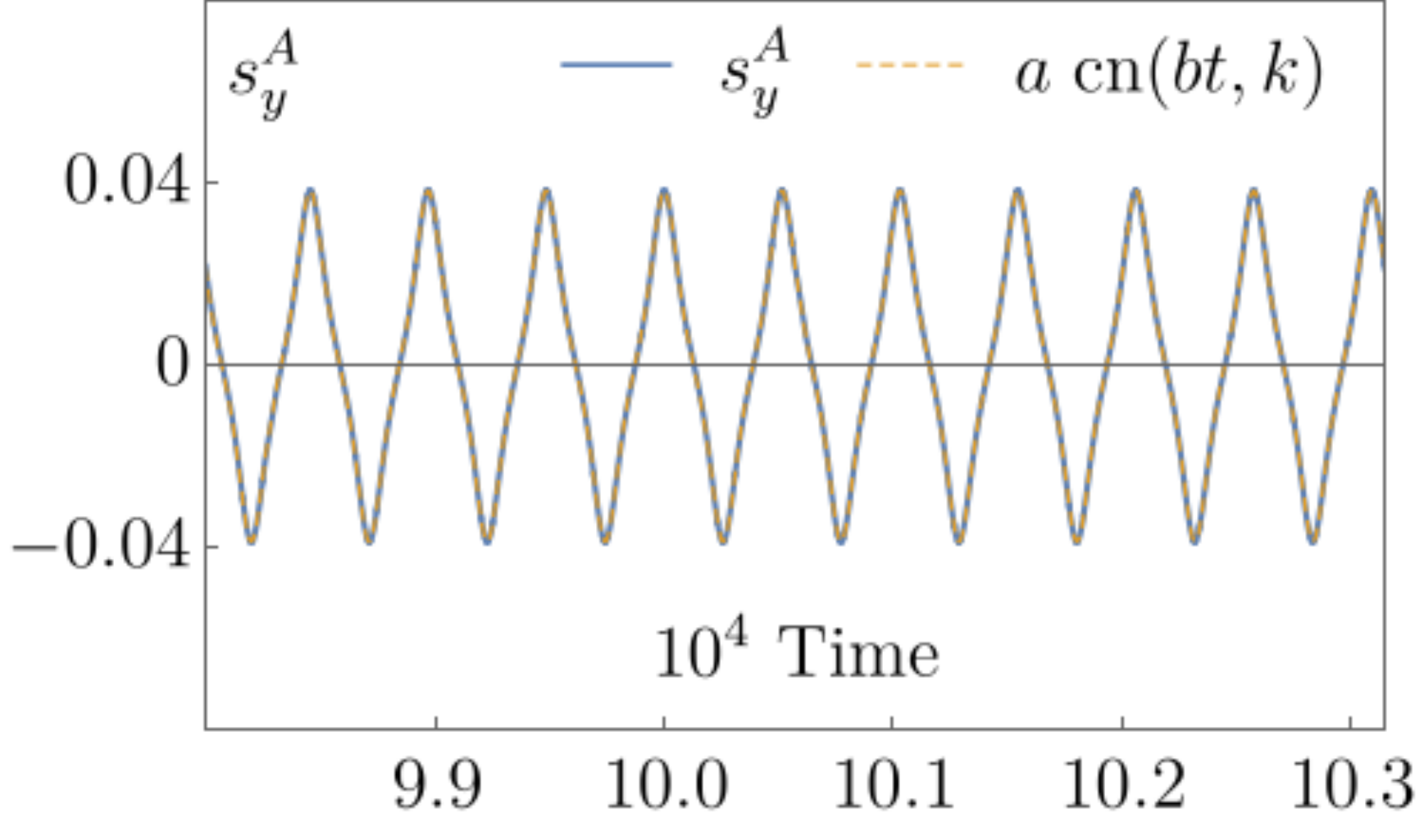}}
\caption{Comparison between numerics and  perturbative solutions \re{Symm_LC_Soln} and \re{s_y_CN}  for the $\Z2$ symmetric limit cycle.  We show the $y$-components only, the agreement for other components is similarly good.  \textbf{(a)} $\delta = 5.0, W = 10^{-3}.$ \textbf{(b)} $\delta = 5.0, W = 1 - 10^{-3}.$ \textbf{(c)} $\delta = 500, W = 0.5.$ \textbf{(d)} $\delta = 1 - (500/755)\times 10^{-3}, W = 1- 10^{-3}.$ Note
the highly anharmonic nature of the oscillations in \textbf{(d)}. }
\label{Comparison_With_Perturbation}
\end{figure*}

\subsubsection{Harmonic Solution}
\label{Harmonic_Soln}

Here we work out a simple solution for the $\Z2$-symmetric limit cycle valid in various limits.
Expressing $s_{x}$ through $s_{y}$ and  $\dot s_{y}$ in \eref{ReducedY} and substituting the result into  \eref{ReducedX}, we find  
\beg
\ddot{s}_{y} + \dot{s}_{y}\big(W - s_{z}\big) + s_{y}\bigg(\frac{\delta^{2}}{4} + \frac{W^{2}}{4} -\frac{W}{2}s_{z}\bigg) = 0.
\label{Symm_One_Spin_Eqn_s_y}  
\en%
We interpret this as a damped harmonic oscillator with a complicated feedback. To cancel the damping term, we assume $s_{z} \approx W$.
We will later check the self-consistency of this assumption. 
For a constant $s_z$,   \eref{Symm_One_Spin_Eqn_s_y} is simple to solve and using additionally \eref{ReducedY}, we derive
 \beg
s_{y} = a \sin{\omega t}, \qquad s_{x} =  a \cos{(\omega t - \alpha)},
\label{Symm_LC_Pre-Soln}  
\en%
where 
\beg
\omega = \frac{ \sqrt{\delta^{2} - W^{2} }}{2},\quad \alpha = \arctan\frac{W}{2\omega}.
\en 
Substituting this $s_x$ into \eref{ReducedZ}, we notice that it has solutions of the form  $s_{z} = C_{1} + C_{2}\cos{(2\omega t + \beta)}$,
where we require $C_1=W$ for consistency. This requirement also fixes the constants $a$, $C_2$ and $\beta$, so that (recall that $0<W<1$ in Phase   III)
\begs 
\bea
s_{x} &=& \sqrt{2W(1-W)}\cos{(\omega t - \alpha)},\\[7pt]
s_{y} &=& \sqrt{2W(1-W)}\sin{\omega t}, \\[7pt]
s_{z} &=& W - \frac{W}{\delta}(1-W)\sin{(2\omega t - \alpha)}.\label{zzzzz}
\eea
\label{Symm_LC_Soln}
\ens%
We see that the assumption $s_z\approx W$ is reasonable when  $W(1-W)\ll \delta$.  Substituting $s_z$ from \eref{zzzzz} into the coefficient of the $s_{y}$ term in \eref{Symm_One_Spin_Eqn_s_y},   we obtain  $(\delta^2 - W^2)/4 + \left[W^2(1 - W)/2\delta\right]\sin{(2\omega t - \a)}$. Therefore, to neglect the time-dependent part of $s_z$ in the frequency term in \eref{Symm_One_Spin_Eqn_s_y}, we additionally need
\beg
\omega^2=\frac{  \delta^{2} - W^{2} }{4}\gg \frac{W^2}{2\delta}(1-W).
\label{ell}
\en
These conditions are fulfilled when: (1) $\delta$ is large, (2) $W$ is small and $\delta$ is  of order 1, and (3) $W$ is close to 1 and $\delta$ is not too close to $W$, namely, $\delta-W\gg 1-W$.  In these regimes,  \eref{Symm_LC_Soln} agrees  with numerical results  very well, see \fref{Comparison_With_Perturbation}. 

In particular,  for small $W$  the solution takes the  form
\begs
\bea 
s_{x} &\approx & \sqrt{2W}\cos{(\delta t/2)}, \\ 
s_{y} &\approx & \sqrt{2W}\sin{(\delta t/2)}, \\
s_{z} &\approx & 0,
\eea
\label{W=0_Limit}
\ens%
while just below the Hopf bifurcation $W=1$ line,
\begs
\bea 
s_{x} &\approx & \sqrt{2(1-W)}\cos{(\omega t - \alpha)}, \\
s_{y} &\approx & \sqrt{2(1-W)}\sin{\omega t}, \\
s_{z} &\approx & 1,
\eea
\label{W=1_Limit}
\ens%
where $ \omega \approx \frac{1}{2}\sqrt{\delta^{2} - 1}$. Note that the amplitude and frequency of the   limit cycle in this limit matches  those  in \sref{Stability_of_TSS_Sect}. Generally, if we neglect the small sine term in \eref{zzzzz}, \eref{Symm_LC_Soln} describes an ellipse perpendicular to the $z$-axis. This is what becomes of the potato chip in \fref{Limit_Cycle}   in the harmonic approximation. When $W\ll \delta$ the ellipse turns into a circle.

\subsubsection{Elliptic Solution Close to $W = \delta = 1$ Point}
\label{ellipticsect}

The harmonic approximation breaks down  in the vicinity of the $(\delta, W)=(1, 1)$ point,
where Phases I, II and III merge in \fref{Phase_Diagram}. In this case the frequency $\omega$ of the limit cycle is small and inequality~\re{ell} does not hold. However, now we can exploit the fact that  oscillations are slow. 
As a result,  derivatives  are  suppressed  by a factor of $\omega$ in \eref{Symm_One_Spin_Eqn}, so that
\begs
\bea
s_{x} &=& \frac{W}{\delta}s_{y} + \frac{2}{\delta}\dot{s}_{y}\approx \frac{W}{\delta}s_{y}, \label{W=del=1_s_x}\\
s_{z} &=& 1 - \frac{s^{2}_{x}}{W} - \frac{\dot{s}_{z}}{W} \approx 1 - \frac{s^{2}_{x}}{W}\label{W=del=1_s_z}. \label{szsmder}
\eea 
\label{W=del=1_s_x_z}
\ens 
Let
\beg
W=1-\epsilon,\quad\delta=1+r,
\en
where $\epsilon>0$ and $r$ are small. Substituting $s_z$ from \eref{szsmder} into \eref{Symm_One_Spin_Eqn_s_y}, we obtain
\beg
\ddot s_y -\dot s_y \left( 2s_y^2+\epsilon\right)+\frac{s_y}{2}\left(  r + s_y^2\right)=0,
\label{W=del=1_Full_s_y}
\en 
where we kept only the leading orders in $\epsilon$ and $r$ in the coefficients. 
Assuming $r$ is of the order of $\epsilon$, we see from \eref{W=1_Limit}   that in the harmonic solution $s_y$ and the frequency $\omega$
are both of the order $\sqrt{\epsilon}$.  Then, the $\dot s_y$ term in \eref{W=del=1_Full_s_y} is negligible. We also verified that this as well as the approximations in \eref{W=del=1_s_x_z} are self-consistent regardless of the magnitude of $|r|/\epsilon$ using the Jacobi elliptic solution for $s_y$ we work out below. 

Neglecting the $\dot s_y$ term in \eref{W=del=1_Full_s_y} , we find
\beg
\ddot{s}_{y} + \frac{r}{2}s_{y} +\frac{1}{2} s^{3}_{y} = 0.
\label{W=del=1_s_y}  
\en%
This reduces to the standard equation for the Jacobi elliptic function cn \cite{Gradshteyn_Ryzhik, Abramowitz_Stegun} via a substitution
\beg 
s_{y}(t) = a\;\textrm{cn}(bt,k),
\label{s_y_CN}
\en%
where
 \beg
b^{2} =  \frac{r}{2(1 - 2k^{2})},\quad
a^{2} = \frac{2k^{2}r}{1-2k^{2}}. 
\label{W=del=1_Constraints_1,2} 
\en
\eref{W=del=1_Constraints_1,2}   provides two constraints for three undetermined parameters  $a, b$ and $k$. We derive  one more constraint by minimizing the effect of the the $\dot s_y$ term that we neglected in \eref{W=del=1_Full_s_y}. Specifically, we  multiply \eref{W=del=1_Full_s_y} by $\dot{s}_{y}$ and integrate over one period. 
  Since   the first and the last terms   turn into complete derivatives, we are left with  
\beg 
 \epsilon \mean{\dot{s}^{2}_{y}} = 2\mean{s^{2}_{y}\dot{s}^{2}_{y}}.
\label{W=del=1_Constraints_3}
\en%
 Evaluating the integrals  on both sides of this equation, we derive
 \beg 
a^{2} = \frac{\epsilon}{2}Y(k),
\label{Elliptic_Soln_Constraints_Final}
\en%
where 
\beg 
Y(k)  = \frac{5k^{2}\big[(2k^{2}-1)E  + (1-k^{2})K \big]}{2(k^{4}-k^{2}+1)E  -(2-k^{2})(1-k^{2}) K },
\label{g(k)}
\en 
and $K\equiv K(k)$ and $E\equiv E(k)$ are complete elliptic integrals of the first and second kind, respectively. Matching \eref{Elliptic_Soln_Constraints_Final} to $a^2$ in   \eref{W=del=1_Constraints_1,2} yields an equation  for $k$,
\beg  
Z(k) \equiv \frac{4k^{2}}{5(1-2k^{2})Y(k)} =   \frac{\epsilon}{r}.
\label{W=del=1_Constraints_3_k}
\en%
 This equation along with  \eref{W=del=1_Constraints_1,2}  specify all three constants of $s_y(t)$ in \eref{s_y_CN}. A plot of
 $Z(k)$ for $r<0$ appears in \fref{f(k)_vs_k}. We see that when $\eps/r$ is between  $-8/5$ and the maximum of $Z(k)$ at approximately $-1.50$,
 there are two solutions for $k$. Numerically we observe that  the evolution with \eref{Symm_One_Spin_Eqn} picks up the solution with smaller $k$ in this case. 
 
 The oscillation period, 
 \beg
 T= \frac{4K(k)\sqrt{ 2|1-2k^2|}}{\sqrt{|r|}},
 \label{ellperiod}
 \en
 diverges when $r\to 0$ at fixed $\epsilon/r$.

 For small $r$ and $\epsilon$, the condition \re{ell} of validity of the harmonic solution reads $r\gg \epsilon>0$.  Hence,   the elliptic and harmonic solutions must agree in the limit $\epsilon/r\to 0^+$, which we now check.  The solution of \re{W=del=1_Constraints_3_k} to the first  order in $\epsilon/r$ is $k^2\approx\epsilon/r$. Elliptic functions become harmonic when $k\to 0$. In particular, $s_{y} = a\,\cn (bt) \approx a\cos{(bt)}$ \cite{Elliptic}. Further, \eref{W=del=1_Constraints_1,2} yields $a^2=2\epsilon$ and $b^2=r/2$, which agrees exactly with $a^2$ and $\omega^2$ for the harmonic solution for small $\epsilon$, $r$, and $\epsilon/r$. We also checked that this agreement does not go beyond the leading order in $\epsilon/r$.

 \subsubsection{Gauging the Taper of Coexistence Region Near $W=\delta = 1$}
 \label{taper}
 
 \begin{figure}[tbp!]
\begin{center}
\includegraphics[scale=0.44]{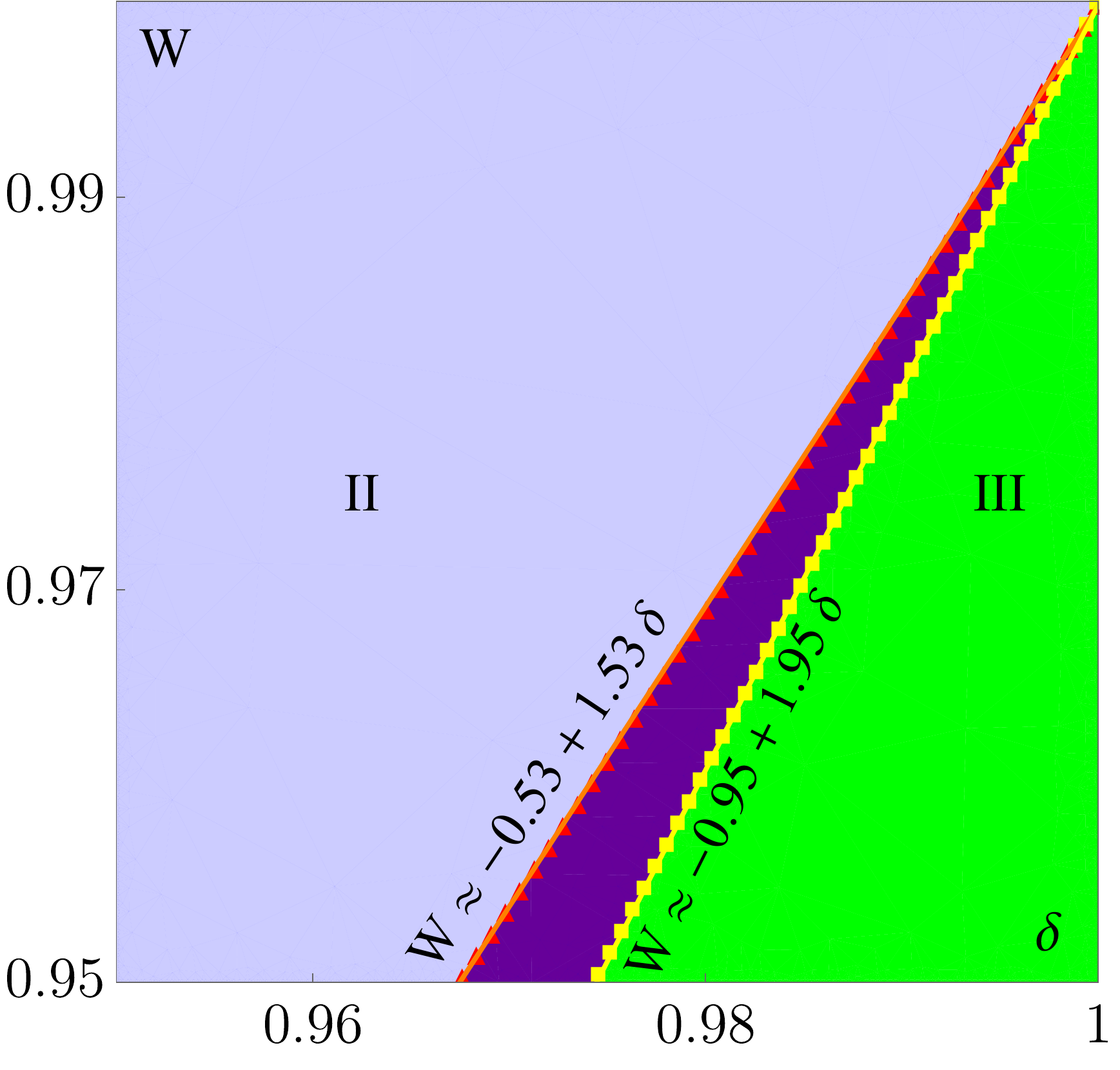}
\caption{Linear fit to the boundaries of the coexistence region (purple) near $\delta = W = 1$.  The right boundary  is the subcritical Hopf bifurcation line. Yellow squares is the numerical result for this line and   solid yellow line  with the slope $1.95$   is the best linear fit to these points. Orange triangles depict points on the left boundary of the coexistence region. The best linear fit to these points (solid orange line) has a slope 1.53.}
\label{Taper_Coexistence}
\end{center}
\end{figure}

\begin{figure}[tbp!]
\begin{center}
\includegraphics[scale=0.45]{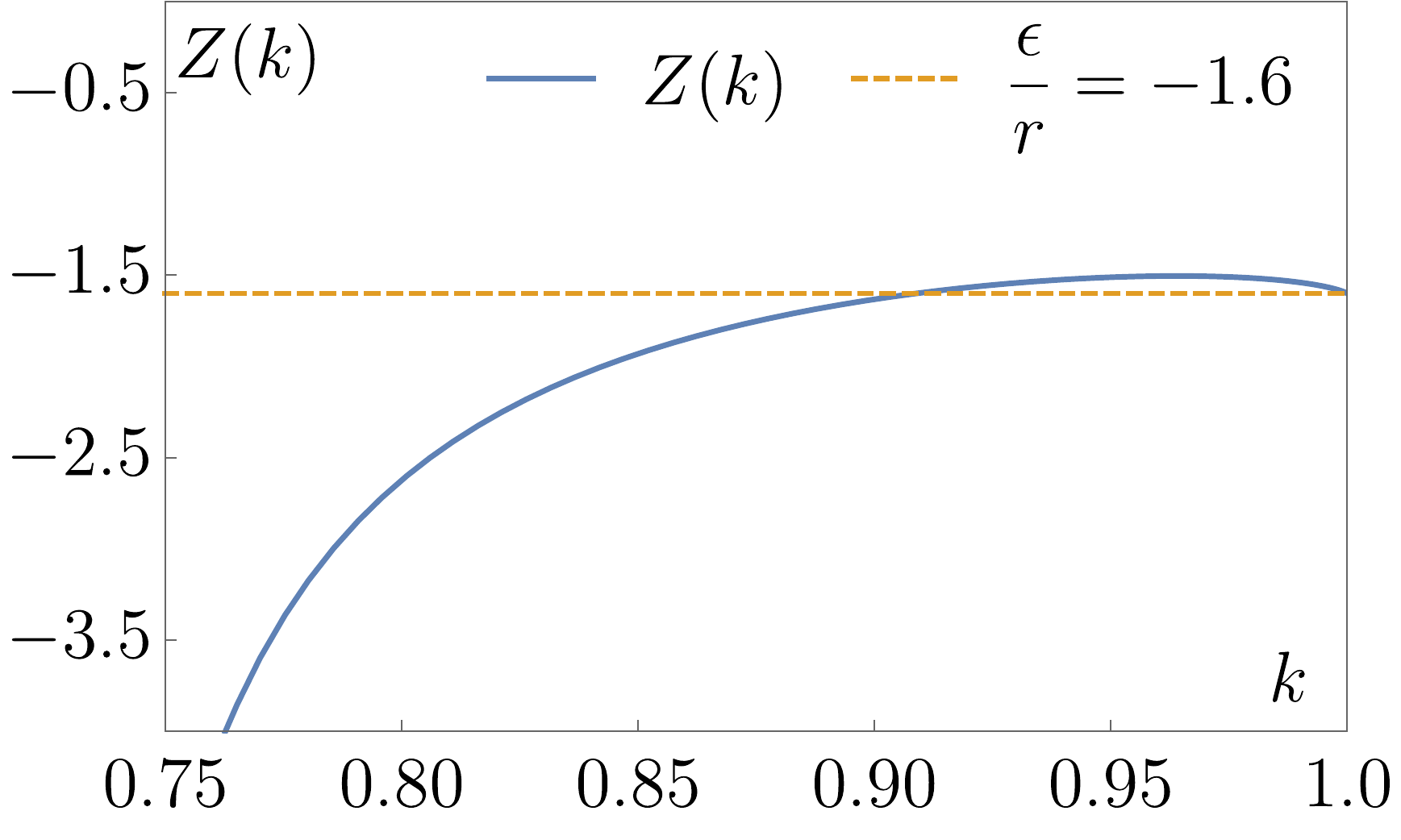}
\caption{ The function $Z(k)$  defined in \eref{W=del=1_Constraints_3_k}  in the domain $1/\sqrt{2}<k<1$. We show $Z(1) = -1.6$ by the dashed line. In this interval $Z(k)$ reaches its maximum of -1.50 at $k\approx 0.96$.}
\label{f(k)_vs_k}
\end{center}
\end{figure}

 The coexistence region gradually tapers off to a point   $(\delta, W) = (1, 1)$, see \fref{Taper_Coexistence}.  Let us determine  the angle with which the coexistence region approaches its pinnacle. To that end, we utilize the elliptic solution for the $\Z2$-symmetric limit cycle valid near $\delta = W = 1$ point. The right boundary of the coexistence region is the subcritical Hopf bifurcation line~\re{d-W}. Linearizing \eref{d-W} in $\epsilon$ and $r$, we find $\epsilon=-2r$. 
 Therefore, the line tangential to the  right boundary at $(\delta, W) = (1, 1)$ has a slope  
 \beg
 \tan{\theta_{R}}  =2.
 \label{thR}
 \en
  We also see that $r<0$ in the coexistence region, because $\epsilon=1-W>0$.
 
 The left boundary of the coexistence region near $\delta = W = 1$ point is the line where the elliptic solution \re{s_y_CN} ceases to exist.
Since $r<0$,   in order for $b^2$ in  \eref{W=del=1_Constraints_1,2}  to be positive, we need $k>1/\sqrt{2}$.  Thus, for the elliptic solution   to exist inside the coexistence region,   the solution of \eref{W=del=1_Constraints_3_k}  must satisfy  $1/\sqrt{2}<k<1$. We    plot $Z(k)$   in this interval  in \fref{f(k)_vs_k}. Observe that $Z(1) = - 8/5=-1.6$ and the maximum value of $Z(k)$   is approximately $-1.50$.  Therefore, \eref{W=del=1_Constraints_3_k}  has no solutions in the desired range when $\epsilon \le -1.50r$ and the slope   of the tangent to the left boundary of the coexistence region at $(\delta, W) = (1, 1)$ is 
\beg
\tan{\theta_{L}} \approx 1.50.
\label{thL}
\en
\esref{thR} and \re{thL} agree well with the results of numerical analysis shown in  \fref{Taper_Coexistence}. The taper angle of the coexistence region near  $\delta = W = 1$  is $\theta_{R} - \theta_{L} \approx \arctan{(2)} - \arctan{(1.5)} \approx 7^{\circ}$.

\subsection{Stability of the $\Z2$-Symmetric Limit Cycle: Floquet Analysis}
\label{Floquet}

\begin{table}[tbt!]
\centering 
\begin{tabular}{| c | c | c |}
\hline
$(\delta, W)$& \begin{tabular}{ c }$\delta_{>},$ \\ $\rho_{1}$, $\rho_{2}$, $\rho_{3}$ \end{tabular} & \begin{tabular}{ c }$\delta_{<},$ \\ $\rho_{1}$, $\rho_{2}$, $\rho_{3}$ \end{tabular} \\
[0.5ex]
\hline\hline
(0.40, 0.051) & \begin{tabular}{ c } 0.42, \\ 1.0, 0.99, 0.27 \end{tabular} & \begin{tabular}{ c } 0.41, \\ 1.0, 1.1, 0.24 \end{tabular} \\
\hline
(0.52, 0.10) & \begin{tabular}{ c } 0.54, \\ 1.0, 0.97, 0.14 \end{tabular} & \begin{tabular}{ c } 0.53, \\ 1.0, 1.0+, 0.066 \end{tabular} \\
\hline
(0.54, 0.15) & \begin{tabular}{ c } 0.56, \\ 1.0, 1.0-, 0.069 \end{tabular} & \begin{tabular}{ c } 0.55, \\ 1.0, 1.0+, 0.031 \end{tabular} \\
\hline
(0.53, 0.20) & \begin{tabular}{ c } 0.53, \\ 1.0, 0.99, 0.033 \end{tabular} & \begin{tabular}{ c } 0.52, \\ 1.0, 1.0+, 0.031 \end{tabular} \\
\hline
(0.55, 0.25) & \begin{tabular}{ c } 0.55, \\ 1.0, 1.0-, 0.017 \end{tabular} & \begin{tabular}{ c } 0.54, \\ 1.0, 1.0+, 0.016 \end{tabular} \\
\hline
(0.54, 0.30) & \begin{tabular}{ c } 0.58, \\ 1.0, 1.0-, 0.00+ \end{tabular} & \begin{tabular}{ c } 0.57, \\ 1.0, 1.0+, 0.00+ \end{tabular} \\
\hline
(0.60, 0.35) & \begin{tabular}{ c } 0.61, \\ 1.0, 1.0-, 0.00+ \end{tabular} & \begin{tabular}{ c } 0.60, \\ 1.0, 1.0+, 0.00+ \end{tabular} \\
\hline
(0.64, 0.40) & \begin{tabular}{ c } 0.64, \\ 1.0, 0.97, 0.00+ \end{tabular} & \begin{tabular}{ c } 0.63, \\ 1.0, 1.1, 0.00+ \end{tabular} \\
\hline
(0.64, 0.45) & \begin{tabular}{ c } 0.67, \\ 1.0, 0.90, 0.00+ \end{tabular} & \begin{tabular}{ c } 0.66, \\ 1.0, 1.0+, 0.00+ \end{tabular} \\
\hline
(0.67, 0.50) & \begin{tabular}{ c } 0.69, \\ 1.0, 0.93, 0.00+ \end{tabular} & \begin{tabular}{ c } 0.68, \\ 1.0, 1.1, 0.00+ \end{tabular} \\
\hline
(0.70, 0.55) & \begin{tabular}{ c } 0.71, \\ 1.0, 0.98, 0.00+ \end{tabular} & \begin{tabular}{ c } 0.70, \\ 1.0, 1.1, 0.00+ \end{tabular} \\
\hline
\end{tabular}
\caption{As discussed in the text, we determine points $(\delta, W)$ where the $\Z2$-symmetric limit cycle  loses stability by requiring that $|s_{z}^{A} - s_{z}^{B}|$ exceeds a certain threshold at large times. The 1st column shows some of these points. Here we compare them to the results of Floquet stability analysis. The 2nd and 3rd columns show Floquet multipliers $\rho_1$, $\rho_2$, and $\rho_3$ for the same $W$ as in the 1st column and two values of $\delta$, $\delta=\delta_>$ and $\delta_<$. We see that one of $\rho_i$ exceeds 1, signaling instability, at approximately the same value of $\delta$ as in the 1st column. We indicate values in the intervals  $(0.00, 0.01)$, $(1.00, 1.01)$ and $(0.99, 1.00)$ as $0.00+$, $1.0+$ and $1.0-$.}
\label{Tab_Z_2_Symmetry_Breaking}
\end{table}

In this section, we analyze the stability of  $\Z2$-symmetric limit cycles using the Floquet theory.  As before, it is convenient to rotate the coordinate system by a fixed angle so  that  the $\Z2$-symmetric limit cycle obeys the constraints \re{Z2_Expl}. 
We   introduce  symmetric coordinates covering the  $\Z2$-symmetric submanifold,
\beg 
s_{x} = \frac{ s^{A}_{x} + s^{B}_{x}}{2},  \quad s_{y} = \frac{ s^{A}_{y} - s^{B}_{y}}{2},\quad s_{z} = \frac{ s^{A}_{z} + s^{B}_{z}}{2},
\label{Floquet_Symm_Var}
\en%
 and transverse coordinates that take the dynamics away from  it,
\beg 
q_{x} =  s^{A}_{x} - s^{B}_{x}, \quad
q_{y} =  s^{A}_{y} + s^{B}_{y}, \quad
q_{z} =  s^{A}_{z} - s^{B}_{z}. 
\label{Floquet_Asymm_Var}
\en
Recasting \eref{Mean-Field_2} in terms of the new variables, we have,
\begs
\bea
\dot{s}_{x} &=& - \frac{\delta}{2}s_{y}- \frac{W}{2}s_{x} + s_{z}s_{x}, \\ [7pt]
\dot{s}_{y} &=& \frac{\delta}{2}s_{x} - \frac{W}{2}s_{y} - \frac{1}{4}q_{y}q_{z}, \\ [7pt]
\dot{s}_{z} &=&  W\big(1 - s_{z}\big) -  s_{x}^{2} - \frac{1}{4}q_{y}^{2}\\[7pt]
\dot{q}_{x} &=& - \frac{\delta}{2}q_{y}- \frac{W}{2}q_{x} + s_{x}q_{z}, \\ [7pt]
\dot{q}_{y} &=& \frac{\delta}{2}q_{x} + \left(s_{z} - \frac{W}{2}\right)q_{y}, \\[7pt] 
\dot{q}_{z} &=& - W q_{z} - s_{x}q_{x} - s_{y}q_{y}. 
\eea 
\label{Floquet_Mean-Field}
\ens%
 For the unperturbed $\Z2$-symmetric limit cycle      $q_{x} = q_{y} = q_{z} = 0$ and   $\bm{s}$ obeys the reduced spin equations   \re{Symm_One_Spin_Eqn}.   To analyze the linear stability with respect to symmetry-breaking perturbations, we   linearize \eref{Floquet_Mean-Field} in $\bm{m}$, 
 \begs
\bea
\dot{q}_{x} &=& - \frac{\delta}{2}q_{y}- \frac{W}{2}q_{x} + s_{x}(t)q_{z}, \\ [7pt]
\dot{q}_{y} &=& \frac{\delta}{2}q_{x} + \left(s_{z}(t) - \frac{W}{2}\right)q_{y}, \\[7pt] 
\dot{q}_{z} &=& - W q_{z} - s_{x}(t)q_{x} - s_{y}(t)q_{y}, 
\eea 
\label{Floquet_Eqn}
\ens%
where $s_{x}(t), s_{y}(t)$ and $s_{z}(t)$ are the  spin components  for the $\Z2$-symmetric limit cycle, which we obtain  separately by simulating \eref{Symm_One_Spin_Eqn}. They play the role of   periodic in time coefficients for the linear system~\re{Floquet_Eqn}. 

 By Floquet theorem the general solution of \eref{Floquet_Eqn} is
\beg 
\bm{q}(t) = \sum_{i=1}^3  d_{i} e^{\varkappa_{i}t}\bm{p}_{i}(t),
\label{Floquet_Gen_Perturb}
\en%
where $\bm{p}_{i}(t)$ are   linearly independent vectors periodic with the same period $T$ as the limit cycle, $d_i$ are arbitrary constants, and $\varkappa_i$ are the Floquet exponents.. To evaluate
  $\varkappa_i$, we first compute the monodromy matrix $\mathbb{M} = [\mathbb{S}(0)]^{-1} \mathbb{S}(T)$, where
$\mathbb{S}(t)$ is a $3\times3$ matrix whose columns are any three linearly independent solutions $\bm q_i(t)$ of \eref{Floquet_Eqn}. We determine $\bm q_i(t)$ by simulating  \eref{Floquet_Eqn} for one period for three randomly chosen initial conditions. The eigenvalues of the
monodromy matrix   $\rho_i=e^{\varkappa_{i}T}$ are known as characteristic multipliers.  If one of the   them is greater than one, the corresponding 
$\varkappa_i$ is positive and the $\Z2$-symmetric limit cycle is unstable.

\begin{figure}[tbp!]
\begin{center}
\includegraphics[scale=0.46]{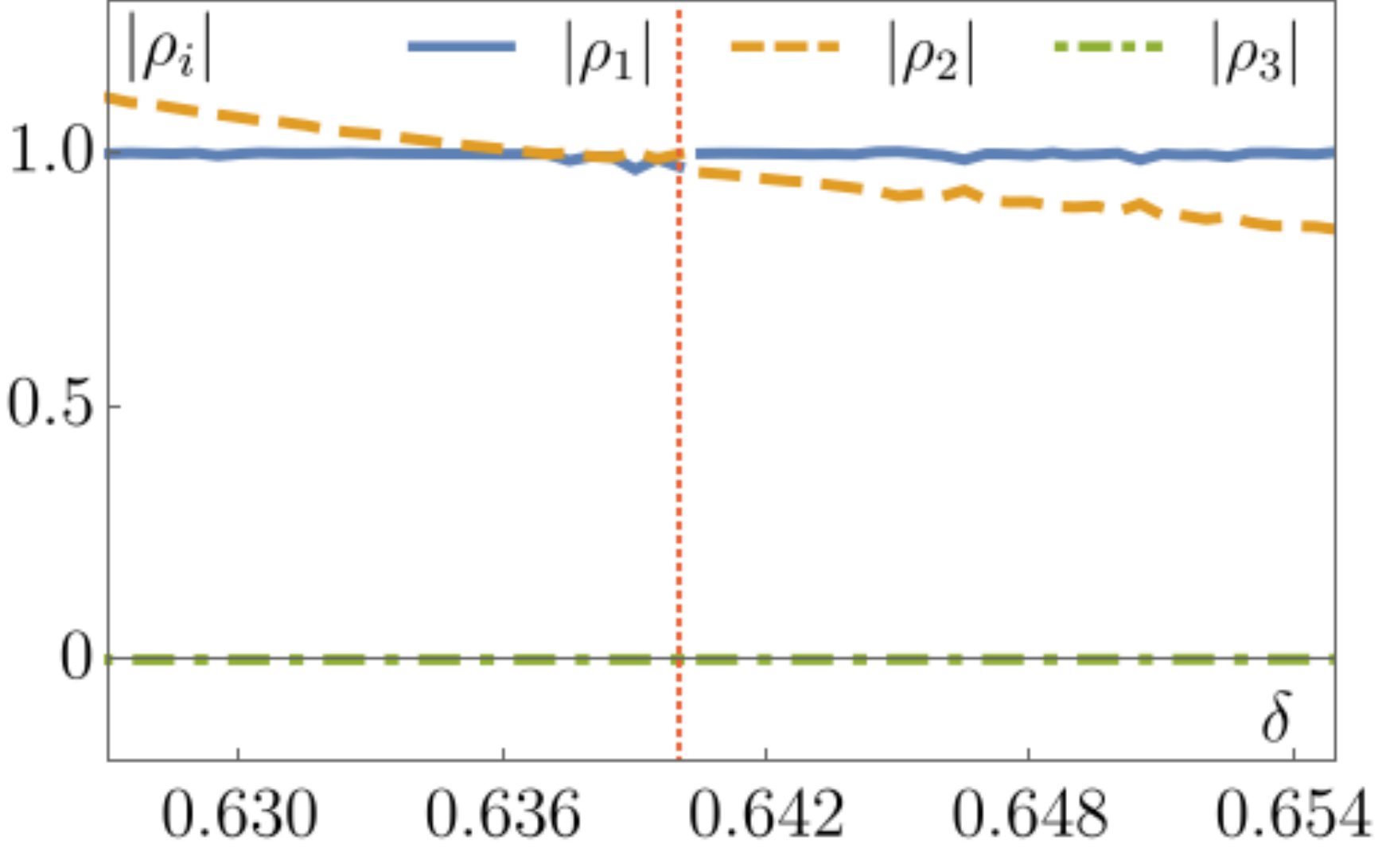}
\caption{Absolute values of  Floquet multipliers $|\rho_i|$ as functions of $\delta$ for the $\Z2$-symmetric limit cycle at $W = 0.40$. The red dotted line marks the point of symmetry breaking (loss of stability)  according to the criterion $|s_{z}^{A} - s_{z}^{B}|>0.01$ at large times.  Across criticality $|\rho_{1}| = 1$  indicates the presence of nearby $\Z2$-symmetric limit cycles related by overall rotations about the $z$-axis, whereas $|\rho_{3}|<1$ corresponds to a stable direction.}
\label{Floquet_Ex_Pic}
\end{center}
\end{figure}

\begin{figure*}[tbp!]
\centering
\subfloat[\large (a)]{\includegraphics[scale=0.37]{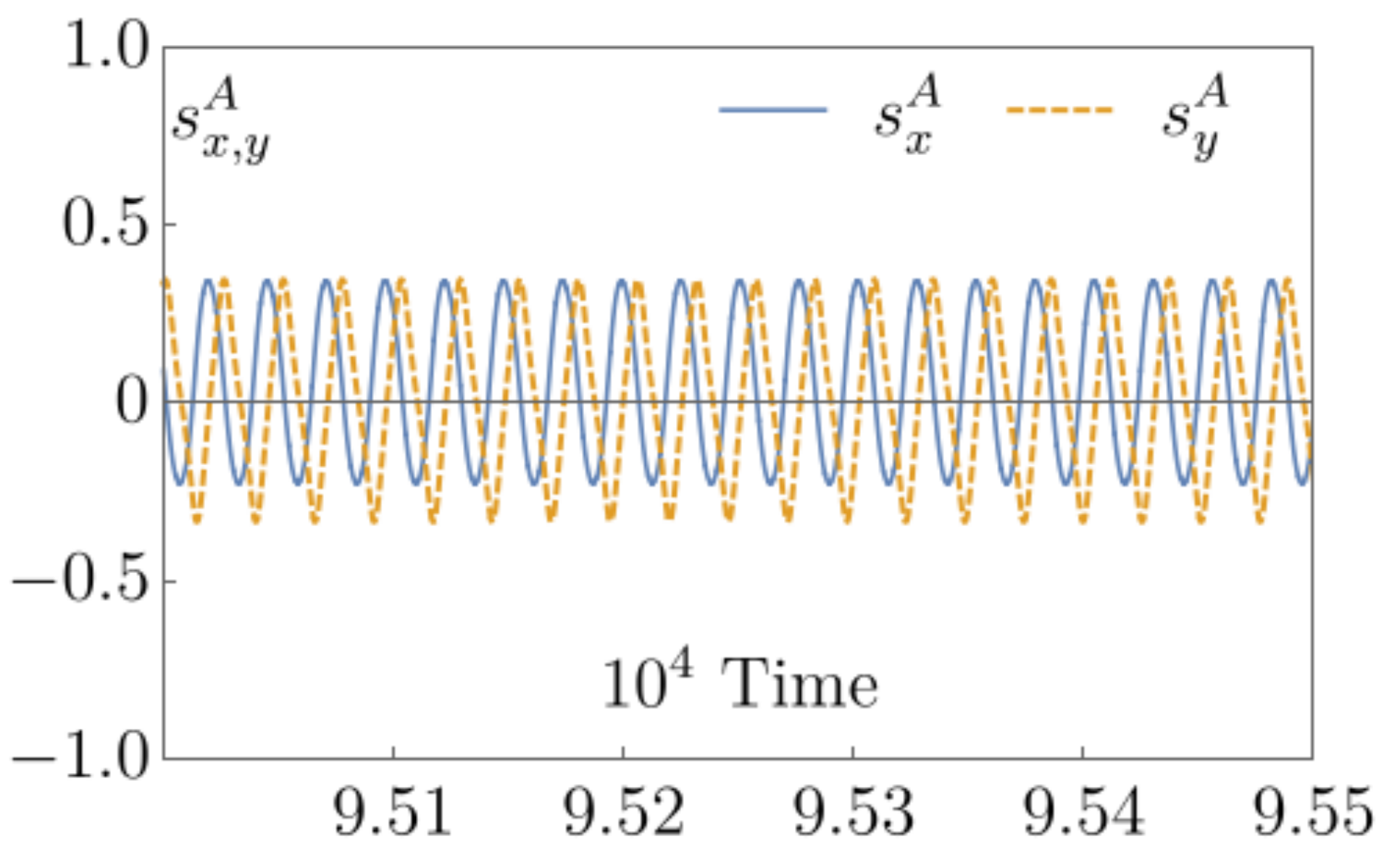}}\qquad\qquad
\subfloat[\large (b)]{\includegraphics[scale=0.37]{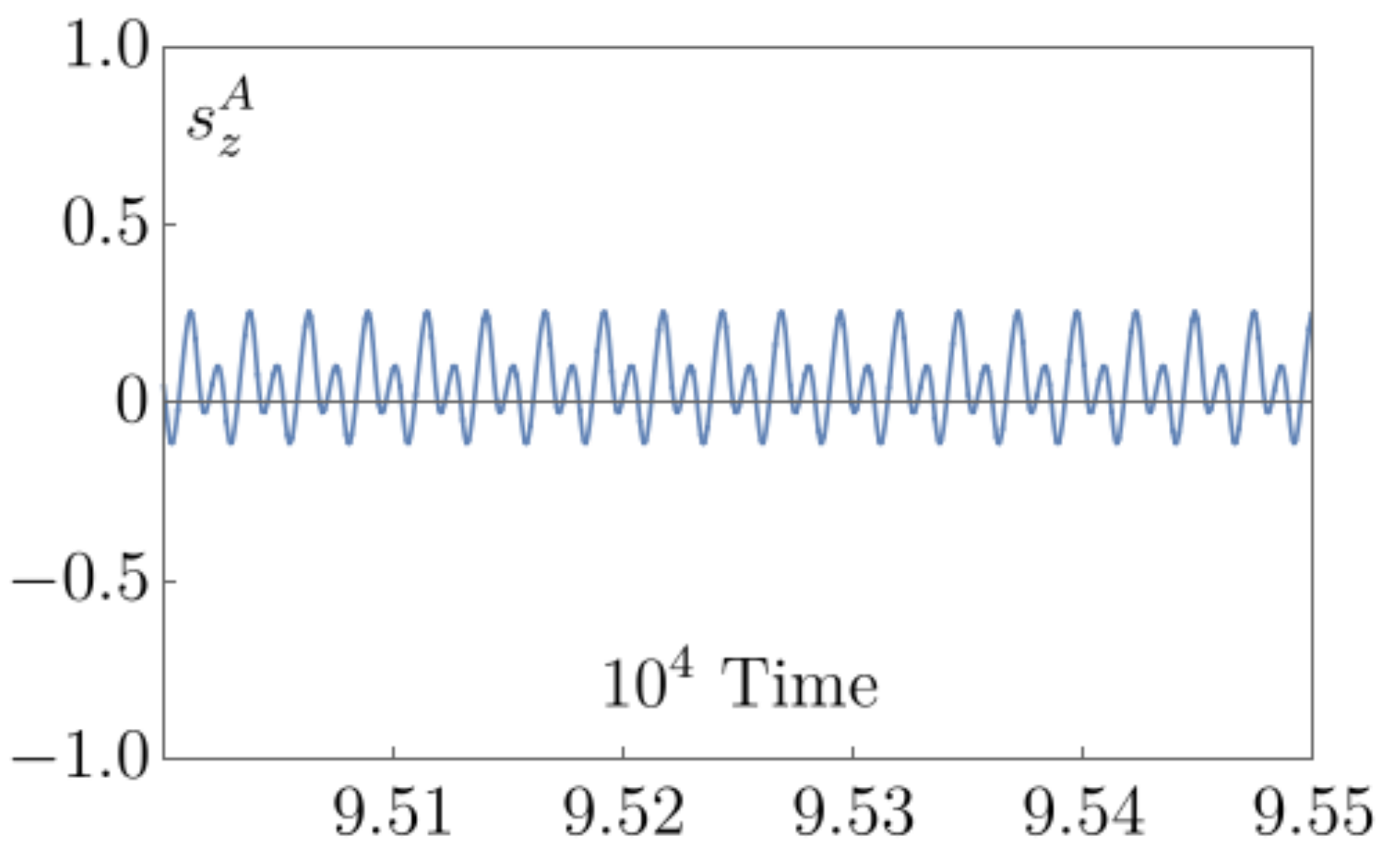}}\\
\subfloat[\large (c)]{\includegraphics[scale=0.37]{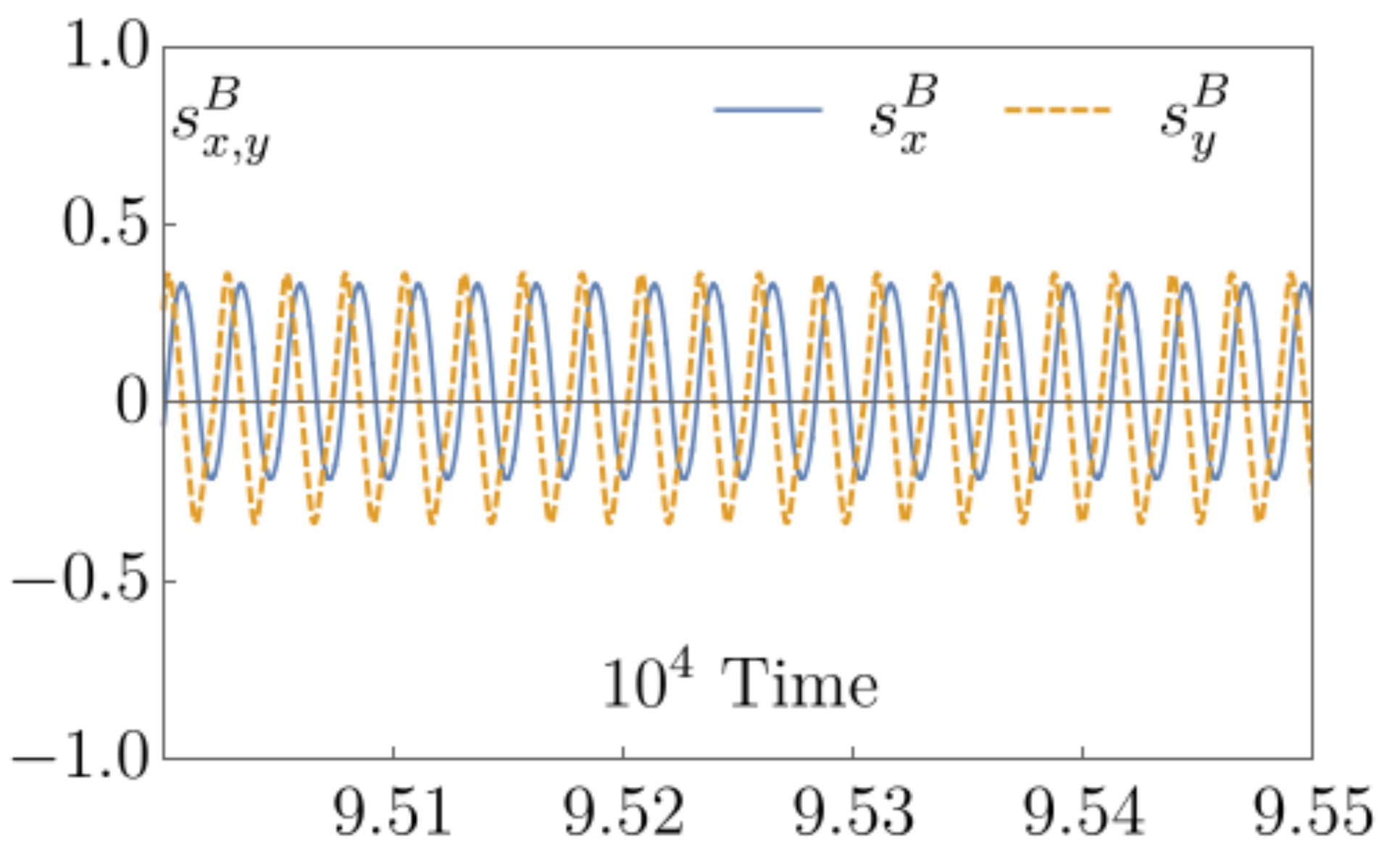}}\qquad\qquad
\subfloat[\large (d)]{\includegraphics[scale=0.37]{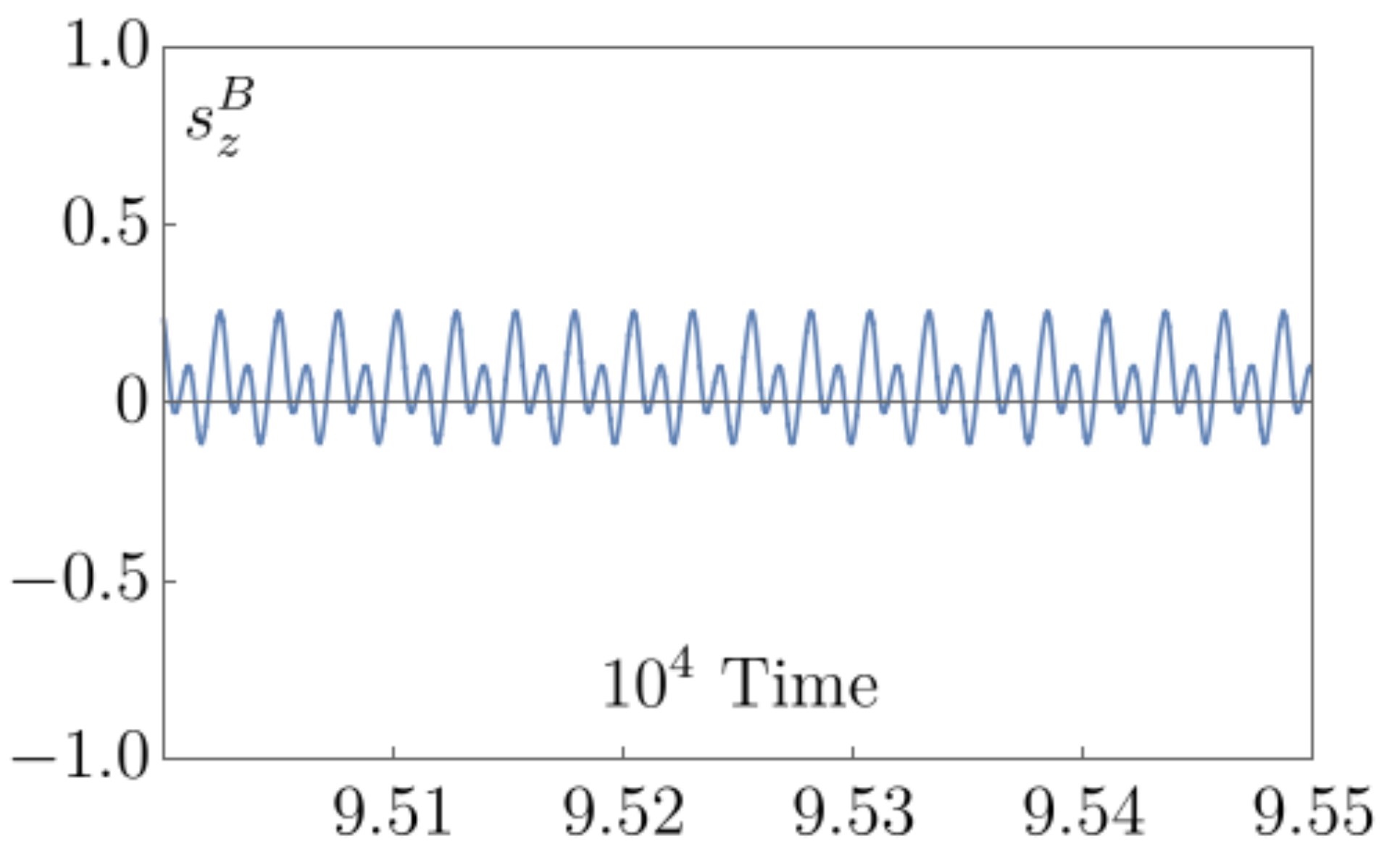}}\\
\subfloat[\large (e)]{\includegraphics[scale=0.37]{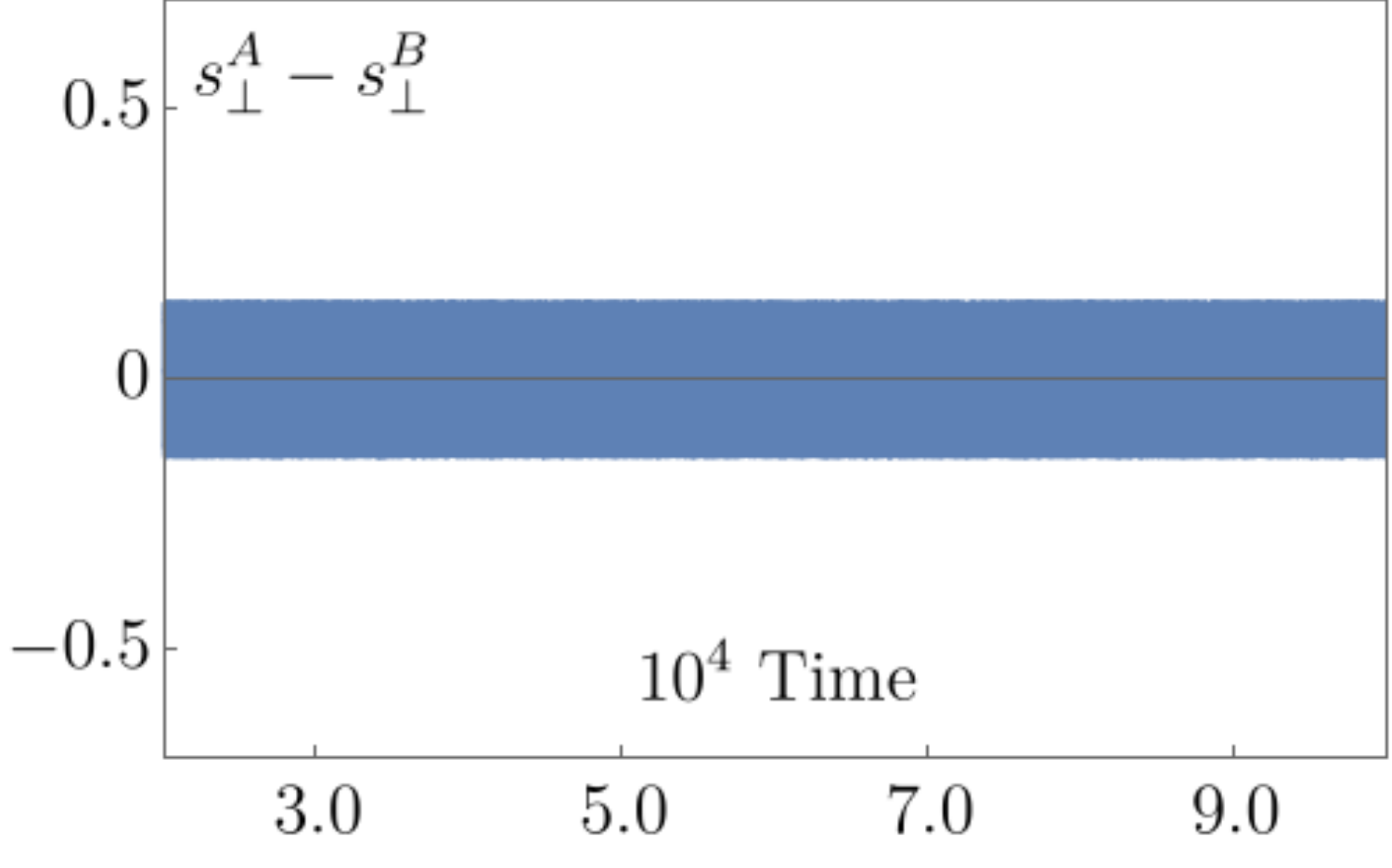}}\llap{\raisebox{2.8cm}{\includegraphics[height=1.8cm]{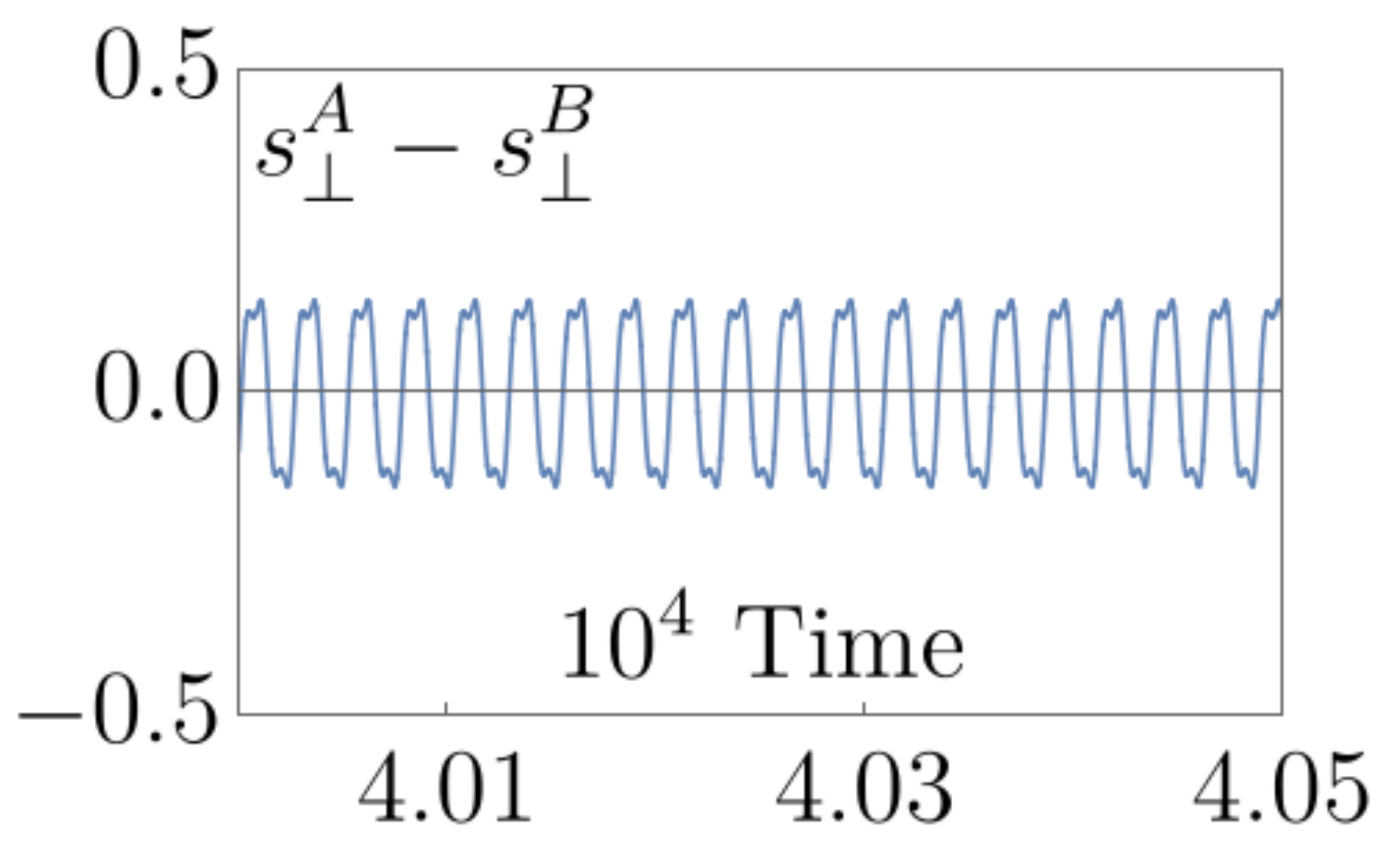}}}\qquad\qquad
\subfloat[\large (f)]{\includegraphics[scale=0.37]{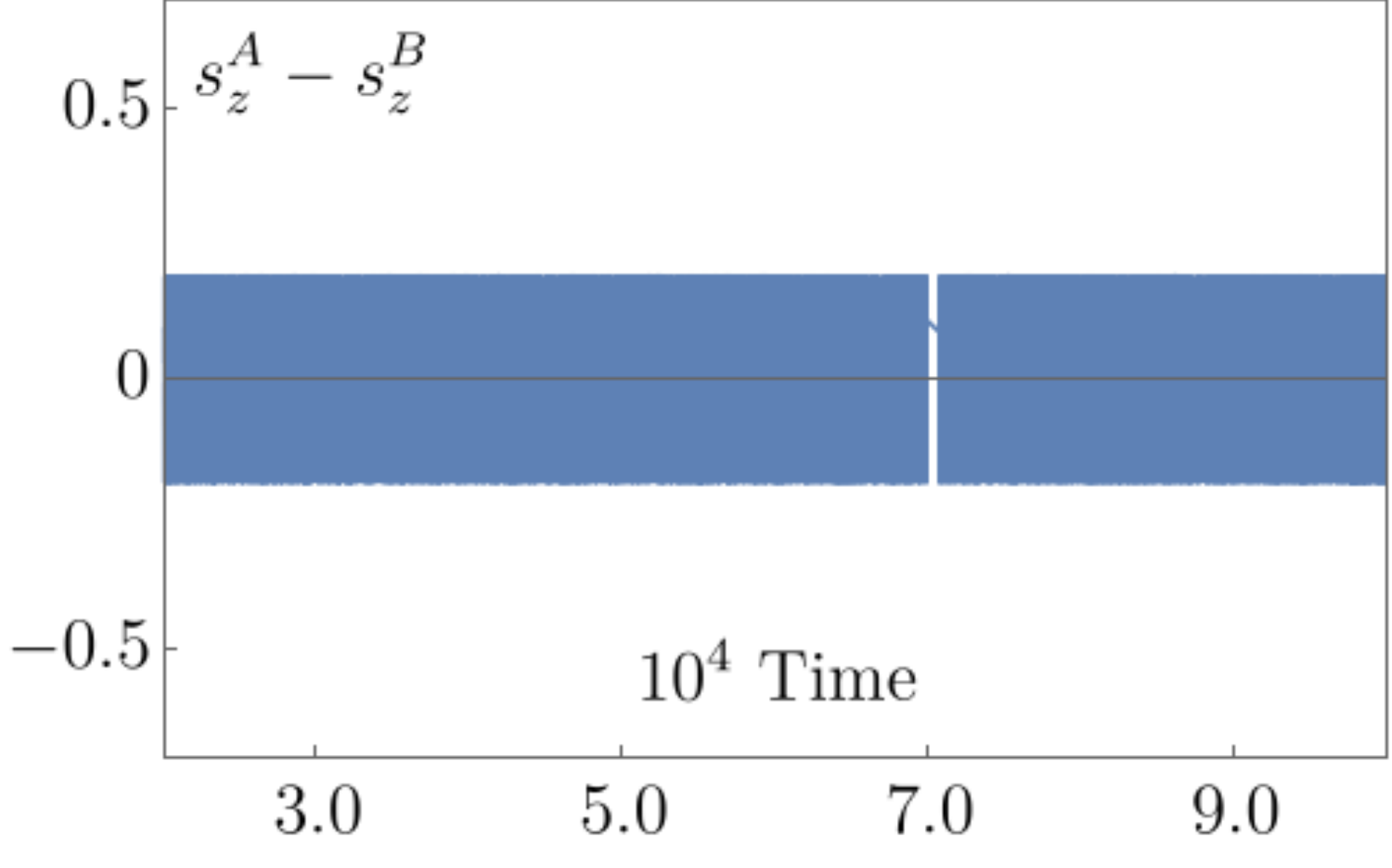}}\llap{\raisebox{2.9cm}{\includegraphics[height=1.8cm]{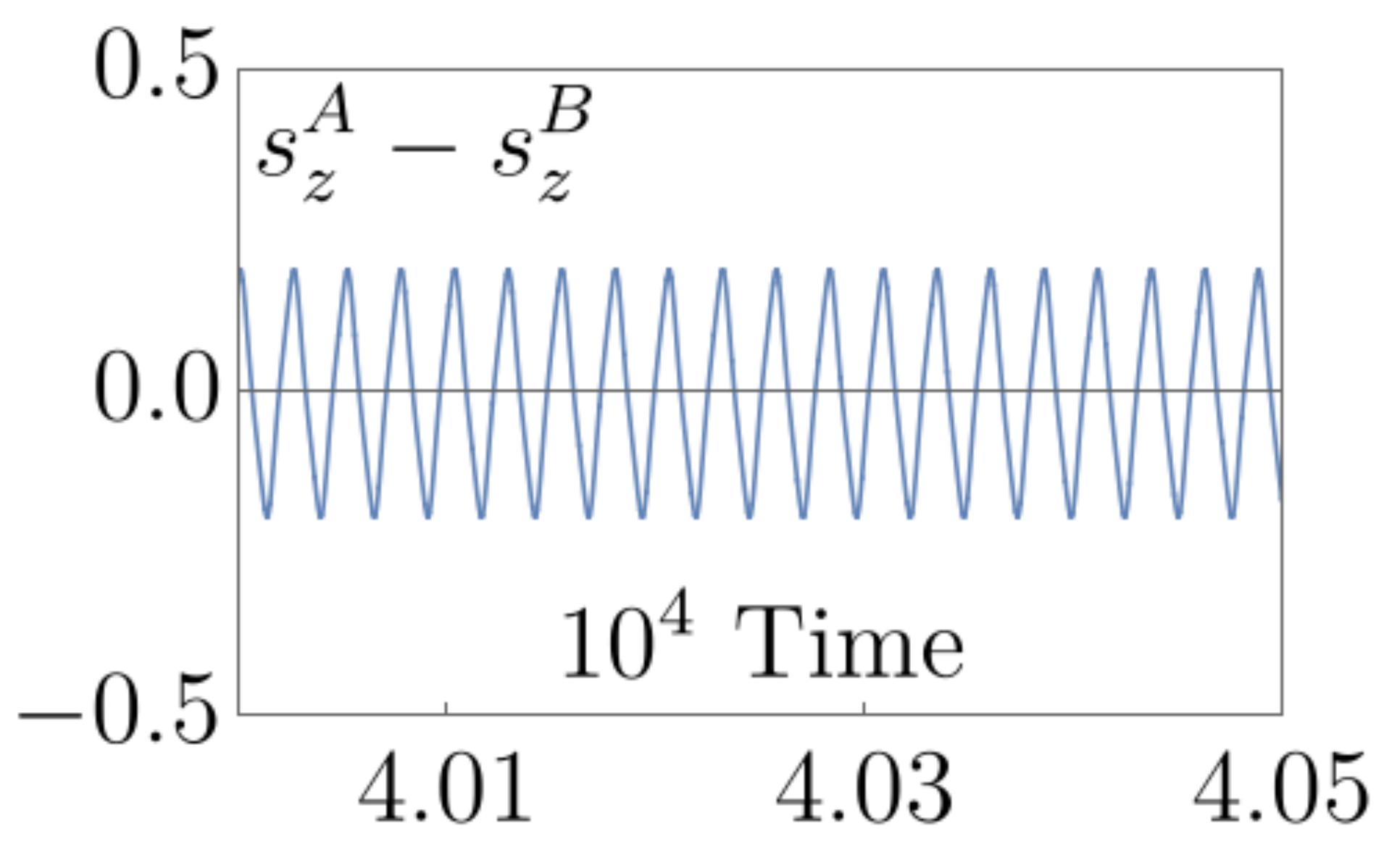}}}
\caption{A   limit cycle at $(\delta, W) = (0.42, 0.056)$. Classical spins $\bm s^A$ and $\bm s^B$ describe the mean-field dynamics of two atomic ensembles A and B coupled to a heavily damped cavity mode. This limit cycle breaks the $\Z2$-symmetry because  $s_\perp^A-s_\perp^B$ and $s_z^A-s_z^B$ are nonzero and time-dependent (see also \fref{Symmetry-broken_LC_X_Sect}), unlike for the $\Z2$-symmetric limit cycle at a nearby point  $(\delta, W) = (0.44, 0.056)$ shown in \fref{Symmetric_LC}. For these values of the detuning $\delta$ between the two ensembles and pumping $W,$ the cavity radiates the frequency comb shown in \fref{Power_Spectrum_Asymmetric_LC}.}
\label{Symmetry-broken_LC}
\end{figure*}

\begin{figure*}[tbp!]
\centering
\subfloat[\large (a)]{\includegraphics[scale=0.34]{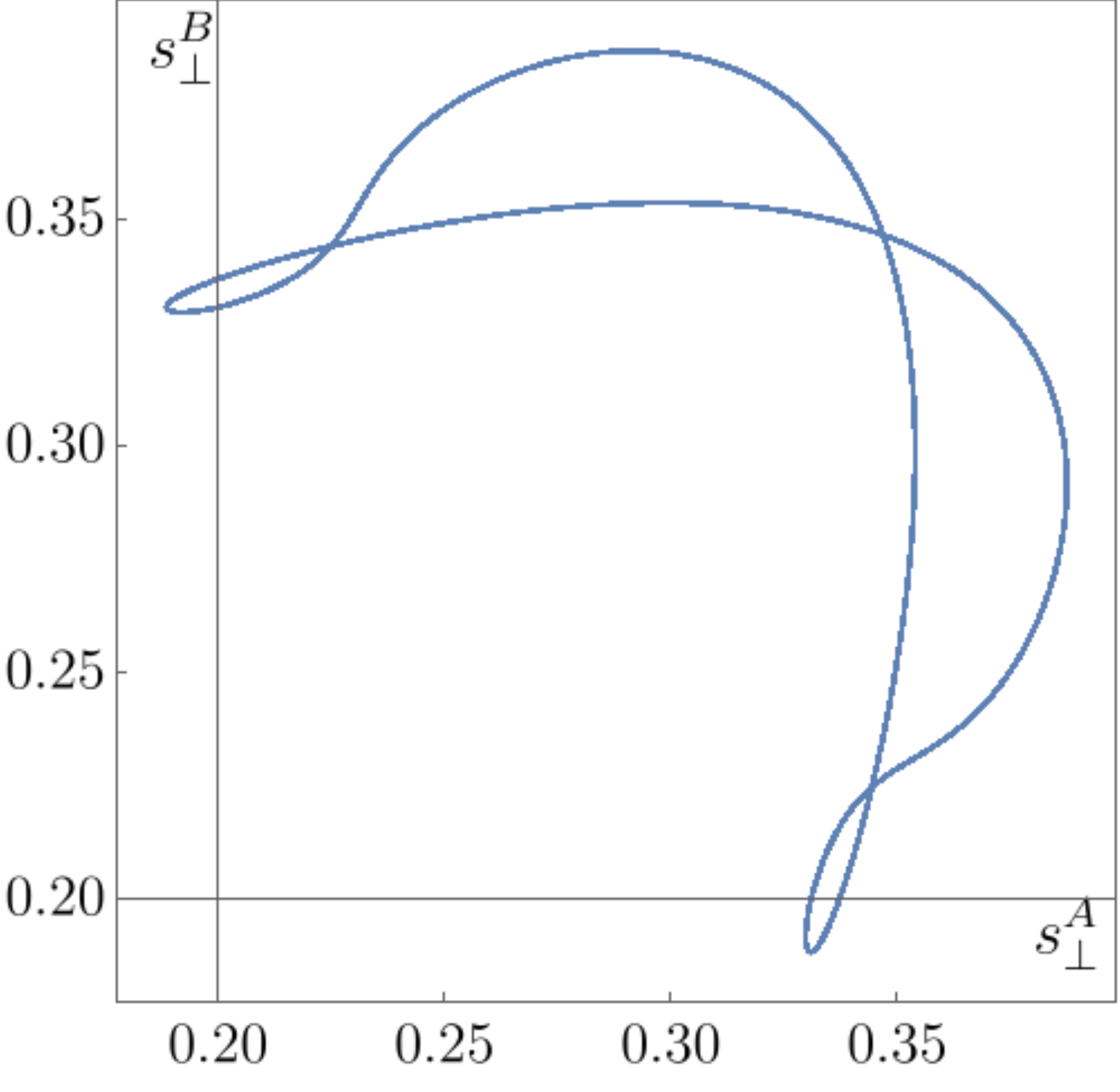}}\qquad\qquad\qquad\qquad
\subfloat[\large (b)]{\includegraphics[scale=0.34]{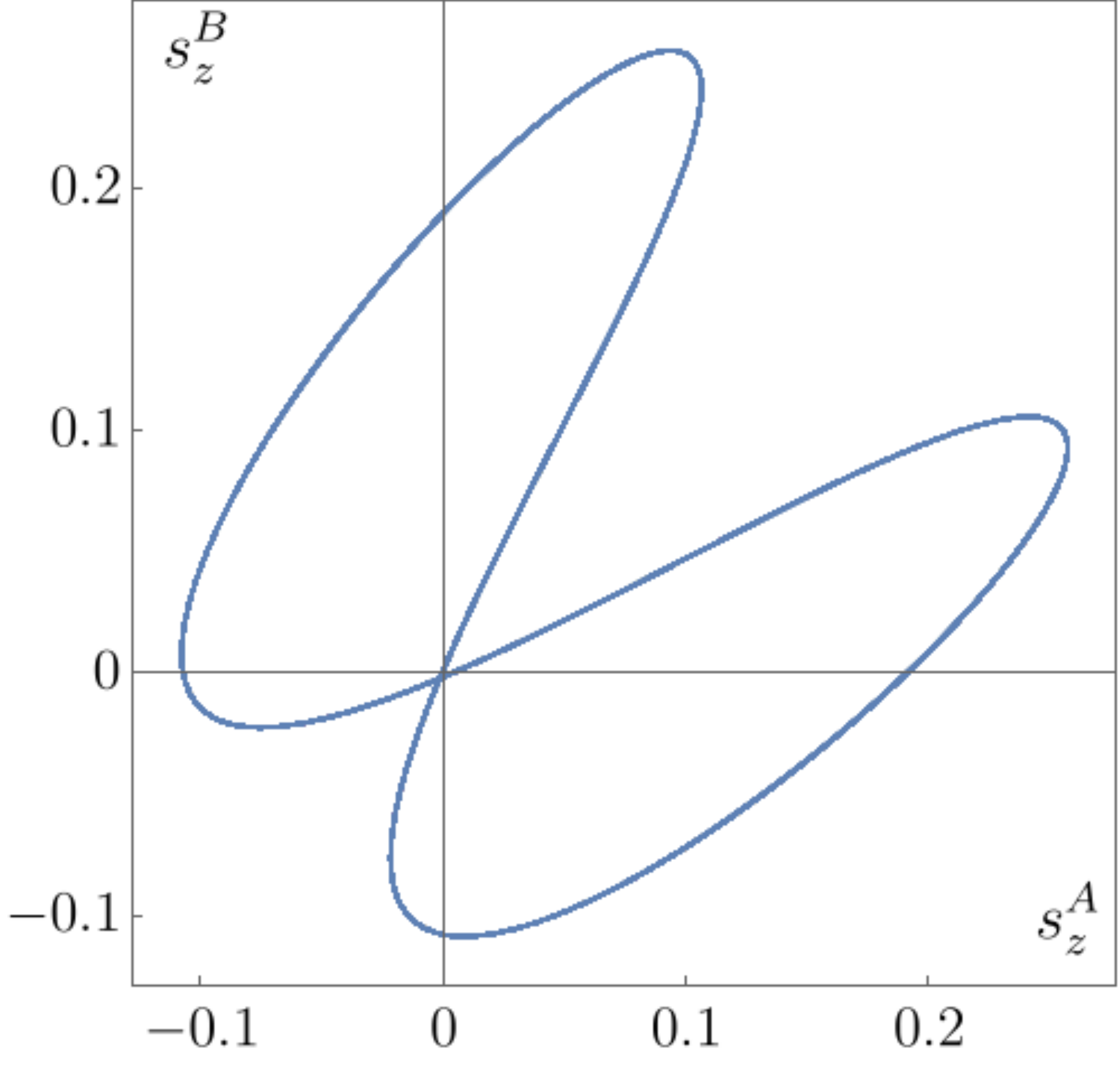}}
\caption{$s^{A}_\perp$ vs. $s^{B}_\perp$ and $s^{A}_z$ vs. $s^{B}_z$  projections  for the  limit cycle      in \fref{Symmetry-broken_LC}.  The   difference with the $\Z2$-symmetric limit cycle  at a nearby point $(\delta, W) =(0.44, 0.056)$ shown in \fref{Symmetric_LC_X_Sect} is dramatic.
Nevertheless,  both plots are symmetric with respect to  reflection through the diagonal.   Note also  self-intersections that result from projecting a multi-dimensional (5D) curve onto 2D planes.  }
\label{Symmetry-broken_LC_X_Sect}
\end{figure*}

 In our case, one of the characteristic multipliers is identically equal to one. This is because, as we discussed below \eref{Theta},  each limit cycle is a member of a one parameter family of limit cycles related to each other by rotations around the $z$-axis.  An infinitesimal rotation of the original   limit cycle produces an identical limit cycle just in a `wrong' coordinate system, where $q_x$ and $q_y$ are infinitesimal  but nonzero. This must be a periodic solution of \eref{Floquet_Eqn}. Specifically, $\Delta s_x^\tau=-\Delta\Phi s_y^\tau$ and
 $\Delta s_y^\tau=\Delta\Phi s_x^\tau$ for a rotation by $\Delta\Phi$. \eref{Floquet_Asymm_Var} then implies $q_x=-2\Delta\Phi s_y$ and $q_y=2\Delta\Phi s_x$. Therefore,
 \beg 
\bm{p}_{1} = \bpm s_{y} \\ -s_{x} \\ 0 \epm,
\en 
is a solution of \eref{Floquet_Eqn} with the same period $T$. We verify this directly using also \eref{Symm_One_Spin_Eqn}. Thus, \eref{Floquet_Gen_Perturb} simplifies to
\beg 
\bm{q}(t) =    d_{1}  \bm{p}_{1}(t)+d_{2} e^{\varkappa_{2}t}\bm{p}_{2}(t)+d_{3} e^{\varkappa_{3}t}\bm{p}_{3}(t).
\label{Floquet_Gen_Perturb1}
\en%
 We observe that the $\Z2$-symmetric limit cycle becomes unstable as we cross the dashed line in \fref{Phase_Diagram} and  the asymptotic dynamics of the two ensembles loose the $\Z2$ symmetry in the yellow region to its left.
 Near this line of symmetry breaking    the other two characteristic multipliers ($\rho_{2}$ and $\rho_{3}$) are also real.  Both  are less than one to the immediate right of this line, while one of them becomes greater than one as we cross the line and enter the yellow subregion, see, e.g., \fref{Floquet_Ex_Pic}.

 In practice, we find it more convenient to determine the line of symmetry breaking by monitoring the asymptotic value of $|s_{z}^{A} - s_{z}^{B}|$, which is zero for a $\Z2$-symmetric limit cycle. When this quantity exceeds a certain threshold (0.01 at   $t>2 \times 10^{4}$ in our algorithm),
 we declare the symmetry broken and the  $\Z2$-symmetric limit cycle unstable. This method agrees with our Floquet stability analysis to within
 few percent as evidenced by Table~\ref{Tab_Z_2_Symmetry_Breaking} and \fref{Floquet_Ex_Pic}.

\subsection{ Limit Cycles without $\Z2$ symmetry}
\label{SBLC}

\begin{figure*}[tbp!]
\centering
\subfloat[\qquad\large (a)]{\includegraphics[scale=0.25]{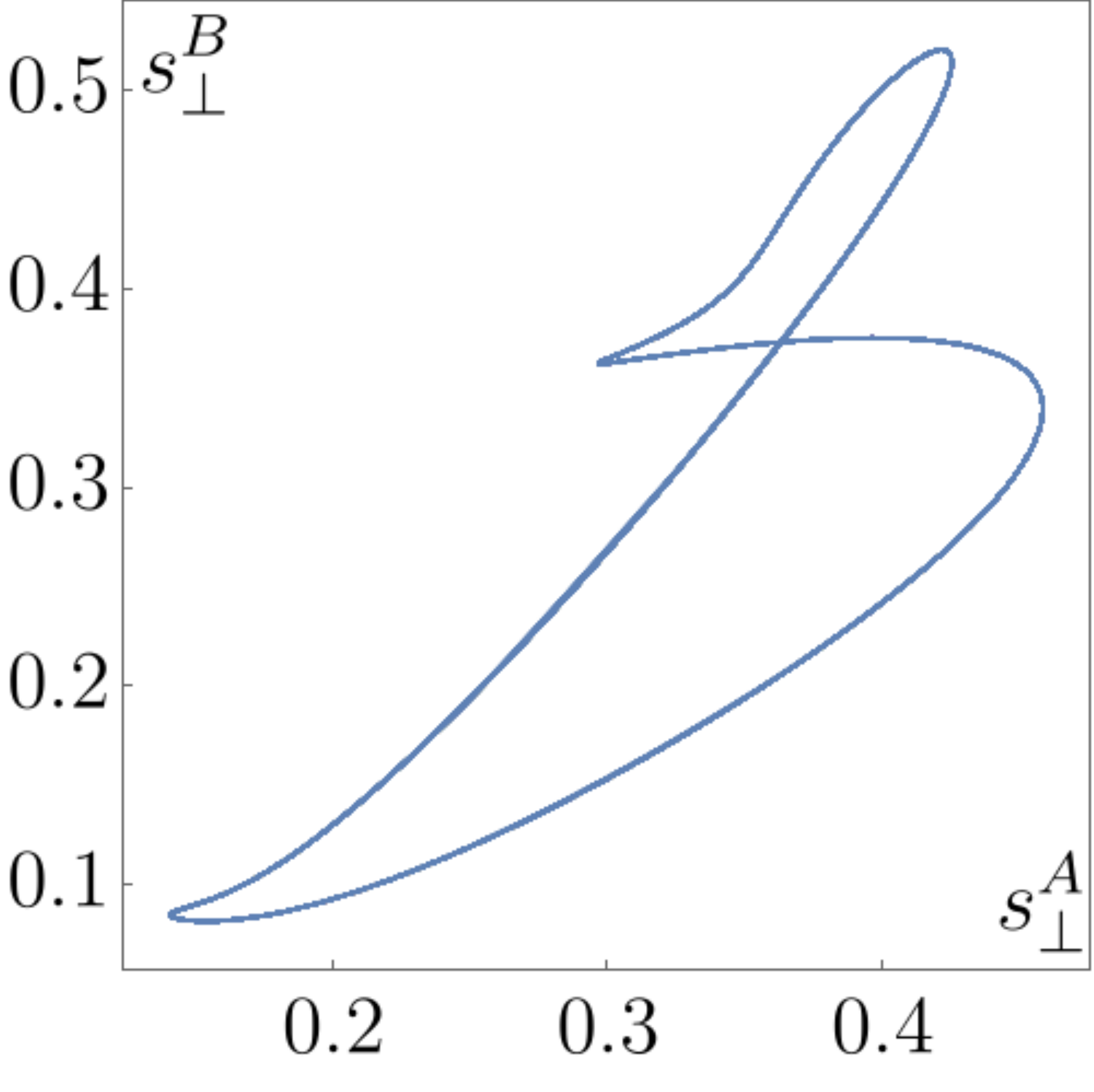}}\qquad \qquad
\subfloat[\qquad\large (b)]{\includegraphics[scale=0.26]{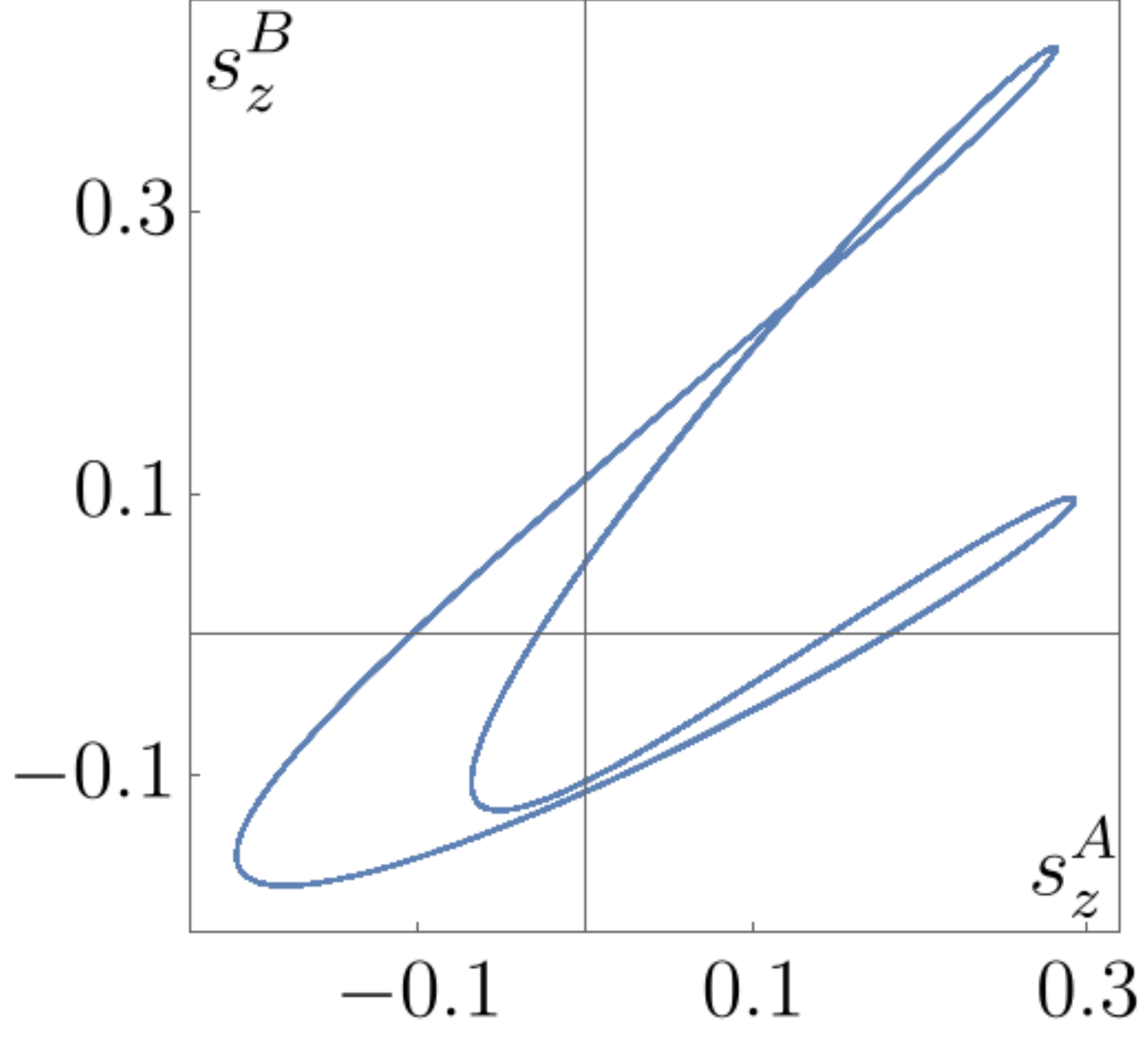}}
\qquad 
\subfloat[\qquad\qquad\qquad\qquad\qquad\large (c)]{\label{annulus}}{\includegraphics[scale=0.27]{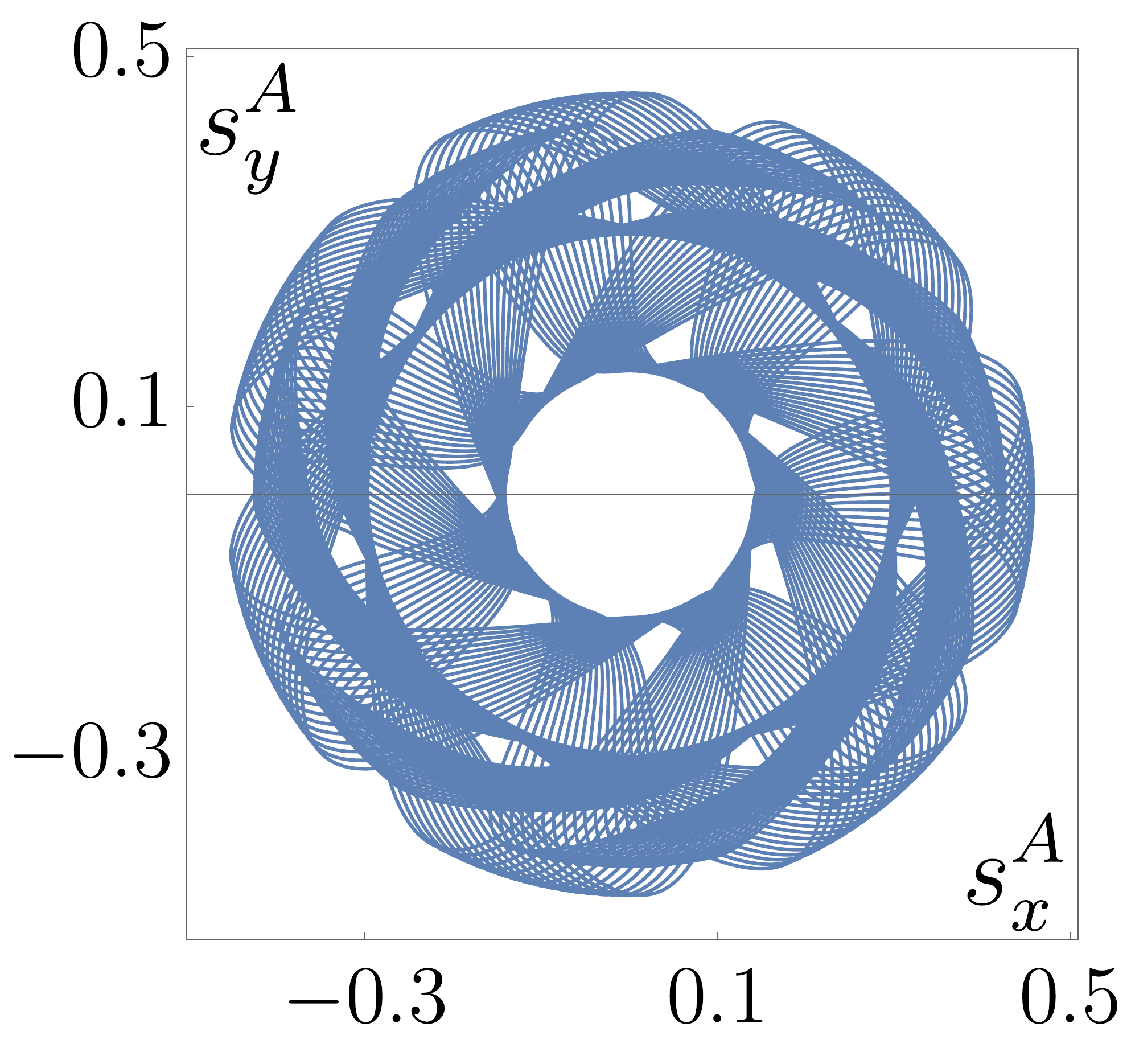}}
\caption{An  asymmetric   limit cycle  at $(\delta, W) = (0.225, 0.05)$ from the dark blue subregion near the origin of  \fref{Phase_Diagram}.
This is a periodic attractor when we factor out the net phase $\Phi(t)$ as   in \eref{Mean-Field_Group4}. Otherwise,  $\Phi(t)$ brings about another frequency $\omega_q$  transforming it into a 2-torus in the full phase-space. Plot \textbf{(c)} is the projection of this torus onto the $s_x^A - s_y^A$ plane. Given enough time the trajectory densely fills an annulus with
radii $s^{A}_{\perp,\max}$ and $s^{A}_{\perp,\min}$. The additional frequency disappears if we move into a rotating frame or consider rotationally invariant quantities as in  \textbf{(a)} and \textbf{(b)}, but shows up as
an overall shift by $\omega_q/2\pi$  in the power spectrum   of the light radiated by the cavity, see   \fref{Power_Spectrum_No_Reflection_Intro}. }
\label{Refection_Z2_Symmetry-broken_Pseudo_LC}
\end{figure*}

\begin{figure*}[tbp!]
\centering
\subfloat[\large (a)]{\label{Original_LC}\includegraphics[scale=0.42]{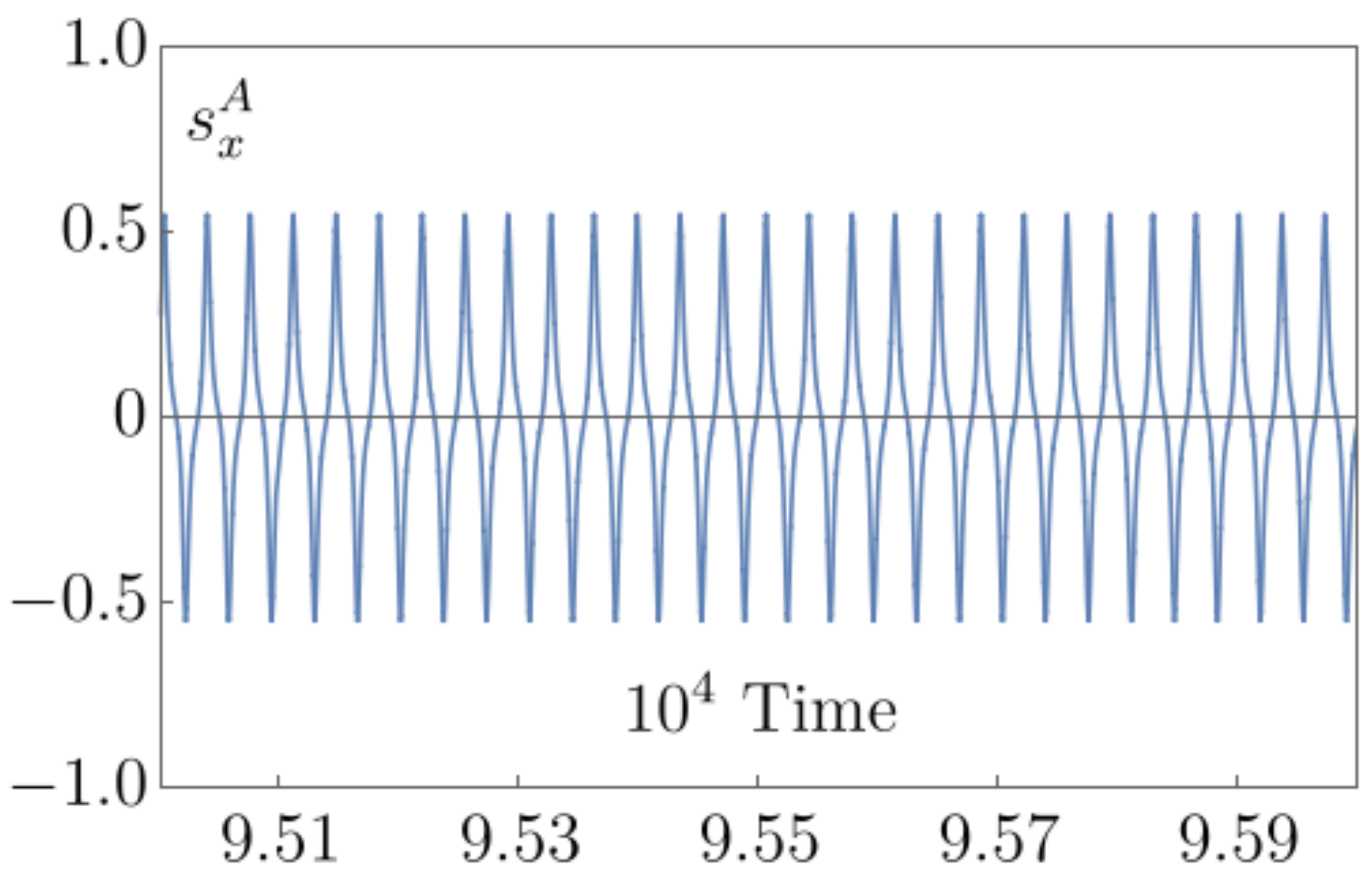}}\llap{\raisebox{4 cm}{\includegraphics[height=2.5cm]{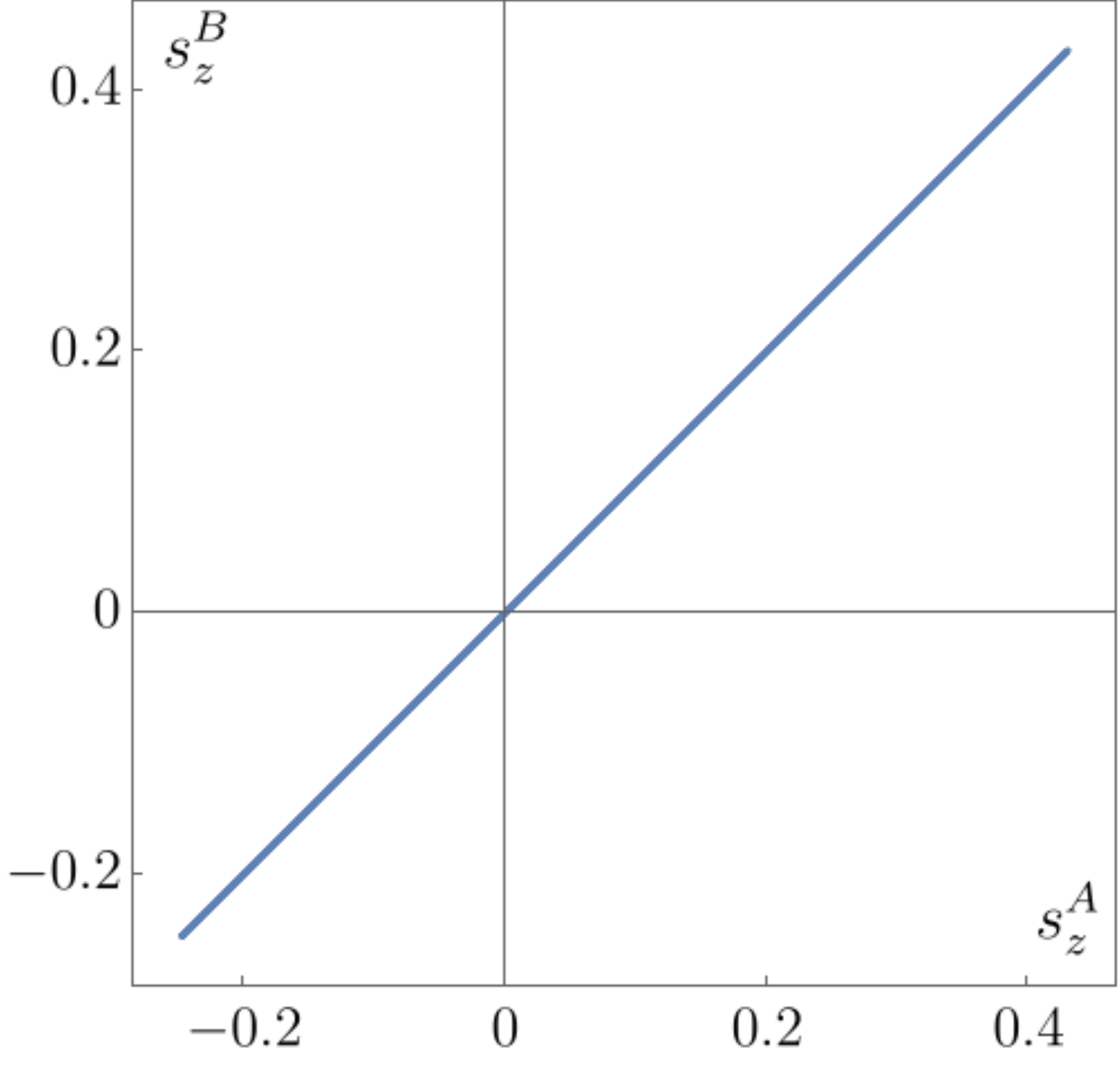}}}\qquad\qquad
\subfloat[\large (b)]{\label{Imposter_LC}\includegraphics[scale=0.42]{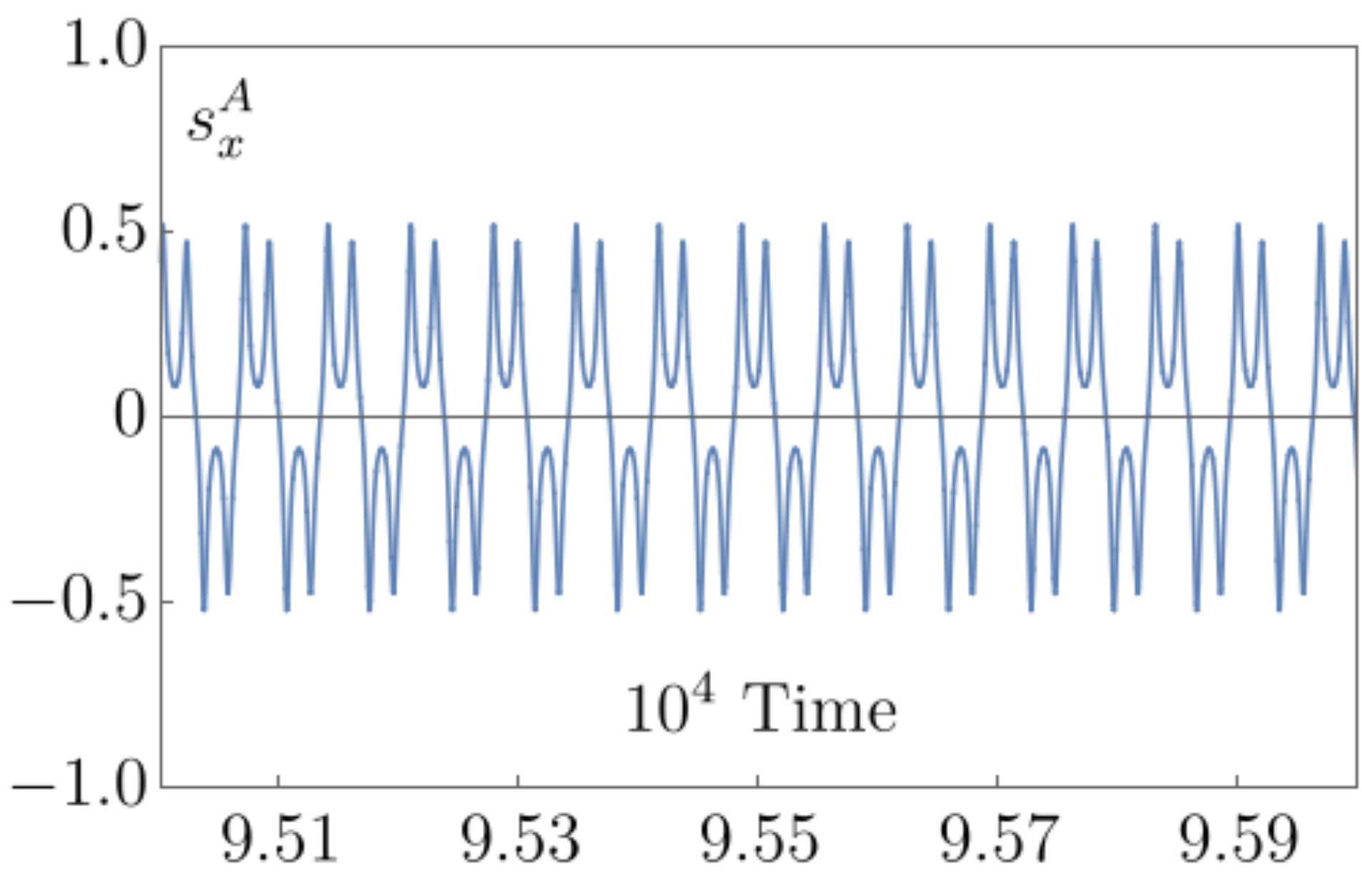}}\llap{\raisebox{4 cm}{\includegraphics[height=2.5cm]{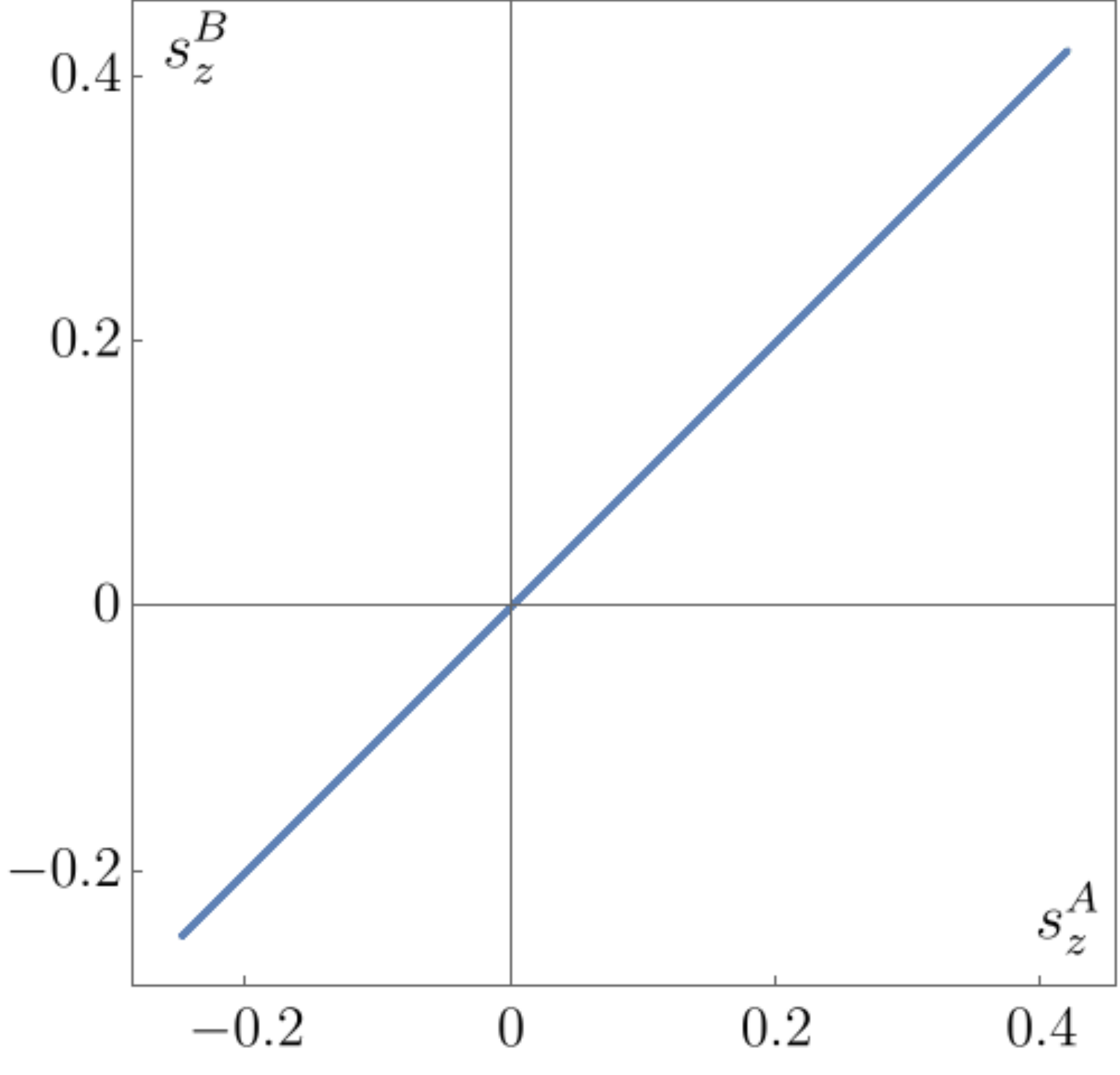}}}\\
\subfloat[\large (c)]{\label{Coexist_SLC_QP_1}\includegraphics[scale=0.42]{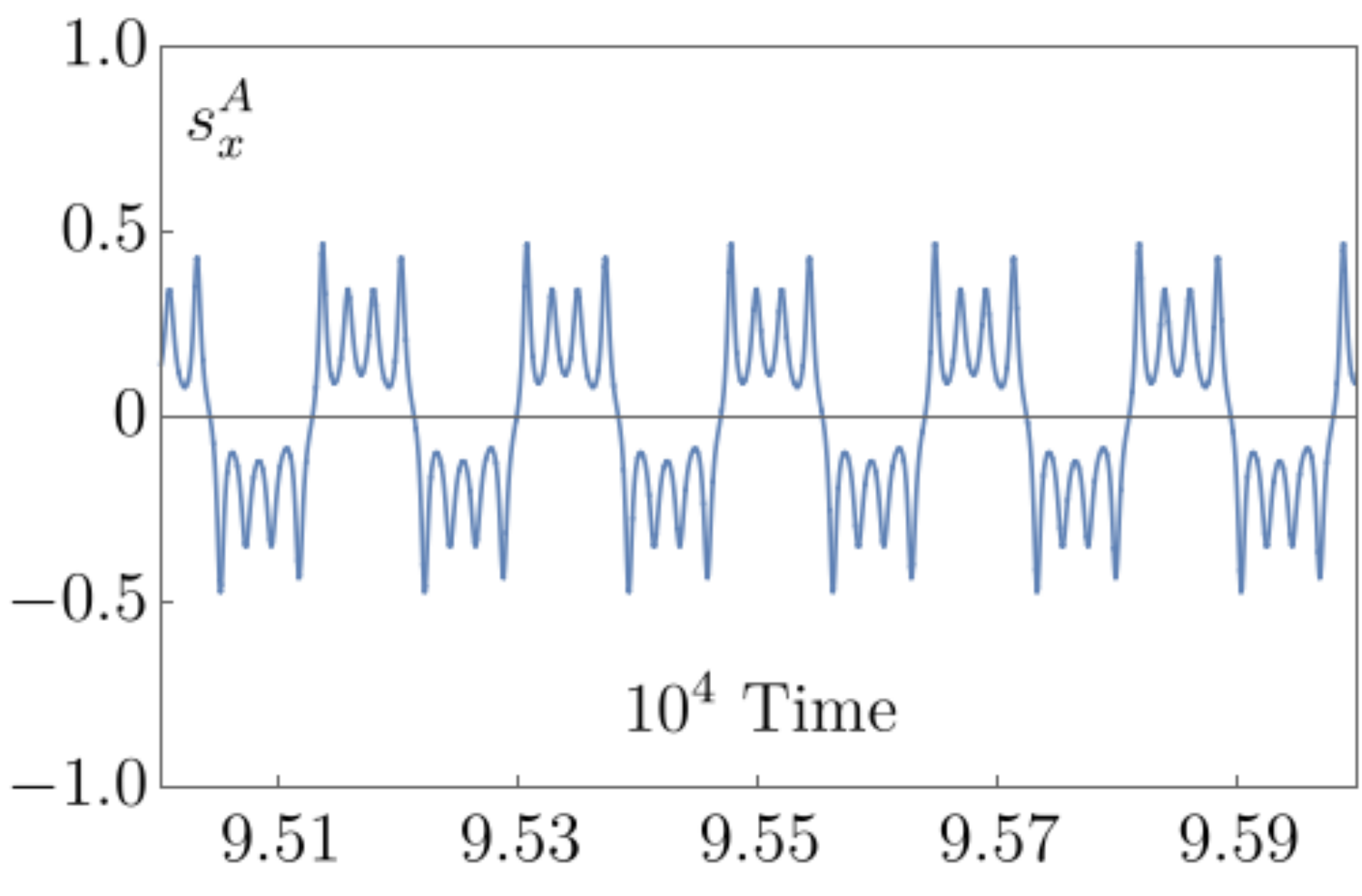}}\llap{\raisebox{4 cm}{\includegraphics[height=2.5cm]{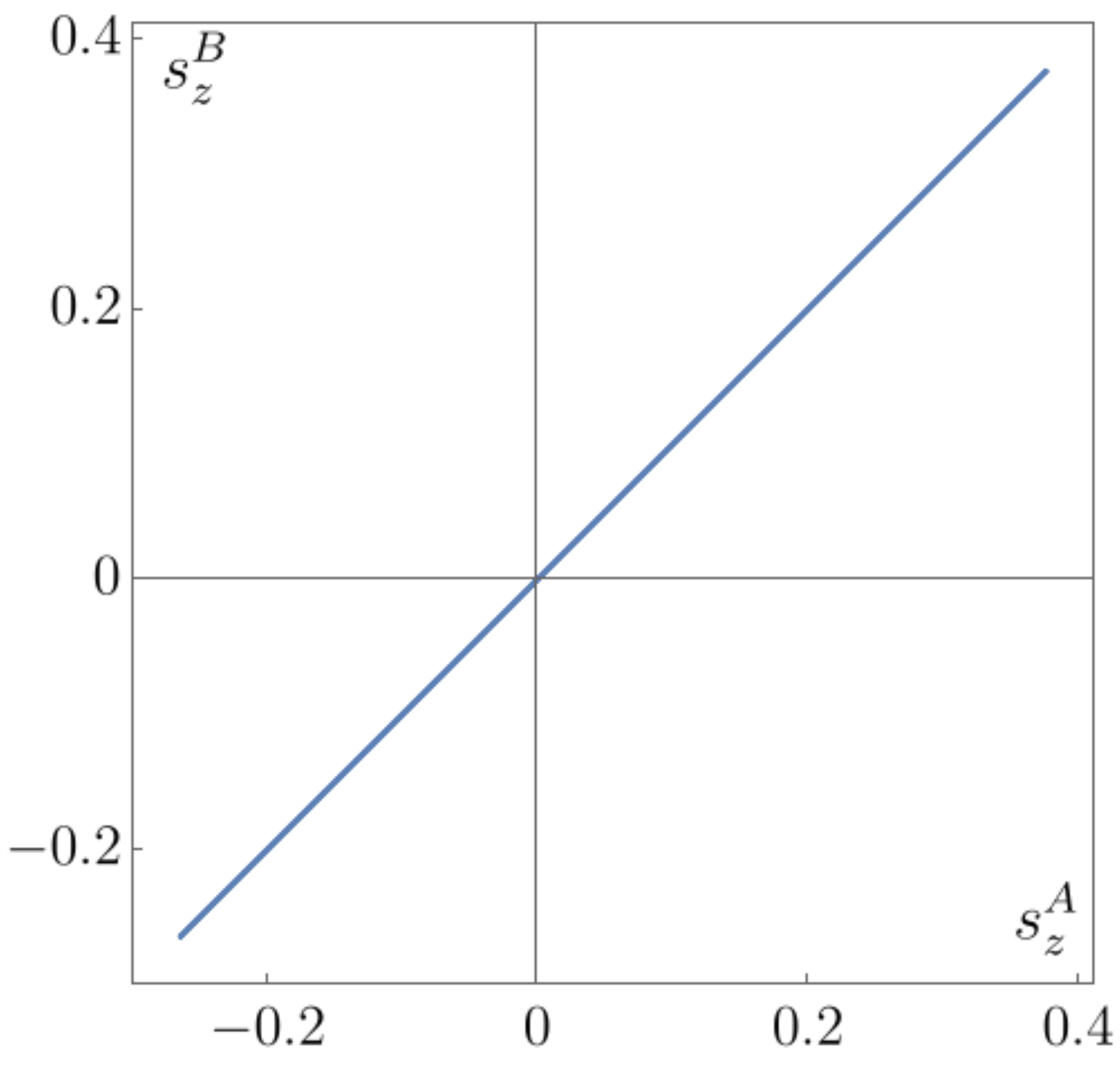}}}\qquad\qquad
\subfloat[\large (d)]{\label{Coexist_SLC_QP_2}\includegraphics[scale=0.42]{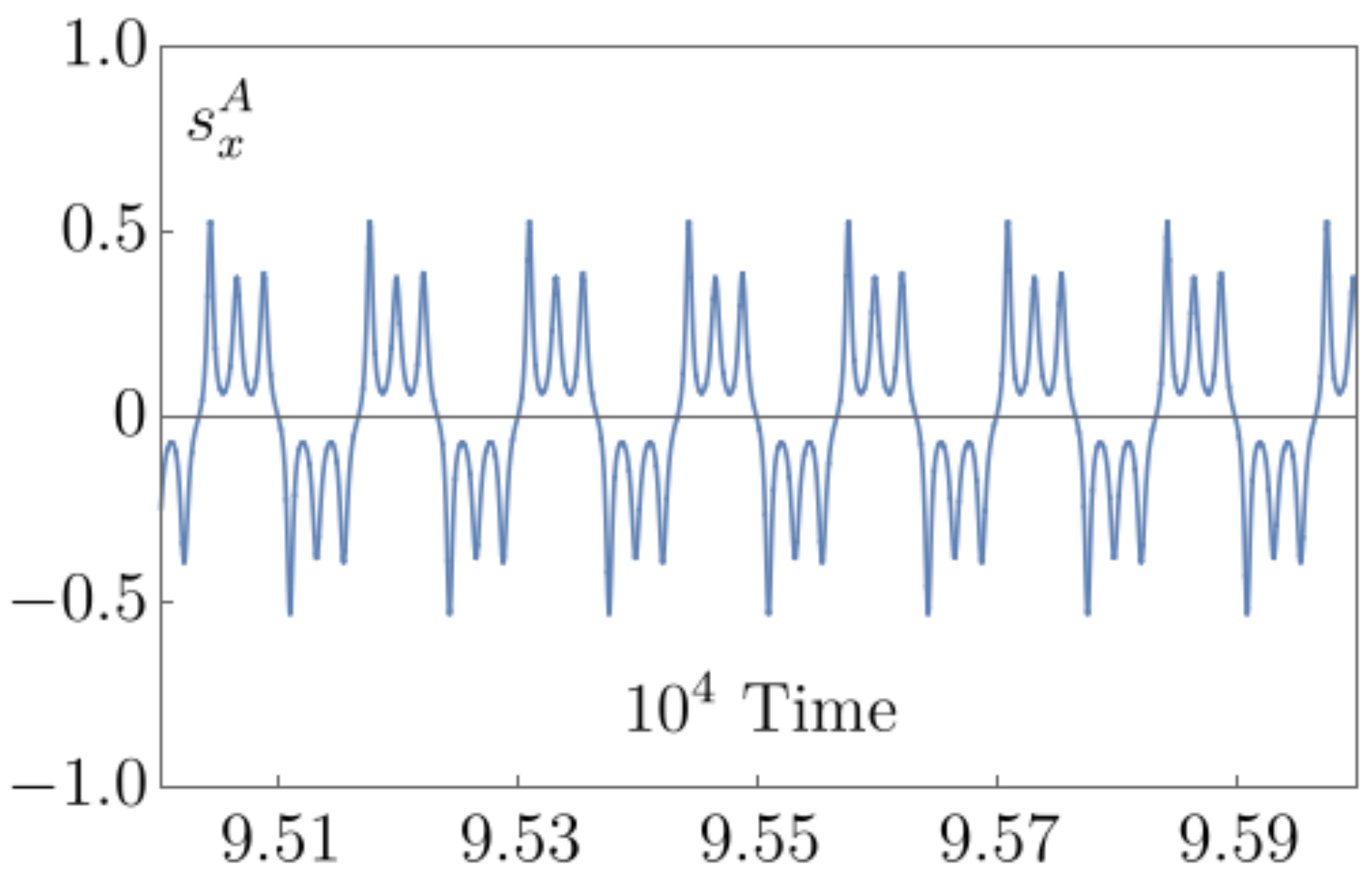}}\llap{\raisebox{4 cm}{\includegraphics[height=2.5cm]{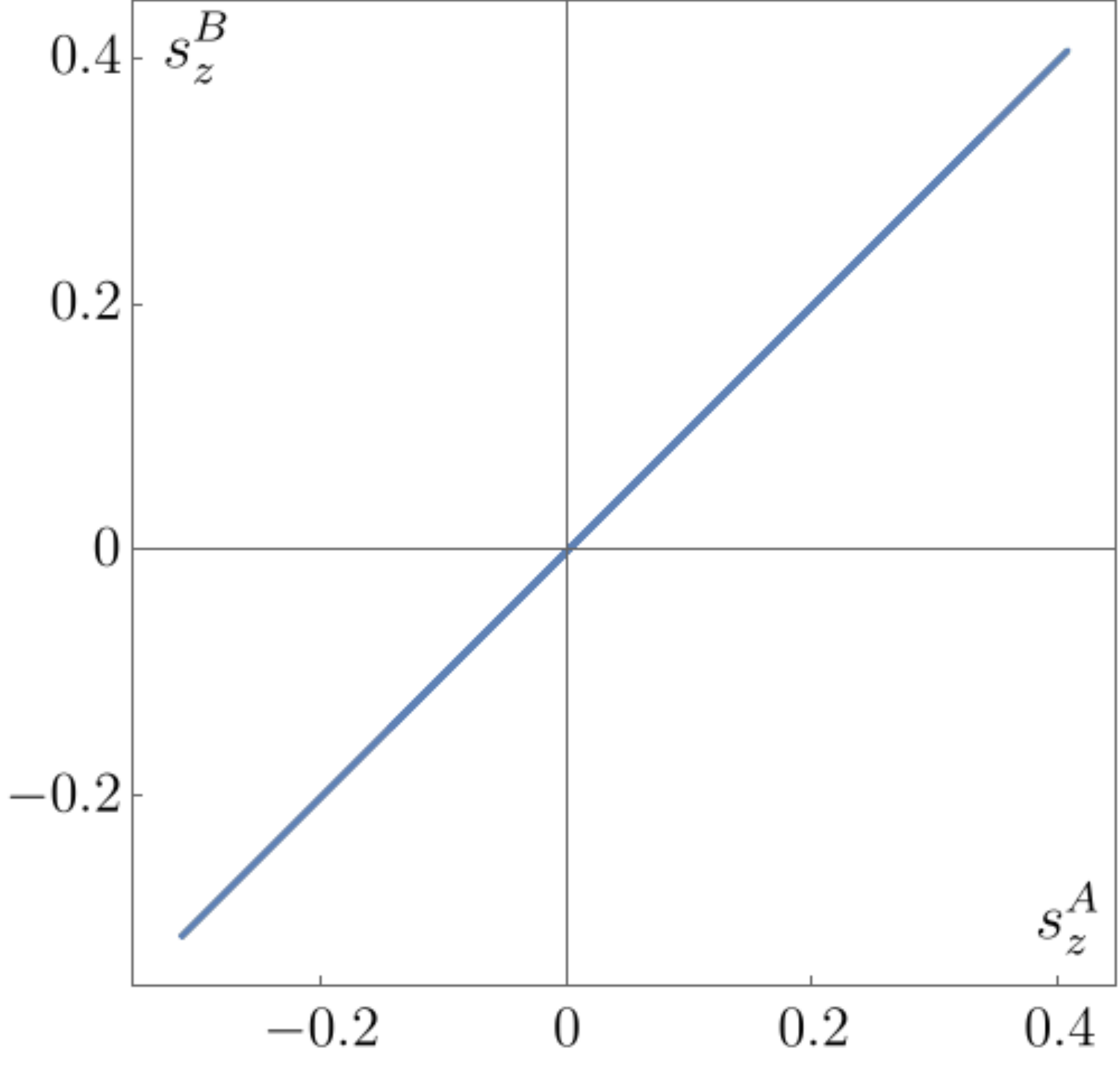}}}
\caption{ Different limit cycles in the $\Z2$-symmetric green island to the left of the symmetry breaking line  in \fref{Phase_Diagram};   $(\delta, W) = (0.23, 0.0802)$ and $(0.12,  0.055)$ in the first and second rows, respectively. In the main figures, we plot $s_{x}^{A}$ vs. time, whereas in the insets we corroborate the $\Z2$-symmetry in the $s_{z}^{A}-s_{z}^{B}$ projections. The initial condition for the two limit cycles in the first column, which are unstable in the full 6D phase-space, is $(\bm{s}_{0}^{A}, \bm{s}_{0}^{B}) = (0.4, -0.469, 0.7, 0.4, 0.469, 0.7)$. The ones in the second column are stable in the full phase-space. The initial condition for \textbf{(b)} is $(\bm{s}_{0}^{A}, \bm{s}_{0}^{B}) = (0.4, 0.63, 0.7, 0.4, -0.64, 0.7)$, whereas    \textbf{(d)} has $(\bm{s}_{0}^{A}, \bm{s}_{0}^{B}) = (0.471036, -0.423628, -0.566317, 0.471036, 0.43, -0.566317)$.}\label{Coexist_SLC_Z2}
\end{figure*}

 Upon losing stability, each  $\Z2$-symmetric limit cycle gives birth to two limit cycles with broken $\Z2$-symmetry in the yellow subregion to the left of the dashed line in  \fref{Phase_Diagram}. The two are related by the $\Z2$ transformation $\S \circ \R(\phi_{0})$ [see the text below \eref{Z2}]. We verified the existence of these two distinct limit cycles numerically by simulating the equations of motion \re{Mean-Field_1} for two initial conditions similarly related by the $\Z2$ transformation. Thus,  the $\Z2$-symmetric limit cycle undergoes a supercritical pitchfork bifurcation on the dashed line, see the text at the end of Sect.~\ref{symmstab} keeping in mind that limit cycles correspond to fixed points on the Poincar\'e section -- intersection of the attractor with a hyperplane transverse to the flow \cite{Hilborn}. Examples of limit cycles without $\Z2$ symmetry appear in \fsref{Symmetry-broken_LC}, \ref{Symmetry-broken_LC_X_Sect} and~\ref{Refection_Z2_Symmetry-broken_Pseudo_LC}. 
 
 While the limit cycles  in the yellow subregion of  \fref{Phase_Diagram} break the $\Z2$ symmetry, e.g., $s_z^A(t)\ne s_z^B(t)$, we find that   a  weaker version of this symmetry specific to periodic solutions of \eref{Mean-Field_Group4}  still survives. Namely,
\beg
\bm{s}^{\tau}(t) = \mathbb{\Sigma}\circ \R(\phi_{0}) \cdot \bm{s}^{\tau}\left(t+\frac{T}{2}\right),
\label{newz2half}
\en
 and in particular,
\beg
s_\perp^A(t)= s_\perp^B\left(t+\frac{T}{2}\right), \quad s_z^A(t)= s_z^B\left(t+\frac{T}{2}\right).
\label{halfperiodsymm}
\en
  In \fref{Symmetry-broken_LC_X_Sect} this property manifests itself as the symmetry with respect to reflection through the diagonal.  In contrast, the limit cycle in \fref{Refection_Z2_Symmetry-broken_Pseudo_LC} does not have this symmetry. This limit cycle is from the dark blue subregion near the origin of  \fref{Phase_Diagram}, where it coexists with a quasiperiodic solution
of  \eref{Mean-Field_Group4}, see Ref.~\onlinecite{Patra_2} for details.

The property \re{newz2half} ensures that the offset frequency $\omega_q$ in \eref{Theta} vanishes even in the absence of the true $\Z2$ symmetry. Consider \eref{Mean-Field_Group1} and recall that by definition $\omega_q$ is the zeroth harmonic of
\beg
G(t)= \frac{\sin{2\varphi}}{4}\bigg(\frac{s^{A}_{z}s^{B}_\perp }{s^{A}_\perp} - \frac{s^{B}_{z}s^{A}_\perp }{s^{B}_\perp }\bigg).
\label{Gt}
\en%
 \eref{halfperiodsymm} implies that the term in the brackets on the right hand side changes sign when shifted by half a period and \eref{newz2half} shows that $\sin 2\varphi$ is periodic with a period $T/2$. Therefore,
\beg
G(t)= -G\left(t+\frac{T}{2}\right).
\en
This in turn means that the Fourier series of $G(t)$ contains only odd harmonics, i.e., $\omega_q=0$.

On the other hand, $\omega_q\ne0$ for the limit cycle in \fref{Refection_Z2_Symmetry-broken_Pseudo_LC}. Like all limit cycles in region III of the phase diagram, this limit cycle is  a periodic solution \eref{Mean-Field_Group4}. However, the net phase $\Phi$ introduces the second fundamental frequency $\omega_q$ as discussed below \eref{Theta}. We can eliminate this frequency by going to an appropriate  rotating frame. If we stay in the frame  where $ \omega_A=-\omega_B$ as we did throughout this paper, then $\Phi=\omega_q t+F(t)$ and the two frequency motion  in the full 6D phase-space of components of $\bm s^A$ and $\bm s^B$ traces out a 2-torus rather than a closed curve. Consider, for example, the projection of this attractor onto the $s_x^A - s_y^A$ plane. The relation  $s^{A}_{+} = s^{A}_\perp e^{\imath(\Phi+\varphi)}$ implies
\beg
\begin{split}
s_x^A(t) =s^{A}_\perp(t) \cos\left[\omega_q t+ F_1(t)\right],\\ 
s_y^A(t) =s^{A}_\perp(t) \sin\left[\omega_q t+ F_1(t)\right],
\label{F1}
\end{split}
\en
where $F_1(t)=F(t)+\varphi(t)$ and $s^{A}_\perp(t)$  are periodic with the period of the limit cycle [recall that $\varphi$ is defined modulo $\pi$]. 
For time-independent $s^{A}_\perp$ and $F_1$, \eref{F1} describes a motion on a circle of radius $s^{A}_\perp$. In the present case, the radius 
of the circle $s^{A}_\perp(t)$ oscillates periodically between certain $s^{A}_{\perp,\max}$ and $s^{A}_{\perp,\min}$ with a frequency that is in general incommensurate with $\omega_q$. The $s_x^A - s_y^B$ projection then  fills an annulus of inner radius   $s^{A}_{\perp,\min}$ and outer radius $s^{A}_{\perp,\max}$  as seen in \fref{annulus}.

Similarly, $l_-=s_-^A+s_-^B$ contains two frequencies,
\beg
l_-(t)=e^{-\imath \omega_q t}\left[s_\perp^A e^{-\imath(\varphi +F)}+s_\perp^A e^{-\imath(\varphi -F)}\right],
\label{lmin2f}
\en
where we used \eref{New_Var_Rot}. The term in  square brackets is periodic with the period of the limit cycle, while $e^{-\imath \omega_q t}$
in front introduces the second period $2\pi/\omega_q$. 

\subsection{Reentrance  of $\Z2$-Symmetric Limit Cycles}

  $\Z2$-symmetric limit cycles remerge as stable attractors of the equations of motion \re{Mean-Field_1} in the green  island to the left of the symmetry breaking line   in \fref{Phase_Diagram}, see \fsref{Imposter_LC} and 
  \ref{Coexist_SLC_QP_2}.  It turns out that the  stable  $\Z2$-symmetric limit cycle living in this island is unrelated to that in the unbounded green subregion to the right of the dashed line.   The latter limit cycle remains unstable in the full 6D phase-space, but is stable in the $\Z2$-symmetric submanifold  $s^{A}_\perp=s^{B}_\perp$ and $s^{A}_z=s^{B}_z$  well past the symmetry breaking line.  
  Therefore, restricting the dynamics to the above  submanifold, we are able to continuously follow this  limit cycle  into the green island (see \fref{Original_LC})  and observe that it is distinct from the stable one shown in \fref{Imposter_LC} . In fact, there are more than one such limit cycles stable in the $\Z2$-symmetric submanifold, but unstable in the full phase-space  in various parts of the green island. For example,  unstable limit cycles in \fsref{Original_LC} and \ref{Coexist_SLC_QP_1} are not related by a continuous deformation. We ascertained  the stability or instability of these limit cycles using  Floquet analysis.

\section{Experimental Signatures}\label{Experimental_Signatures} 
\label{expsign}

A key experimental observable is the autocorrelation function of the radiation electric field outside of the cavity, measurable with a Michelson interferometer. Its Fourier transform  is  the power spectrum of the radiated light. In \aref{Expt}, we show that
within mean-field approximation this quantity is proportional to $|l_{-}(f)|^{2}$ -- the Fourier transform of the transverse part of the total classical spin  $\bm l(t)$, see \eref{l_Minus_f}. In other words,
\beg
\mbox{Power spectrum $\propto | l_{-}(f)|^{2}$}.
\en
 
\subsection{Fixed Points: Time Independent superradiance}
\label{sec: Time Independent superradiance}

In the TSS both spins are along the $z$-axis and $l_-=0$, see \eref{TSS}. Therefore, $|l_{-}(f)|^{2} = 0$ and no light is radiated by the cavity 
when $(\delta, W)$ is in Phase   I.

On the other hand, the power spectrum of the NTSS has a single peak at $f = 0$. Here we must recall that   we are working in a rotating frame, where all  frequencies are shifted by $f_\mathrm{mc}=(\omega_{A} + \omega_{B})/4\pi$.  Thus, the NTSS  produces monochromatic superradiance with this frequency. For example, if  we take $\prescript{1}{}{\textrm{S}}^{}_{0}$ and $\prescript{3}{}{\textrm{P}}^{}_{0}$ levels of $\prescript{87}{}{\textrm{Sr}}^{}_{}$ atoms to be the ground and excited states of our two-level atoms, the monochromatic superradiance  frequency is $f_\mathrm{mc}\approx4.3 \times 10^{5}$  GHz \cite{3P0-1S0_87Sr}. 
Also, note that in this phase we have
\beg 
|\mean{\hat{J}_{-}(f)}|^{2} = \frac{ |l_{-}(f) N|^{2}}{4} \propto N^{2},
\en%
where $\hat{J}_{-}(f)$ is the Fourier transform of $\hat{J}_{-}(t)$ defined below \eref{Full_Master_Conv}. This provides high intensity light (recall that in good cavity lasers the intensity is proportional to $N$). Moreover, such lasers have relatively high Q-factors. These observations motivated  the proposal for accurate atomic clocks utilizing this kind of superradiance \cite{Holland_One_Ensemble_Theory_1, Holland_One_Ensemble_Theory_2}.

\subsection{Limit Cycles: Frequency Combs}
\label{sec: Frequency Combs} 

In Phase   III, the ensembles  synchronize nontrivially to emit a frequency comb, such as the one in \fref{Power_Spectrum_Symmetric_LC}, rather than a single frequency. This behavior corresponds to  the limit cycle that comes to pass after the TSS loses stability via a supercritical Hopf bifurcation on the boundary between Phases I and III in  \fref{Phase_Diagram}.  We will see that  the distance between  consecutive peaks in the  comb can take arbitrary values depending on $\delta$ and $W.$ For typical experimental parameters and $\delta$ and $W$ of order 1, this distance    is many orders of magnitude smaller than  the frequency $f_\mathrm{mc}$  of the monochromatic superradiance in the NTSS.  By filtering out one of the peaks, we can therefore  fine-tune the laser frequency to a high precision.  

\subsubsection{$\mathbb{Z}_{2}$-Symmetric Limit Cycle} 
\label{Expt_Z2}

\begin{figure}[tbp!]
\begin{center}
\includegraphics[scale=0.45]{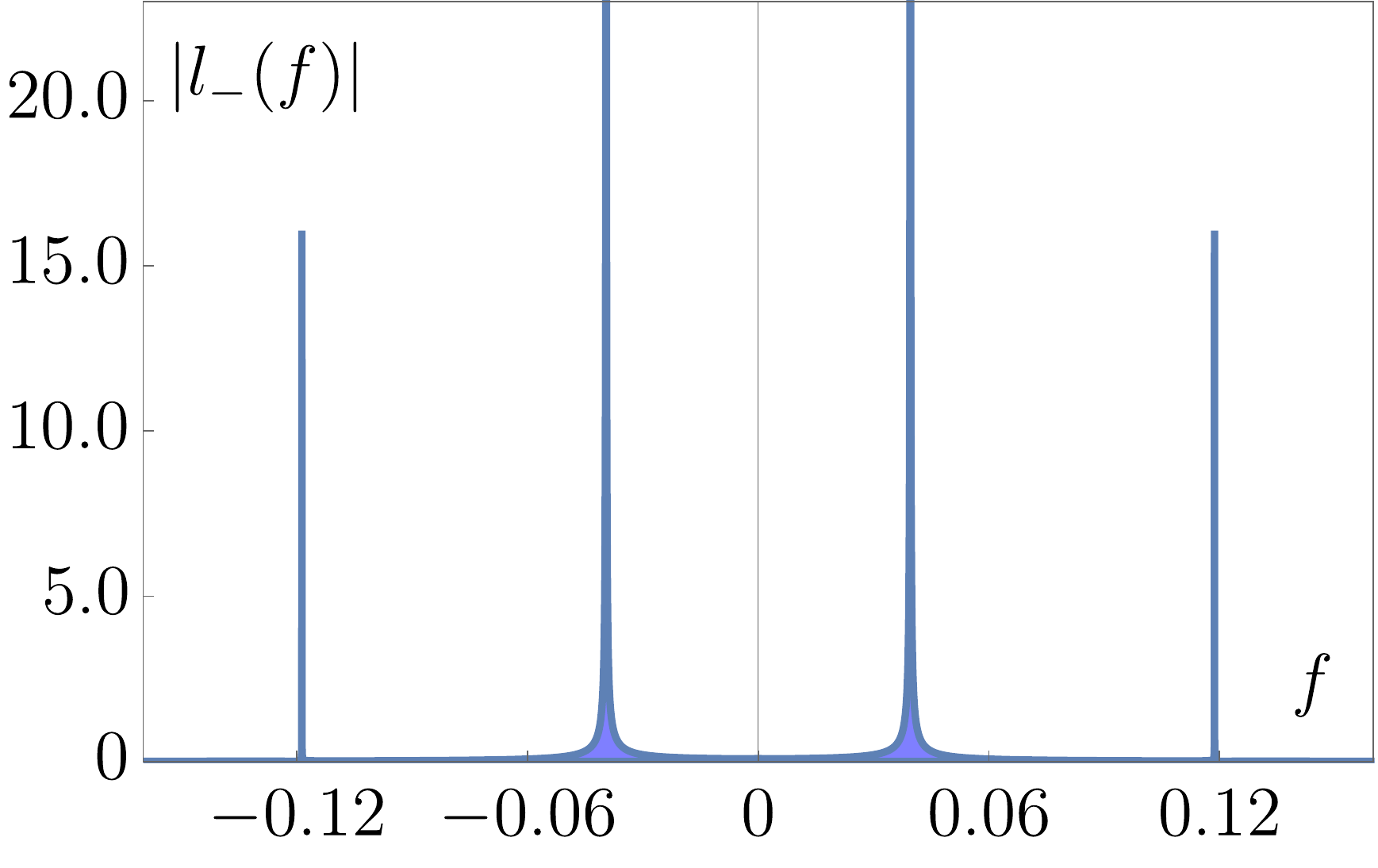}
\caption{Power spectrum for a $\Z2$-symmetric limit cycle at $(\delta, W) = (0.44, 0.056)$. Note, only odd peaks with fundamental frequency $f_{0} \approx 0.040$, are present. In particular, there is no peak at the origin.} 
\label{Power_Spectrum_Symmetric_LC}
\end{center}
\end{figure}

 \fref{Power_Spectrum_Symmetric_LC}  shows the power spectrum for a representative $\mathbb{Z}_{2}$-symmetric limit cycle at $\delta = 0.44$ and $W = 0.056$ in the rotating frame.  This frequency comb  has peaks at $\pm f_{0}, \pm 3f_{0}, \pm 5f_{0}, \cdots$, where $f_{0}\approx 0.040$ is the fundamental frequency.  To estimate the value of $f_0$ in SI units, recall that  in our units  $N\Gamma_{c} = 1$.  In a typical experiment there are about $N = 10^{6}$ atoms inside the optical cavity. Representative values of the Rabi frequency $\Omega$ and  the cavity decay rate $\kappa$ are   $\Omega=$37 Hz and $\kappa = 9.4\times 10^{5}$ Hz according to Ref.~\onlinecite{Holland_One_Ensemble_Theory_1}. Using these numbers, we calculate, 
 \beg
 \mbox{$N\Gamma_{c} = 1.4$ kHz}, \quad \mbox{$f_{0} =0.040 N\Gamma_{c} = 56$ Hz,}
 \label{56hz}
 \en
 which is indeed 4 orders of magnitude smaller than $f_\mathrm{mc}$.
 
We numerically verify that  $\mathbb{Z}_{2}$-symmetric limit cycles have the following time translation property:
\beg 
s_{x,y}\left(t \pm \frac{T}{2}\right) = -s_{x,y}(t),\quad s_{z}\left(t \pm \frac{T}{2}\right) = s_{z}(t).
\label{Time_Transl_Prop_Symm_LC}
\en%
 This property explains why the power spectrum consists only of odd harmonics. We prove this by Fourier transforming the two sides of the  equation $l_-(t)=l_-(t+T/2)$.  \eref{Time_Transl_Prop_Symm_LC} also holds for the analytical solutions in \esref{W=1_Limit} and \re{s_y_CN}. Since the $\mathbb{Z}_{2}$ symmetric limit cycle elsewhere is topologically connected to the one near the $W = 1$ line, it also retains the above property. However, note that   \eref{Time_Transl_Prop_Symm_LC} is different from a related property \re{newz2half} of limit cycles without $\Z2$ symmetry. \eref{Time_Transl_Prop_Symm_LC} is valid for any choice of $x$ and $y$-axes. On the other hand,  \eref{newz2half}  implies that
 $s_y(t)$  changes sign when shifted by half a period, while $s_x(t)$  does {\it not}, in a special coordinate frame rotated by $\phi_0$ around the
 $z$-axis.

Moreover, because of the $\mathbb{Z}_{2}$ symmetry $l_{y}(t) = 0$ and $l_x(t)=2s_x(t)$ in a suitable coordinate system [see the text above \eref{Symm_One_Spin_Eqn}], i.e., $l_{-}(t)$ is a real function. As a result the power spectrum has a reflection symmetry about $f = 0$,   
\beg
|l_{-}(f)|^{2} = |l_{-}(-f)|^{2}.
\label{Refl_Symm_PS}
\en

One can infer more about the power spectra where analytical solutions exist. The harmonic solution from \sref{Harmonic_Soln} entails prominent peaks at $\pm f_{0}$, where $f_{0}$  in various limits  is,
\begs
\bea
W &\rightarrow & 1: \quad f_{0} = \frac{\sqrt{\delta^{2} - 1}}{4\pi}, \\
W &\rightarrow & 0: \quad f_{0} = \frac{\delta}{4\pi}, \\ 
\delta &\gg & 1: \quad f_{0} = \frac{\sqrt{\delta^{2} - W^{2}}}{4\pi}. 
\eea 
\label{Harmonic_Peaks}
\ens%
Near $\delta = W = 1$ the solution is  in terms of the Jacobi elliptic function cn.  According to \esref{s_y_CN} and \re{W=del=1_s_x_z}, $l_x(t)=2s_x(t)=\frac{2aW}{\delta}\mathrm{cn } (bt, k)$. The function cn has the following series expansion \cite{Abramowitz_Stegun, Gradshteyn_Ryzhik}:
\beg
\textrm{cn}u = \frac{2\pi}{kK(k)}\sum_{n = 1}^{\infty} \frac{q^{n - \frac{1}{2}}}{1 - q^{2n - 1}}\cos{\bigg[(2n - 1)\frac{\pi u}{2K(k)}\bigg]}. 
\label{Jacobi_Elliptic_CN_Expansion} 
\en%
where $q = e^{-\frac{\pi K(k^{\prime})}{K(k)}}$ and $k^{\prime} = \sqrt{1 - k^{2}}$.  In our case,
\beg 
u \equiv bt = \frac{\sqrt{|r|}t}{\sqrt{2|1-2k^{2}|}},
\en%
where $r=\delta - 1$ and we used \eref{W=del=1_Constraints_1,2}. This again corroborates the appearance of only odd harmonics in the power spectrum at $\pm f_{0}, \pm 3f_{0}, \pm 5f_{0}, \cdots$, where $f_{0}$ is
\beg 
f_{0} = \frac{\sqrt{|r|}}{4K(k)\sqrt{2\left|1-2k^{2}\right|}}.
\label{f_0_CN} 
\en 
Expressions \re{Harmonic_Peaks} and \re{f_0_CN} demonstrate that the frequency $f_0$ and, therefore, the spacing between peaks in the power spectrum, can  take any value from 0 to $\infty$ depending on $\delta$ and $W$. In particular, for $\delta$ and $W$ of order 1, $f_0\sim 0.1 N\Gamma_c$ close to the value in \eref{56hz}.
 
\subsubsection{$\Z2$-Symmetry-Broken Limit Cycle}
\label{Experiment_SBLC}
 
\begin{figure}[tbp!]
\begin{center}
\includegraphics[scale=0.45]{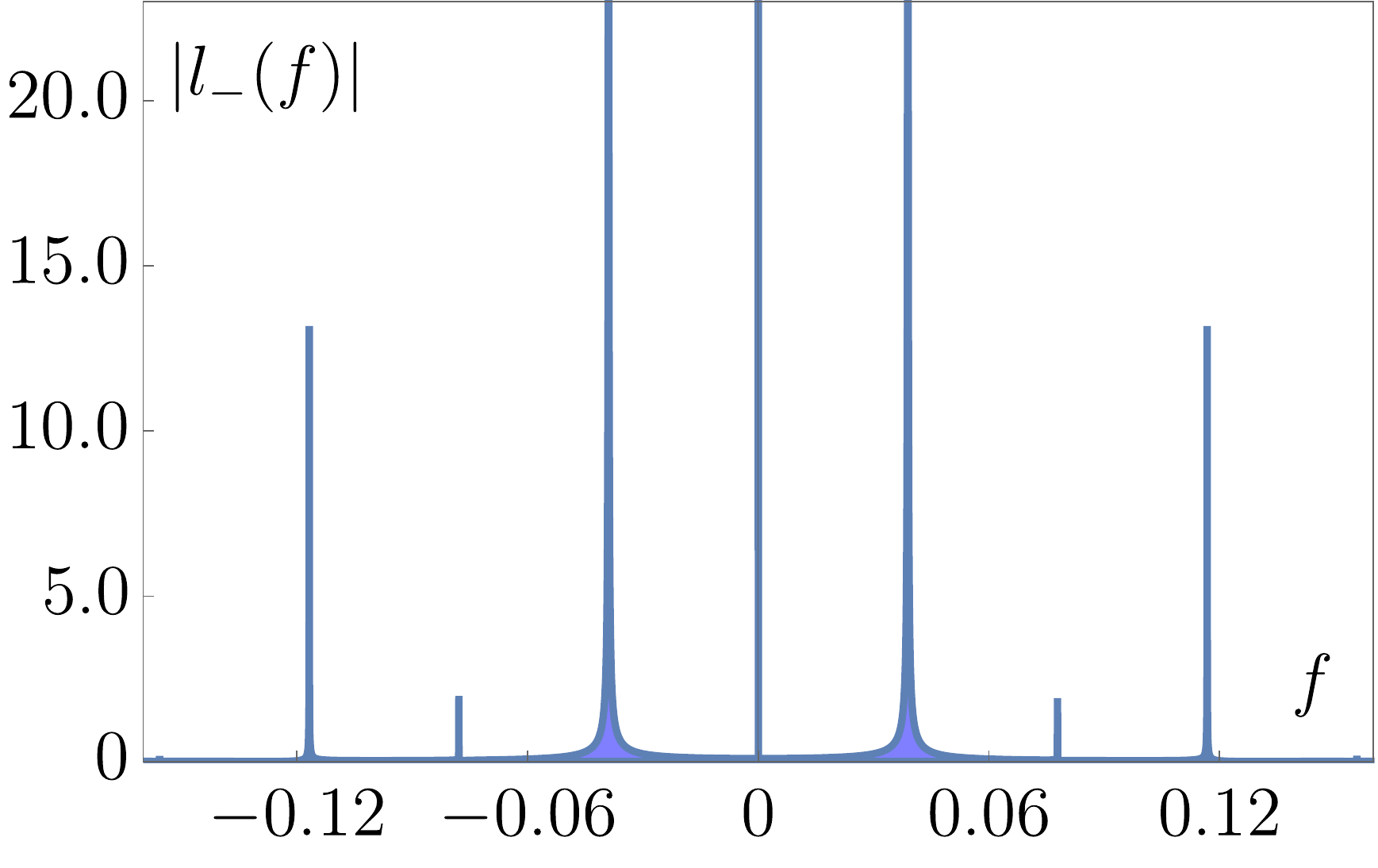}
\caption{Power spectrum for a $\Z2$-symmetry-broken limit cycle at $(\delta, W) = (0.42, 0.056)$. Unlike in the $\Z2$-symmetric spectrum, both even and odd peaks are present. The fundamental frequency is $f_{0} \approx 0.038$. The most pronounced peak is at $f = 0$.}
\label{Power_Spectrum_Asymmetric_LC}
\end{center}
\end{figure}

\begin{figure}[tbp!]
\begin{center}
\includegraphics[scale=0.45]{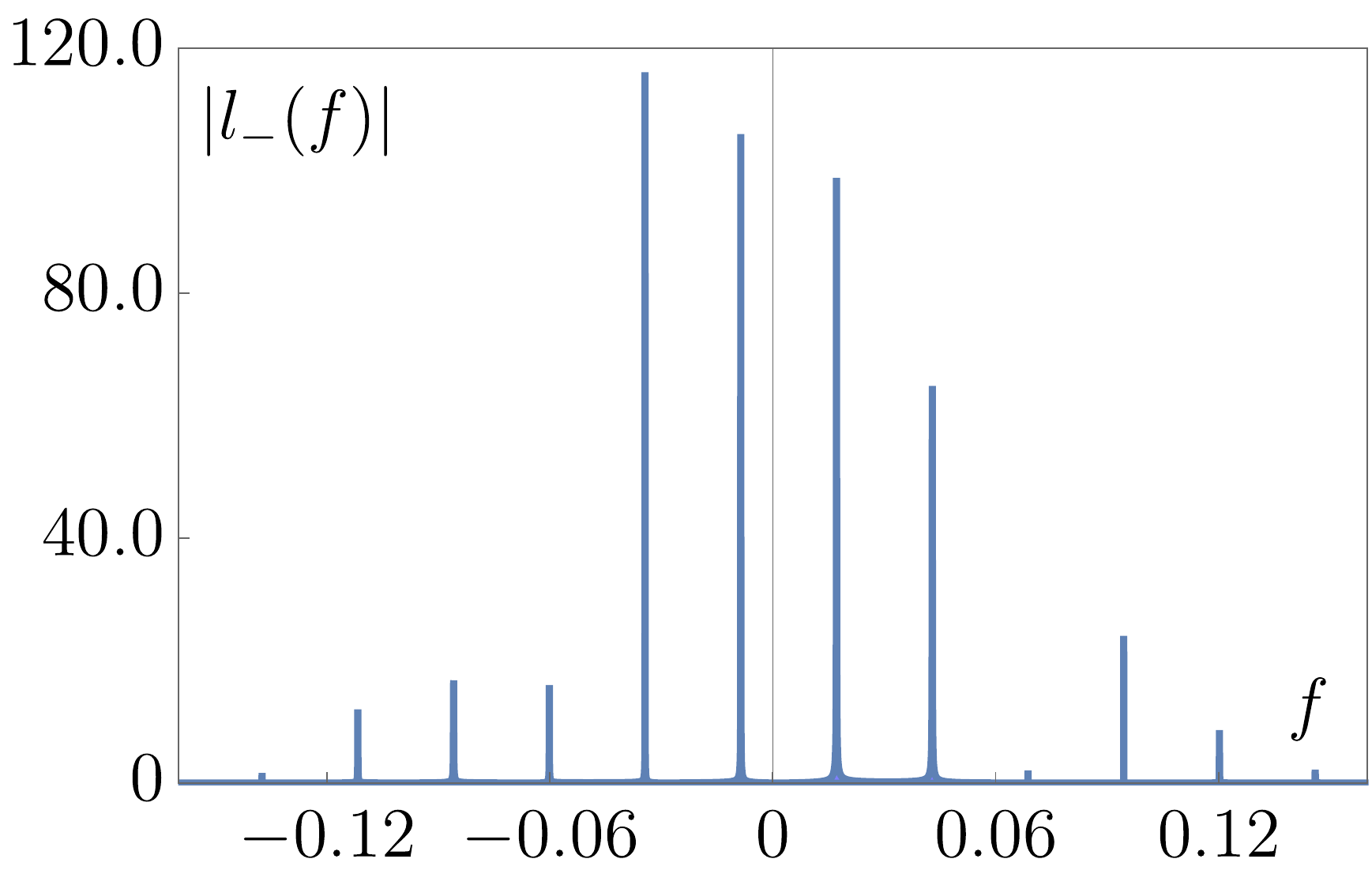}
\caption{Power spectrum for a $\Z2$-symmetry-broken limit cycle at $(\delta, W) = (0.23, 0.056)$. In contrast to \fref{Power_Spectrum_Asymmetric_LC}, the reflection symmetry of the power spectrum is completely lost and the carrier frequency is shifted to a negative value from $f = 0$.}
\label{Power_Spectrum_Asymmetric_LC_Coexists_With_Quasiperiodic_Fully_Symmetry_Broken_Power_Spectrum}
\end{center}
\end{figure} 
 
In \fref{Power_Spectrum_Asymmetric_LC}, we show the power spectrum of a $\Z2$-symmetry-broken limit cycle at $\delta = 0.42$ and $W = 0.056$ in the rotating frame. Unlike for the $\Z2$-symmetric limit cycle,  both odd and even harmonics are present, i.e., peaks are at $0, \pm f_{0}, \pm 2f_{0}, \cdots$  The most pronounced peak is at zero. This is  because of the loss of the time translation property \re{Time_Transl_Prop_Symm_LC}.  

Moreover, here $l_{y}(t) \neq 0$ and as a result  $l_{-}(t)$ is complex. Thus, $l_-(f)$  does not obey \eref{Refl_Symm_PS} and the spectra no longer have reflection symmetry about the $f=0$ axis. However, notice that the spectrum in \fref{Power_Spectrum_Asymmetric_LC} still seems to have retained this   symmetry. We explain this based on our numerical observation that for these values of  $\delta$ and $W,$ in a suitably rotated frame,  
\beg 
l_{y}(t) = l_{y0} + \textrm{small oscillations},
\label{Stokes-Antistokes} 
\en%
 where $ l_{y0}$ is a constant complex number.  In the Fourier transform of $l_-(t)=l_x(t)+\imath l_y(t)$, $l_x(t)$ produces a symmetric spectrum, 
 $l_{y0}$ contributes only to the peak at $f=0$, and the small oscillations lead to  a small asymmetry.
 As a result, although a careful analysis of the peak heights shows that the reflection symmetry of the power spectrum is in fact broken, this is hard to discern from \fref{Power_Spectrum_Asymmetric_LC}. 
 
 In contrast, the $\Z2$-symmetry-broken limit cycle in  \fref{Power_Spectrum_Asymmetric_LC_Coexists_With_Quasiperiodic_Fully_Symmetry_Broken_Power_Spectrum} is visibly asymmetric with respect to the $f=0$ axis. Furthermore, it features an offset of all frequencies originating from the $\omega_q t$ term (overall precession) in the net phase $\Phi(t)$ discussed in detail in \sref{SBLC}. Specifically, \eref{lmin2f} implies that the power spectrum of such limit cycles is $f_q+p f_0,$ where $p$ is an arbitrary integer, $f_q=\omega_q/2\pi$, and $f_0$ is the frequency of the limit cycle.

\section{Discussion} 

 In this paper, we studied the long time dynamics of two atomic  ensembles (clocks) in an optical cavity and  constructed the  nonequilibrium phase diagram for this system shown in \fref{Phase_Diagram}. In the extreme bad cavity regime, we adiabatically eliminated the cavity degrees of freedom to obtain an effective master equation in terms of the atomic operators only. Further, we performed a   consistent system size expansion for the master equation to
 derive  the mean-field equations of motion and  the Fokker-Planck equation governing quantum fluctuations. Mean-field time evolution  is 
 in terms of two collective classical spins  representing  individual ensembles. Each nonequilibrium phase in \fref{Phase_Diagram} corresponds
 to a distinct attractor (asymptotic solution) of the mean-field dynamics.  
 
 Mean-field equations of motion  for two ensembles have two symmetries:  an axial symmetry about the $z$-axis and  a $\Z2$ symmetry  with respect to an interchange of the two ensembles.  The phase diagram  features spontaneous breaking of one or both of these symmetries.
 
There are two types of fixed points -- the trivial steady state or TSS  (normal nonradiative phase), and the nontrivial steady state  or NTSS  (monochromatic superradiance). Using linear stability analysis, we obtained their basins of attraction as Phases I and II in \fref{Phase_Diagram}. Both of them lose stability via Hopf bifurcations. Going beyond the linear stability analysis, by deriving the Poincar\'{e}-Birkhoff normal form, we proved that the TSS goes through a supercritical Hopf bifurcation on the boundary of Phases I and III, whereas the NTSS undergoes a subcritical Hopf bifurcation on the II-III boundary. Thus, II to III and I to III transitions are analogous to the first and second order phase transitions, respectively. This analysis also explains the coexistence region near the boundary of Phases II and III.

After bifurcation, the TSS gives rise to a  $\Z2$-symmetric limit cycle  (periodically modulated superradiance). We were able to derive analytical solutions for this limit cycle in terms of harmonic or Jacobi elliptic functions in several  parts of Phase    III. Moreover, we have shown with Floquet analysis that the $\Z2$-symmetric limit cycle becomes unstable on the symmetry breaking line (the dashed line in \fref{Phase_Diagram}) to bring about two distinct $\Z2$-symmetry-broken limit cycles.

Experimentally, one distinguishes between different dynamical phases of the two  ensembles by measuring the power spectrum of the light radiated by the cavity. In particular, the NTSS emits monochromatic light at a certain frequency $f_\mathrm{mc}$. Limit cycles emit frequency combs --  series of equidistant   peaks at $f_\mathrm{mc}+ p f_0$, where $p$ is an integer and $f_0$ is the limit cycle frequency.  For a  $\Z2$-symmetric limit cycle, $p$ is always odd, while for  $\Z2$-symmetry-broken limit cycles it takes arbitrary integer values.  Certain symmetry-broken limit cycles also  renormalize the value of $f_\mathrm{mc}$ relative to the NTSS  and  produce power spectra that are asymmetric about the $f=0$ axis. We estimated typical values of $f_0$ from available experimental data and found that it is several orders of magnitude smaller than $f_\mathrm{mc}$. Therefore, an interesting feature of limit cycles from the point of view of applications to ultrastable lasers is that they provide access to a range of frequencies drastically different from the atomic transition (lasing) frequency.

 Here we have not analyzed more complicated time-dependent solutions to the left of the dashed line in \fref{Phase_Diagram} marking the spontaneous breaking of the $\Z2$ symmetry. After the loss of the symmetry  between the two ensembles, one needs to consider all six mean-field equations of motion (three for each classical spin) together. According to the Poincar\'{e}-Benedixson theorem, a system of three or more coupled first-order ordinary differential equations   admits chaos.  Indeed, we show in \Rsref{Patra_2, Patra_3}  that chaos emerges by way of   quasiperiodicity in our system. Moreover, eventually the chaotic time dependence of one of the clocks   synchronizes with that of the other via on-off intermittency.  The transition from chaos to chaotic synchronization is  an example of spontaneous restoration of the $\Z2$ symmetry.  

Making system parameters, such as  pump rates and numbers of atoms, unequal for the two ensembles naturally destroys  the $\Z2$ symmetry
of the mean-field equations of motion with respect to the interchange of the ensembles. Nevertheless, we verified numerically that for a weak asymmetry in system parameters the long time dynamics of the two ensembles remain close to that in our nonequilibrium phase diagram in 
 \fref{Phase_Diagram} and exhibit the same main phases. For example,  a nearly $\Z2$-symmetric limit cycle replaces the $\Z2$-symmetric limit cycle in Phase    III etc.
 
  Let us also briefly discuss how the mean-field dynamics changes when the cavity decay rate $\kappa$ is not extremely large. Note that although one cannot adiabatically eliminate the photon degrees of freedom in this case, the fixed points are identical to the ones obtained in the bad cavity limit. The dynamics, however, lead to higher dimensional phase diagrams. For example, in the single ensemble setup, the semiclassical equations for fixed $N$  (number of atoms in the ensemble)  have two dimensionless parameters and can be mapped to the Lorenz equation \cite{Carmichael_1}. Thus, even the single ensemble equations lead to periodic and chaotic asymptotic solutions. For two ensembles, we  do not anticipate any new kinds of asymptotic solutions other than fixed points, limit cycles, quasiperiodicity, and chaos. However, the mechanisms that give rise to different phases (especially chaos and chaotic synchronization) and the corresponding stability analyses are expected to be more complicated. 
 
  It would be interesting to explore the many-body version of our system with $n\gg1$ atomic ensembles identically coupled to a heavily damped cavity mode. It is simple to check that the TSS and NTSS survive in the many-body case. At zero  pumping, the  evolution equations \re{Mean-Field_1} for $n$ ensembles resemble mean-field equations of motion for the $s$-wave Bardeen-Cooper-Schrieffer (BCS) superconductor in terms of classical Anderson pseudospins. Here too individual spins couple through the $x$ and $y$ components of the total spin. The main difference is that   BCS dynamics are Hamiltonian and integrable \cite{yuzbashyan2005}.  Nevertheless, there are many similarities between the nonequilibrium phase diagram in \fref{Phase_Diagram} and many-body quantum quench phase diagrams of BCS superconductors \cite{ydgf}. In particular, the latter  contain three phases closely analogous to Phases I-III in \fref{Phase_Diagram}.  Here the amplitude of the superconducting order parameter, which is the analog of $|l_-(t)|$, either asymptotes to zero (Phase I) or to a finite constant (Phase II) or oscillates periodically (Phase III), see Ref.~\onlinecite{jasen} for  more on this  similarity between the phase diagrams.

 Another interesting problem, especially in the many-body context,  is to analyze the dynamics beyond mean-field and  determine if the full master equation  supports truly quantum  attractors  inaccessible  to the semiclassical dynamics \cite{Vakulchyk}. Recent work has also pointed out  an interpretation of limit cycles in atom-cavity systems as boundary time crystals \cite{saro}.   Alternatively, one can consider the same two atomic ensembles, but place them inside a multimode cavity to see if new types of correlated behaviors emerge in this setup  \cite{Multimode}. 

\begin{acknowledgments}

 We thank  S. Denisov, S. Gopalakrishnan, V. Gurarie, K. Mischaikow and Y. G. Rubo for helpful discussions. This work was supported by the National Science Foundation Grant DMR-1609829.  

\end{acknowledgments}
       
\onecolumngrid
\appendix

\section{Nonequilibrium Phase Diagram for a Single Atomic Clock}
\label{singleclock}

Here we show that the nonequilibrium phase diagram for a single atomic clock maps to the $\delta = 0$ axis of the two-clock diagram in \fref{Phase_Diagram}. Therefore, there are only two phases  in this case -- the normal phase  with no radiation (TSS), and  monochromatic superradiance  (NTSS), see also Refs.~\onlinecite{Holland_One_Ensemble_Theory_1, Holland_One_Ensemble_Theory_2, Holland_One_Ensemble_Expt_1}. We will also show that the mean-field evolution equations in this case reduce to damped Toda oscillator. 

For  one ensemble, the evolution equations \re{Mean-Field_1}  read,
 \begs
\bea
\dot{s}_{-} &=& \biggl(-\imath \omega_{1}- \frac{W}{2}\biggr)s_{-} + \frac{1}{2}s_{z}s_{-},  \label{REOMPM1en}\\
\dot{s}_{z} &=&  W\big(1 - s_{z}\big) - \frac{1}{2}s_{+}s_{-}. \label{REOMz1en} 
\eea 
\label{Mean-Field_1ens}%
\ens%
Going to a uniformly rotating frame, $s_-\to s_- e^{-\imath\omega_1 t}$ and $s_z\to s_z$, we eliminate $\omega_1$ from \eref{REOMPM1en}, i.e.,
\begs
\bea
\dot{s}_{-} &=&  - \frac{W}{2} s_{-} + \frac{1}{2}s_{z}s_{-},  \label{REOMPM1en2}\\ 
\dot{s}_{z} &=&  W\big(1 - s_{z}\big) - \frac{1}{2}s_{+}s_{-}. \label{REOMz1en2} 
\eea 
\label{Mean-Field_1ens2}%
\ens%
Now consider \eref{Mean-Field_1} for two ensembles with $\omega_A=-\omega_B=\delta/2$ at detuning $\delta=0$. Summing these equations
over $\tau$, we obtain
 \begs
\bea
\dot{l}_{-} &=&  - \frac{W}{2} l_{-} + \frac{1}{2}l_{z}l_{-}, \\ \label{REOMPMlm}
\dot{l}_{z} &=&  W\big(2 - l_{z}\big) - \frac{1}{2}l_{+}l_{-}. \label{REOMzlm} 
\eea 
\label{Mean-Field_1lm}%
\ens%
 After   rescaling $\bm{l}\to 2\bm{s}, W\to 2W$ and $2t\to t$, these equations coincide with \eref{Mean-Field_1ens2}. The scaling factor of 2 arises because \eref{Mean-Field_1lm}  describes a single ensemble with $2N$ atoms, while \eref{Mean-Field_1ens2}  is for $N$ atoms.
 
 On the other hand, \eref{Mean-Field_1lm} corresponds to two ensembles at $\delta=0$. Thus, the phases are those on the vertical $\delta=0$ axis in \fref{Phase_Diagram}, i.e., $l_-=2s_-$  asymptotes to its value in the TSS or NTSS for $\delta=0$. Therefore, the nonequilibrium phase diagram for a single atomic clock is 1D with the following two phases:
\beg
\begin{array}{llll}
\mbox{TSS:} & s_-=0, &\quad s_z=1& \mbox{for $W>1;$}\\ 
\mbox{NTSS:} & s_{-} =  e^{-\imath \Phi}\sqrt{2W(1-W)},&\quad   s_{z} = W, &  \mbox{for $W<1,$ }\\
\end{array}
\label{One_NTSS}
\en%
 where $\Phi$ is arbitrary. We derive these expressions directly from \esref{TSS} and \re{NTSS} by replacing $l_\perp\to 2s_\perp$, $W\to 2W$, and setting $\delta=0$.
 
Let us also analyze the transient mean-field dynamics of a single atomic clock.  Using $s_-=s_\perp e^{-\imath\Phi}$ in \eref{REOMPM1en2} and separating it into real and imaginary parts, we find that $\dot\Phi=0$. \eref{Mean-Field_1ens2} becomes,
\beg
s_z=\frac{2\dot s_\perp}{s_\perp}+W,\quad \dot{s}_{z} =  W\big(1 - s_{z}\big) - \frac{s_\perp^2}{2}.
\label{szsppp}
\en
Making the substitution $s_\perp=e^{X/2}$, we obtain $s_z=\dot X+W$ and the second order differential equation for $X$,
\beg
\ddot X+W \dot X+\frac{e^X}{2}+W(W-1)=0.
\label{dampedT}
\en
This equation describes the damped Toda oscillator \cite{Toda}. It has been  studied with Painlev\'e analysis and argued to be nonintegrable  unless
the last term in \eref{dampedT} is twice the square of the damping coefficient \cite{toda1}. In our case, this condition of integrability reads $W(W-1)=2W^2,$ 
i.e., $W=0$ or $W=-1$. The case $W=0$ is straightforward to solve. It corresponds to the origin, $\delta=W=0$, of the two clock phase diagram
and we solve it in Appendix~\ref{origin}. The case $W=-1$ is unphysical in our context. Thus,  dynamics  in the presence of pumping are nonintegrable already for one bad cavity ensemble.

\section{Derivation of the Poincar\'{e}-Birkhoff Normal Form}
\label{Derivation_PB_Normal_Form_appdx}

We established in Sect.~\ref{symmstab} that  the reduced equations of motion \re{Symm_One_Spin_Eqn}  determine the stability of the TSS and NTSS. 
In this appendix we  derive the corresponding Poincar\'{e}-Birkhoff normal forms, i.e., the right hand side of  \eref{Normal_Form_Intro}, starting from \eref{Symm_One_Spin_Eqn}.
We closely follow the steps in Ref.~\onlinecite{Intro_bifurc}, but fix a number of mistakes along the way and in the final answer. 

A key ingredient in this construction is the center manifold. 
Recall that in a Hopf bifurcation two complex conjugate characteristic values cross the imaginary line and acquire positive real parts.   In this case,   it is sufficient to study the  dynamics projected onto a 2D center manifold.    Imagine all the limit cycles as they continuously change their shape and size   upon changing a parameter, such as $\delta$  or $W$.   Heuristically, the center manifold  is the 2D sheet (this can be sufficiently warped away from the bifurcation) made by putting these limit cycles one after the other.  This manifold    is tangent to the  plane defined by the two unstable characteristic directions at the bifurcation.  For a pitchfork bifurcation, where  a real characteristic value becomes positive and there is a single unstable characteristic direction, the center manifold is 1D.

 \subsection{Hopf Bifurcation of the TSS and NTSS}\label{Appendx_Hopf}

The main steps of the derivation of the  Poincar\'{e}-Birkhoff normal form for a Hopf bifurcation are as follows:

\begin{enumerate}

\item Shift the origin of the coordinate system to the fixed point. 

\item Perform a linear  transform $\bm s\to{\bm s}'$ such that \eref{Symm_One_Spin_Eqn} takes the  form
\beg 
\frac{d{\bm s}^{\prime}}{dt} = \big[\underbrace{\textrm{Block diagonal linear part}}_{(2 \times 2) \bigoplus (1 \times 1)}\big]\cdot {\bm s}^{\prime}
 +  \textrm{ Second or higher order terms},
\en 
where the $2 \times 2$ part corresponds to the dynamics in the center manifold ($s_{x}^{\prime}, s_{y}^{\prime}$).

\item  Since the center manifold is 2D,  parameterize $s_{z}^{\prime}$ in terms of the other two spin components near criticality, 
\beg 
s_{z}^{\prime} = h(s_{x}^{\prime}, s_{y}^{\prime}).
\label{Approx_center_manifold}
\en%
Using this, produce the effective part of the dynamics projected onto the center manifold. This is still not the Poincar\'{e}-Birkhoff  normal form. It contains \textit{all} (including the non-essential)  nonlinear terms.

\item Write the equations in terms of $s^{\prime}_{\pm} = s^{\prime}_{x}\pm \imath s^{\prime}_{y}$.

\item Perform the ``near identity transformation".  This is a smooth change of variables  $s'_\pm \to s^{\prime\prime}_{\pm}$  that simplifies the $k$th and higher order terms in  projected dynamical equations. For Hopf bifurcations, it is possible to   eliminate   even order nonlinear terms in this way (nonessential terms for this bifurcations). Carrying out this transformation up to the second order, we obtain  the coefficient $a_{1}$ in front of $ s^{3}$ in the Poincar\'{e}-Birkhoff normal form as  written in \eref{Normal_Form_Intro}.  

\end{enumerate}

\subsubsection{Center Manifold Reduction}\label{Center_Manifold}

We start by shifting the origin of the coordinate system to the fixed point $\bm{s}_{0}$ in \eref{Symm_One_Spin_Eqn},  \begs
\bea
\dot{s}_{x}  &=& \bigg(s_{z0}-\frac{W}{2}\bigg)s_{x}  - \frac{\delta}{2}s_{y} + s_{x0}s_{z} + s_{z}s_{x}, \\ \label{ShiftedReducedX}
\dot{s}_{y} &=& \frac{\delta}{2}s_{x} - \frac{W}{2}s_{y}, \\ \label{ShiftedReducedY}
\dot{s}_{z} &=& -2s_{x0}s_{x} - Ws_{z}  -  s^2_{x}. \label{ShiftedReducedZ} 
\eea 
\label{Shifted_Symm_One_Spin_Eqn}
\ens%
Although   the fixed point is now at $(0, 0, 0)$,   the Jacobian matrix   \re{3_by_3_Jacobian} remains the same.     Let $\lambda_{r}$ ($v_1$) and  $ \gamma \pm \imath\omega$ ($v_{r}\pm\imath v_{i}$) be the real and complex characteristic values (vectors) at a Hopf bifurcation. Explicitly, for the bifurcation of the TSS on the $W=1, \delta\ge 1$ half-line, we read off $\lambda_r$, $\gamma$, and $\omega$ from \eref{3values} and the   characteristic  vectors are,
\beg 
 v_{1} = \bpm[1.5] 0 \\ 0 \\ 1\epm  , \qquad   v_{r} = \bpm[1.5]  1\\ \delta \\ 0\epm, \qquad  v_{i} = \bpm[1.5]  \sqrt{\delta^{2} - 1} \\ 0 \\ 0\epm.
\en
Similarly, for the bifurcation of the NTSS on the II-III boundary $\lambda_{r}$   and  $ \gamma \pm \imath\omega$ are the roots of the polynomial $P_3(\lambda)$ in \eref{P3}, while the   characteristic  vectors read,
\beg
  v_{1} = \bpm[1.5] \lambda_{r} + W \\ \delta\bigg(\frac{\lambda_{r} + W}{2\lambda_{r} + W}\bigg)\ \\ -2s_{x0}\epm  , \qquad   v_{r} = \bpm[1.5] \gamma + W \\ \frac{\delta\big[(2\gamma + W)(\gamma + W) + 2\omega^{2}\big]}{\big[(2\gamma+W)^{2} + 4\omega^{2}\big]} \\ -2s_{x0}\epm, \qquad  v_{i} = \bpm[1.5] \omega \\ \frac{-W\omega\delta}{\big[(2\gamma+W)^{2} + 4\omega^{2}\big]} \\ 0\epm,
\label{Eigenvectors_TSS_NTSS}
\en

Next, we perform a  linear transformation that  block-diagonalizes the Jacobian into  $2\times 2$  and $1\times 1$ blocks,
\beg 
\bpm s_{x} \\ s_{y} \\ s_{z}\epm   =  \mathbb{Q}_{H} \cdot \bpm s^{\prime}_{x} \\ s^{\prime}_{y} \\ s^{\prime}_{z}\epm,
\label{Unitery_Trans}
\en%
 where $ \mathbb{Q}_{H} = (v_{r}  \;  v_{i}  \;  v_{1})$.   
Using the above relation in \eref{Shifted_Symm_One_Spin_Eqn}, we obtain,
\begs
\bea
\bpm[1.5] \dot{s}^{\prime}_{x} \\ \dot{s}^{\prime}_{y} \epm &=& \bpm[1.5] \gamma & \omega \\ -\omega & \gamma \epm \bpm[1.5] s^{\prime}_{x} \\ s^{\prime}_{y} \epm + \bpm[1.5] R_{1}\big(s^{\prime}_{x}, s^{\prime}_{y}, s^{\prime}_{z}\big) \\ R_{2}\big(s^{\prime}_{x}, s^{\prime}_{y}, s^{\prime}_{z}\big)\epm , \\[15pt] \label{Unitery_Trans_xy} 
\dot{s}^{\prime}_{z} &=& \lambda_{r}s^{\prime}_{z} + R_{3}\big(s^{\prime}_{x}, s^{\prime}_{y}, s^{\prime}_{z}\big), \label{Unitery_Trans_z} 
\eea 
\label{Unitery_Trans_Expl}
\ens%
where, 
\beg
R_{i}\big(s^{\prime}_{x}, s^{\prime}_{y}, s^{\prime}_{z}\big) = R_{i1}\big(s^{\prime}_{x}\big)^{2} + R_{i2}\big(s^{\prime}_{y}\big)^{2} + R_{i3}s^{\prime}_{x}s^{\prime}_{y} + R_{i4}s^{\prime}_{x}s^{\prime}_{z} + R_{i5}s^{\prime}_{y}s^{\prime}_{z} + R_{i6}\big(s^{\prime}_{z}\big)^{2}.
\label{Rij_Hopf}
\en%
Note, $R_{ij}$ for all $i$ and $j$ are known functions of $\delta$ and $W$. 

Near the fixed point $(0, 0, 0)$ we parameterize the center manifold through $s^{\prime}_{z}\equiv h\big(s^{\prime}_{x}, s^{\prime}_{y}\big)$. Since the fixed point belongs to  the manifold and the $s^{\prime}_{x}\!\! -\!\!  s^{\prime}_{y}$ plane is tangential to it at $(0,0,0)$, we have,
\beg 
h(0,0) = 0, \qquad \frac{\partial h}{\partial s^{\prime}_{x}}\bigg|_{(0,0)} = 0, \qquad \frac{\partial h}{\partial s^{\prime}_{y}}\bigg|_{(0,0)} = 0.
\label{CM_Conditions}
\en%
This implies,
\beg 
h\big(s^{\prime}_{x}, s^{\prime}_{y}\big) =  h_{1}\big(s^{\prime}_{x}\big)^{2} + h_{2}\big(s^{\prime}_{y}\big)^{2} + h_{3}s^{\prime}_{x}s^{\prime}_{y} + \mathcal{O}\left(|\bm s'_\perp|^3\right).
\label{Stip_h}
\en%
Substituting   $s^{\prime}_{z} =h(s^{\prime}_{x}, s^{\prime}_{y})$ into \eref{Unitery_Trans_Expl}, we derive,
\beg 
\bpm[1.5] \frac{\partial h}{\partial s^{\prime}_{x}}& \frac{\partial h}{\partial s^{\prime}_{y}}\epm  \bem[1.5] \bpm[1.5] \gamma & \omega \\ -\omega & \gamma \epm \bpm[1.5] s^{\prime}_{x} \\ s^{\prime}_{y} \epm + \bpm[1.5] R_{1} \\ R_{2}\epm \enm = \lambda_{r} h\big(s^{\prime}_{x}, s^{\prime}_{y}\big) + R_{3}. 
\label{CM_Main_Eqn}
\en%
 Now using the form of $h\big(s^{\prime}_{x}, s^{\prime}_{y}\big)$  in \eref{Stip_h} and equating the coefficients of $\big(s^{\prime}_{x}\big)^{2}, \big(s^{\prime}_{y}\big)^{2}$ and $s^{\prime}_{x}s^{\prime}_{y}$ on both sides, we solve for $h_{1}, h_{2}$ and $h_{3}$   as follows:
\begin{equation}
h_{1}  = \frac{h_{3}\omega + R_{31}}{2\gamma - \lambda_{r}}, \qquad
h_{2} = \frac{-h_{3}\omega + R_{32}}{2\gamma - \lambda_{r}},\qquad
h_{3} = \frac{2\omega(R_{32} - R_{31}) + (2\gamma - \lambda_{r})R_{33}}{(2\gamma - \lambda_{r})^{2} + 4\omega^{2}}.
\label{CM_Projection}
\end{equation}%
Finally,   the effective equation projected on the center manifold   is,
\begin{multline}
\bpm[1.5] \dot{s}^{\prime}_{x} \\ \dot{s}^{\prime}_{y} \epm = \bpm[1.5] \gamma & \omega \\ -\omega & \gamma \epm \bpm[1.5] s^{\prime}_{x} \\ s^{\prime}_{y} \epm + \bpm[1.5] R_{11}\big(s^{\prime}_{x}\big)^{2} + R_{12}\big(s^{\prime}_{y}\big)^{2} + R_{13}s^{\prime}_{x}s^{\prime}_{y} \\ R_{21}\big(s^{\prime}_{x}\big)^{2} + R_{22}\big(s^{\prime}_{y}\big)^{2} + R_{23}s^{\prime}_{x}s^{\prime}_{y} \epm + \\ + \bigg(h_{1}\big(s^{\prime}_{x}\big)^{2} + h_{2}\big(s^{\prime}_{y}\big)^{2} + h_{3}s^{\prime}_{x}s^{\prime}_{y}\bigg)\bpm[1.5] R_{14}s^{\prime}_{x} + R_{15}s^{\prime}_{y} \\R_{24}s^{\prime}_{x} + R_{25}s^{\prime}_{y} \epm +  \mathcal{O}\left(| s'_-|^4\right).
\label{CM_Final}
\end{multline}

\subsubsection{The Normal Form}\label{Normal_Form}

The next step is to  rewrite \eref{CM_Final} in terms of $s^{\prime}_{\pm} = s^{\prime}_{x}\pm \imath s^{\prime}_{y}$,
\begin{multline}
\bpm[1.5] \dot{s}^{\prime}_{+} \\ \dot{s}^{\prime}_{-} \epm =  \underbrace{\bpm[1.5] (\gamma - \imath\omega)s^{\prime}_{+} \\ (\gamma + \imath\omega)s^{\prime}_{-} \epm}_{V^{(1)}\big(s^{\prime}_{+}, s^{\prime}_{-}\big)} + \underbrace{\bpm[1.5] R^{(2,0)}_{+}\big(s^{\prime}_{-}\big)^{2} + R^{(2,1)}_{+}s^{\prime}_{-}s^{\prime}_{+} + R^{(2,2)}_{+}\big(s^{\prime}_{-}\big)^{2} \\  R^{(2,0)}_{-}\big(s^{\prime}_{-}\big)^{2} + R^{(2,1)}_{-}s^{\prime}_{+}s^{\prime}_{-} + R^{(2,2)}_{-}\big(s^{\prime}_{+}\big)^{2} \epm}_{V^{(2)}\big(s^{\prime}_{+}, s^{\prime}_{-}\big)} + \\ + \underbrace{\bpm[1.5] R^{(3,0)}_{+}\big(s^{\prime}_{-}\big)^{3} + R^{(3,1)}_{+}s^{\prime}_{+}\big(s^{\prime}_{-}\big)^{2} + R^{(3,2)}_{+}\big(s^{\prime}_{+}\big)^{2}s^{\prime}_{-} + R^{(3,3)}_{+}\big(s^{\prime}_{+}\big)^{3} \\  R^{(3,0)}_{-}\big(s^{\prime}_{-}\big)^{3} + R^{(3,1)}_{-}s^{\prime}_{+}\big(s^{\prime}_{-}\big)^{2} + R^{(3,2)}_{-}\big(s^{\prime}_{+}\big)^{2}s^{\prime}_{-} + R^{(3,3)}_{-}\big(s^{\prime}_{+}\big)^{3} \epm}_{V^{(3)}\big(s^{\prime}_{+}, s^{\prime}_{-}\big)} + \mathcal{O}\left(| s'_-|^4\right),
\label{Pre_Normal_Form_Explicit}
\end{multline}%
where $V^{(k)}: \mathbb{R}^{2}\lra \mathbb{R}^{2}$ are homogeneous polynomial maps of degree $k = 1, 2, 3, \ldots$. Hence, for $a\in \mathbb{R}$ one has $V^{(k)}\big(as^{\prime}_{+}, as^{\prime}_{-}\big) = a^{k}V^{(k)}\big(s^{\prime}_{+}, s^{\prime}_{-}\big)$. At this point it is helpful to introduce the following basis functions for $V^{(k)}$:
\begin{equation}
\left.\begin{array}{l}
\xi^{{(k,l)}}_{+}=\bpm[1.5]  s_{+}^{l} s_{-}^{k-l} \\ 0 \epm,\\[25pt]
\xi^{{(k,l)}}_{-}=\bpm[1.5] 0 \\  s_{+}^{l} s_{-}^{k-l} \epm, 
\end{array}
\right\}
\qquad \text{$l$ = 0, 1, \ldots, $k$.}
\label{Eigen_Normal_Form}
\end{equation}%
From the definitions of $R_{\pm}^{(k,l)}$ in \eref{Pre_Normal_Form_Explicit}, it is clear that they are  nothing but the coefficients of different $\xi^{{(k,l)}}_{\pm}$, i.e.,
\beg
V^{(k)}= \sum_{l=0}^k R_{+}^{(k,l)} \xi^{(k,l)}_{+}+\sum_{l=0}^k R_{-}^{(k,l)} \xi^{(k,l)}_{-}.
\label{vkkkk}
\en  
 We read off  these coefficients $R_{\pm}^{(k,l)}$ from \esref{Pre_Normal_Form_Explicit} and (\ref{CM_Final}) as,
\begs
\begin{align}
R_{+}^{(3,0)} = \left[R_{-}^{(3,3)}\right]^{*} &= \frac{1}{8}\big[h_{1}\big(R_{14} - R_{25}\big) + h_{2}\big(R_{25} - R_{14}\big) - h_{3}\big(R_{15} + R_{24}\big)\big] + \nonumber \\[10pt] 
&\qquad\qquad\qquad\qquad\qquad+\frac{\imath}{8}\big[\big(h_{1} - h_{2}\big)\big(R_{15} + R_{24}\big) + h_{3}\big(R_{14} + R_{25}\big)\big], \\[10pt]
R_{+}^{(3,1)} = \left[R_{-}^{(3,2)}\right]^{*} &= \frac{1}{8}\big[3\big(h_{1}R_{14} - h_{2}R_{25}\big) + h_{2}R_{14} + h_{3}R_{15} - h_{3}R_{24} - h_{1}R_{25}\big] + \nonumber \\[10pt] 
&\qquad\qquad\qquad\quad+\frac{\imath}{8}\big[3\big(h_{2}R_{15} + h_{1}R_{24}\big) + h_{3}R_{14} + h_{1}R_{15} + h_{2}R_{24} + h_{3}R_{25}\big], \\[10pt]
R_{+}^{(3,2)} = \left[R_{-}^{(3,1)}\right]^{*} &= \frac{1}{8}\big[3\big(h_{1}R_{14} + h_{2}R_{25}\big) + h_{2}R_{14} + h_{3}R_{15} + h_{3}R_{24} + h_{1}R_{25}\big] + \nonumber \\[10pt] 
&\qquad\qquad\qquad\quad+\frac{\imath}{8}\big[3\big(h_{1}R_{24} - h_{2}R_{15}\big) + h_{2}R_{24} + h_{3}R_{25} - h_{3}R_{14} + h_{1}R_{15}\big], \\[10pt]
R_{+}^{(3,3)} = \left[R_{-}^{(3,0)}\right]^{*} &= \frac{1}{8}\big[h_{3}\big(-R_{15} + R_{24}\big) + h_{1}\big(R_{14} + R_{25}\big) - h_{2}\big(R_{14} + R_{25}\big)\big] + \nonumber \\[10pt] 
&\qquad\qquad\qquad\qquad\qquad+\frac{\imath}{8}\big[-\big(h_{1} - h_{2}\big)\big(R_{15} - R_{24}\big) - h_{3}\big(R_{14} + R_{25}\big)\big], \\[10pt]
R_{+}^{(2,0)} = \left[R_{-}^{(2,2)}\right]^{*} &= \frac{1}{4}\big(R_{11} - R_{12} - R_{23}\big) + \frac{\imath}{4}\big(R_{13} + R_{21} - R_{22}\big), \\[10pt]
R_{+}^{(2,1)} = \left[R_{-}^{(2,1)}\right]^{*} &= \frac{1}{2}\big(R_{11} + R_{12}\big) + \frac{\imath}{2}\big(R_{21} + R_{22}\big), \\[10pt]
R_{+}^{(2,2)} = \left[R_{-}^{(2,0)}\right]^{*} &= \frac{1}{4}\big(R_{11} - R_{12} + R_{23}\big) + \frac{\imath}{4}\big(R_{21} - R_{13} - R_{22}\big). 
\end{align}\label{RPM(k,l)}
\ens%
 Next, we eliminate as many nonlinear terms as possible from \eref{Pre_Normal_Form_Explicit}. To achieve this, we introduce the following   transformation:
\beg 
\bpm[1.5] s^{\prime}_{+} \\ s^{\prime}_{-} \epm = \bpm[1.5] s''_{+} \\ s''_{-} \epm - \bpm[1.5] \phi^{(k)}_{s''_{+}}\big(s''_{+}, s''_{-} \big) \\ \phi^{(k)}_{s''_{-}}\big(s''_{+}, s''_{-}\big) \epm, \qquad \phi^{(k)}\big(s''_{+}, s''_{-}\big) \equiv \bpm[1.5] \phi^{(k)}_{s''_{+}}\big(s''_{+}, s''_{-}\big) \\ \phi^{(k)}_{s''_{-}}\big(s''_{+}, s''_{-}\big) \epm, 
\label{NI_Transform}
\en%
where $\phi^{(k)}_{s''_{\pm}}\big(s''_{+}, s''_{-}\big)$ are small homogeneous polynomials of order $k$.
One needs to perform such near identity transformations iteratively.        In particular, substituting \eref{NI_Transform} into \eref{Pre_Normal_Form_Explicit} and using  Taylor expansions for $V^{(2)}$ and   $(\mathbb{1} - \mathbb{O})^{-1} = \mathbb{1} + \mathbb{O} + \mathbb{O}^{2} +\ldots$, we derive   
\begin{multline}
\bpm[1.5] \dot{s}''_{+} \\ \dot{s}''_{+} \epm = V^{(1)} +   \bigg[V^{(2)} - DV^{(1)}\cdot \phi^{(2)} + D\phi^{(2)}\cdot V^{(1)}\bigg] + \\ 
+ \bigg[V^{(3)} - DV^{(2)}\cdot \phi^{(2)} + D\phi^{(2)}\cdot V^{(2)} -  D\phi^{(2)}\cdot DV^{(1)}\cdot \phi^{(2)} + \left(D\phi^{(2)}\right)^{2}\cdot V^{(1)}\bigg] + \mathcal{O}\left(| s''_-|^4\right),
\label{NI_Normal_Form}
\end{multline}%
where
\beg 
DV^{k} \equiv \bpm[2] \frac{\partial V^{k}_{+}}{\partial s''_{+}} & \frac{\partial V^{k}_{+}}{\partial s''_{-}} \\ \frac{\partial V^{k}_{-}}{\partial s''_{+}} & \frac{\partial V^{k}_{-}}{\partial s''_{-}} \epm, \qquad\qquad D\phi^{k} \equiv \bpm[2] \frac{\partial \phi^{k}_{s''_{+}}}{\partial s''_{+}} & \frac{\partial \phi^{k}_{s''_{+}}}{\partial s''_{-}} \\ \frac{\partial \phi^{k}_{s''_{-}}}{\partial s''_{+}} & \frac{\partial \phi^{k}_{s''_{-}}}{\partial s''_{-}} \epm.
\label{NI_Normal_Form_1}
\en%
The modified nonlinear terms are 
\begs
\bea 
\wt{V}^{(2)} &\equiv & \bigg[V^{(2)} - DV^{(1)}\cdot \phi^{(2)} + D\phi^{(2)}\cdot V^{(1)}\bigg],\\
\wt{V}^{(3)} &\equiv & \bigg[V^{(3)} - DV^{(2)}\cdot \phi^{(2)} + D\phi^{(2)}\cdot V^{(2)} -  D\phi^{(2)}\cdot DV^{(1)}\cdot \phi^{(2)} + \left(D\phi^{(2)}\right)^{2}\cdot V^{(1)}\bigg].\label{NI_Normal_Form_2_V3}
\eea
\label{NI_Normal_Form_2}
\ens%
Note,   a near identity transformation at the $k$th order,    alters terms  of the  $k$th and higher orders.     In particular, the $k$th order term  becomes
\beg 
\wt{V}^{(k)} = V^{(k)}- DV^{(1)}\cdot \phi^{(k)} + D\phi^{(k)}\cdot V^{(1)} \equiv V^{(k)} - L\left(\phi^{(k)}\right), 
\label{Mod_kth_Order}
\en%
where we have introduced a linear operator. A function $\phi^{(k)}$  satisfying
\beg 
V^{(k)} =  L\left(\phi^{(k)}\right),
\label{Vanishing_kth_Order}
\en%
 eliminates the $k$th order nonlinear term. We verify that the eigenfunctions of $L$ are the column vectors $\xi^{{(k,l)}}_{\pm}$ defined in \eref{Eigen_Normal_Form}. The corresponding eigenvalues are
\beg
\lambda^{(k,l)}_{\pm} \equiv \gamma (1-k) - \imath\omega(k - 2l \pm 1),\quad L\left(\xi^{{(k,l)}}_{\pm}\right) = \lambda^{(k,l)}_{\pm}\xi^{{(k,l)}}_{\pm}.
\label{Eigen_Normal_Form_Expl}
\en%
At criticality $(\gamma = 0, \omega > 0)$ one ends up with $\lambda^{(k,l)}_{\pm} = 0$, if and only if $k = 2l\mp 1$, i.e., when $k$ is odd. Moreover, $\lambda^{(k,l)}_{\pm} = 0$ guarantees that one is unable to invert \eref{Vanishing_kth_Order} to obtain $\phi^{(k)}$.  Therefore, \eref{Vanishing_kth_Order} does not have a solution  if $k$ is odd and $V^{(k)}$  contains terms proportional to either $\xi^{{(k,\frac{k+1}{2})}}_{+}$ or $\xi^{{(k,\frac{k-1}{2})}}_{-}$, i.e., when $R_{\pm}^{(k,\frac{k\pm1}{2})}\ne0$ in \eref{vkkkk}. Such  nonlinearities that cannot  be eliminated  with a near identity transformation are known as essential nonlinearities.  We see that any $k$th order polynomial map is of the form   $V^{(k)} = V^{(k)}_{r} + V^{(k)}_{c}$, where $ V^{(k)}_{r}$ and $V^{(k)}_{c}$ are the removable (inessential) and essential nonlinearities. 

 \eref{Eigen_Normal_Form_Expl} implies that to eliminate $V^{(2)}$ by the second order near identity transformation, we need
\beg 
\phi^{(2)} = \sum_{l = 0}^{2} \bigg[ \frac{R_{+}^{(2,l)}}{\lambda_{+}^{2,l}}\xi_{+}^{(2,l)} + \frac{R_{-}^{(2,l)}}{\lambda_{-}^{2,l}}\xi_{-}^{(2,l)} \bigg].
\label{2nd_Order_NI}
\en%
 This introduces extra terms at the third and higher order.  Similarly, the third order near identity transformation, such that    $L\big(\phi^{(3)}\big) = \wt{V}^{(3)}_{r}$, eliminates   the nonessential parts of $\wt{V}^{(3)}$  defined  in \eref{NI_Normal_Form_2}.  This transformation does not affect   $\wt{V}^{(3)}_{c} = \alpha_{1}\xi^{{(3,2)}}_{+} + \alpha^{*}_{1}\xi^{{(3,1)}}_{-}$. Thus,  
\begin{multline} 
\alpha_{1} = R_{+}^{(3,2)} + R_{+}^{(2,1)}\left(\phi_{+}^{(2,2)} - \phi_{-}^{(2,1)}\right) + \phi_{+}^{(2,1)}\left(R_{-}^{(2,1)} - R_{+}^{(2,2)}\right) + 2\left(R_{-}^{(2,2)}\phi_{+}^{(2,0)} - R_{+}^{(2,0)}\phi_{-}^{(2,2)}\right) + \gamma \left[2\phi_{-}^{(2,2)}\phi_{+}^{(2,0)}\right. + \\ 
+ \left.\phi_{+}^{(2,1)}\left(\phi_{-}^{(2,1)} + 3\phi_{+}^{(2,2)}\right)\right] + \imath\omega\left[\phi_{+}^{(2,1)}\left(\phi_{+}^{(2,2)} - \phi_{-}^{(2,1)}\right) - 6\phi_{-}^{(2,2)}\phi_{+}^{(2,0)} \right],
\label{alpha_1}
\end{multline}%
where
\beg 
\phi_{\pm}^{(2,l)} \equiv \frac{R_{\pm}^{(2,l)}}{\lambda_{\pm}^{(2,l)}}.
\label{phi_Appendix}
\en
 Substituting  $s''_{\pm} = se^{\pm\imath\theta}$ into  the resulting equations of motion, we obtain
 \begs
\bea
\dot{s} &=& \gamma s + \mathrm{Re}(\alpha_{1}) s^{3} + \mathcal{O}\big(s^{5}\big),\label{Normal_Form_Final_1} \\ 
\dot{\theta} &=& \omega - \mathrm{Im}(\alpha_{1}) s^{2} + \mathcal{O}\big(s^{4}\big). \label{Normal_Form_Final_2}
\eea 
\label{Normal_Form_Final}
\ens%
 This is \eref{Normal_Form_Intro} with $a_{1}=\mathrm{Re}(\alpha_{1})$. 

 \subsection{Pitchfork Bifurcation of the TSS}
 \label{Appendx_PF}

 As we discussed at the end of \sref {symmstab},   the TSS loses stability via a pitchfork bifurcation in  the reduced equations of motion~\re{Symm_One_Spin_Eqn} on the upper quarter-arc forming the boundary between phases I and II.  In this case,   the center manifold is  1D. We   diagonalize the linear part of \eref{Shifted_Symm_One_Spin_Eqn} with the following linear transformation:
\beg 
\bpm s_{x} \\ s_{y} \\ s_{z}\epm   =  \mathbb{Q}_{P} \cdot \bpm s^{\prime}_{x} \\ s^{\prime}_{y} \\ s^{\prime}_{z}\epm,
\label{Unitery_Trans_PF}
\en%
where  $\mathbb{Q}_{P} = (v_{x^{\prime}} \; v_{y^{\prime}} \; v_{z^{\prime}})$, and $v_{x^{\prime}}, v_{y^{\prime}}$ and $v_{z^{\prime}}$ are the eigenvectors   of the Jacobian  \re{3_by_3_Jacobian}, where now $s_{x0}=0$ and $s_{z0}=1$.  
We  wrote down the corresponding eigenvalues in \eref{3values}, which we now rename as $\lambda_{x^{\prime}}\equiv \lambda_{3,4}$, $\lambda_{y^{\prime}}\equiv \lambda_{5,6}$, and $\lambda_{z^{\prime}}\equiv \lambda_{1,2}$.
Explicitly, the eigensystem of the Jacobian matrix at the TSS reads,
\beg
\begin{aligned}
\lambda_{x^{\prime}} =& \frac{1}{2}\big(1-W+\sqrt{1-\delta^{2}}\big), \qquad &\lambda_{y^{\prime}} &= \frac{1}{2}\big(1-W-\sqrt{1-\delta^{2}}\big), \qquad &\lambda_{z^{\prime}} &= -W,\\
v_{x^{\prime}} =& \bpm[1.5] \frac{1}{\delta}\big(1+\sqrt{1-\delta^{2}}\big) \\ 1 \\ 0\epm  , \qquad  & v_{y^{\prime}} &= \bpm[1.5] \frac{1}{\delta}\big(1-\sqrt{1-\delta^{2}}\big) \\ 1 \\ 0\epm, \qquad & v_{z^{\prime}} &= \bpm[1.5] 0 \\ 0 \\ 1\epm.
\end{aligned}
\en    
Inside Phase    II, $\lambda_{x^{\prime}}$ becomes positive.    After  the linear transformation \re{Unitery_Trans_PF}, \eref{Shifted_Symm_One_Spin_Eqn} becomes
\begs
\bea
\dot{s}^{\prime}_{x} &=& \lambda_{x^{\prime}}s^{\prime}_{x} + R_{1}\big(s^{\prime}_{x}, s^{\prime}_{y}, s^{\prime}_{z}\big)\label{Unitery_Trans_x_PF}, \\ 
\dot{s}^{\prime}_{y} &=& \lambda_{y^{\prime}}s^{\prime}_{y} + R_{2}\big(s^{\prime}_{x}, s^{\prime}_{y}, s^{\prime}_{z}\big)\label{Unitery_Trans_y_PF},\\ 
\dot{s}^{\prime}_{z} &=& \lambda_{z^{\prime}}s^{\prime}_{z} + R_{3}\big(s^{\prime}_{x}, s^{\prime}_{y}, s^{\prime}_{z}\big). \label{Unitery_Trans_z_PF} 
\eea 
\label{Unitery_Trans_Expl_PF}
\ens%
Below we list the nonzero $R_{ij}$ using the same notation as in \eref{Rij_Hopf}:
\beg 
\begin{aligned}
R_{14} = -R_{24} = \frac{1}{2} + \frac{1}{2\sqrt{1-\delta^{2}}}, \quad R_{15} = -R_{25} = -\frac{1}{2} + \frac{1}{2\sqrt{1-\delta^{2}}}, \\
R_{31} = -\frac{1}{\delta^{2}}\big(1+\sqrt{1-\delta^{2}}\big)^{2}, \quad R_{32} = -\frac{1}{\delta^{2}}\big(1-\sqrt{1-\delta^{2}}\big)^{2}, \quad R_{33} = -2.
\end{aligned}
\en  
Since the center manifold is 1D, we  parametrize it as,
\beg 
s^{\prime}_{y} = h_{1}(s^{\prime}_{x})^{2} + \cdots, \qquad s^{\prime}_{z} =  g_{1}(s^{\prime}_{x})^{2} + \cdots.
\label{Paramet_PF} 
\en%
This parameterization guarantees that the fixed point $s^{\prime}_{x}=s^{\prime}_{y}=s^{\prime}_{z}=0$ lies on the center manifold and the $s^{\prime}_{x}$-axis (the unstable characteristic direction) is tangential to it, cf. \eref{CM_Conditions}. Using \eref{Paramet_PF} in \esref{Unitery_Trans_y_PF} and (\ref{Unitery_Trans_z_PF})  and equating the coefficients of $s^{\prime}_{x}$ and $(s^{\prime}_{x})^{2}$ on both sides, we find
\beg 
h_{1}   = 0, \quad g_{1} = \frac{R_{31}}{2\lambda_{x^{\prime}} - \lambda_{z^{\prime}}} = -\frac{1}{\delta^{2}}\big(1+\sqrt{1-\delta^{2}}\big).
\en%
Finally, substituting the resulting $s^{\prime}_{y}$ and $s^{\prime}_{z}$  into \eref{Unitery_Trans_x_PF}, we derive the normal form for this bifurcation as
\beg
\dot{s}^{\prime}_{x} =   \lambda_{x^{\prime}}s^{\prime}_{x} + R_{14} g_{1} (s^{\prime}_{x})^{3}+ \mathcal{O}\big(s^{4}\big).  
\label{NF_PF}
\en%
 The coefficient of $(s^{\prime}_{x})^{3}$ is $ -\frac{\big(1+\sqrt{1-\delta^{2}}\big)^{2}}{2\delta^{2}\sqrt{1-\delta^{2}}}<0$  proving that the TSS goes through a supercritical pitchfork bifurcation at the boundary between Phases I and II.

\section{Dynamics of Two Atomic Ensembles in a Bad Cavity in the Absence of Pumping}
\label{Other Non-radiative Fixed Points}

In this Appendix, we analyze the dynamics of our system without pumping, i.e., on the $W=0$ line of the nonequilibrium phase diagram in \fref{Phase_Diagram}. Only attractors on this line are non-radiative fixed points. In fact, there is a family of such fixed
points for each value of the detuning $\delta$, only one of which is continuously connected to the limit cycle living in the green region of  \fref{Phase_Diagram} at $W\ne0$.
We will also
see that in the absence of pumping,  mean-field equations of motion \re{Mean-Field_1} for an arbitrary number $n$  of ensembles are a  set of generalized Landau-Lifshitz-Gilbert equations
that conserve the lengths of the spins.  Separately, we will study the dynamics for $\delta=W=0$, which reduces to the $W=0$ case of the one ensemble dynamics we considered in 
Appendix~\ref{singleclock}. 
 
 At $W=0$  it is useful to rewrite the mean-field evolution equations  \re{Mean-Field_1} [or equivalently \eref{Mean-Field_2}] in a vector form as
  \beg 
\fd{\bm{s}^{\tau}} = \omega_{\tau}\hat{z}\times\bm{s}^{\tau} + \frac{1}{2} \big(\hat{z} \times \bm{l}\big)\times\bm{s}^{\tau},\quad 
\bm l=\sum_{\t} \bm s^\t.
\label{W=0}
\en%
This is   the Landau-Lifshitz-Gilbert equation \cite{LL,gilbert} for the total spin $\bm l$  when all $\omega_{\tau}$ are the same. Otherwise, it is an inhomogeneous variant of the latter.  The Landau-Lifshitz-Gilbert damping term in \eref{W=0} pushes the spins towards the   $z$-axis whenever
$\bm l\ne0$.  \eref{W=0} conserves the magnitudes of  classical spins $\bm s^{\tau}$ since 
\beg 
\bm{s}^{\tau}\cdot \fd{\bm{s}^{\tau}} = 0.
\en%
 First, consider the case $\delta\neq 0$.   We observe numerically that for a generic initial condition $\big(s^{A}_{x0}, s^{A}_{y0}, s^{A}_{z0}, s^{B}_{x0}, s^{B}_{y0}, s^{B}_{z0}\big)$, the asymptotic solution for two atomic ensembles is $\big(0, 0, -|\bm s^{A}_{0}|, 0, 0, -|\bm s^{B}_{0}|\big)$. By analyzing the eigenvalues of the Jacobian \re{6_by_6_Jacobian}, we also establish that this fixed point is stable, while any other choice of signs of the $z$-components results in an unstable  fixed point. Therefore, there is a family of stable fixed points labeled by $|\bm s^{A}_{0}|$ and $|\bm s^{B}_{0}|$ for each $\delta$ on the $W=0$ axis  of the
phase diagram, which  are not  $\mathbb{Z}_{2}$-symmetric and  different from the TSS.
  According to \eref{W=0_Limit}, the $\Z2$-symmetric limit cycle in $W\to0$ limit turns into a fixed point
$\bm s^A=\bm s^B=0$, which is only one member of the above family of fixed points.

\subsection{The Origin of the Phase Diagram}
\label{origin}

Here we exactly solve the case $W = \delta = 0$. In this case, the total spin satisfies \eref{Mean-Field_1lm} with $W=0$, which is the same
as \eref{Mean-Field_1ens2} at $W=0$. Therefore, it is legitimate to simply replace $\bm s$ with $\bm l$ and set $W=0$ in \eref{szsppp}, i.e.,
\beg
l_z  l_\perp= 2\dot l_\perp,\quad \dot{l}_{z} =   - \frac{l_\perp^2}{2}.
\label{lzsppp}
\en
Let $l_\perp=e^{X/2}$ to obtain  the Toda oscillator equation,
\beg
\ddot X+ \frac{e^X}{2}=0.
\label{undampedT}
\en
As discussed below \eref{dampedT}, this equation is integrable. Its general solution is
\beg 
l_\perp^2 = \frac{2C^2_{1}}{1+\cosh{\big[ C_{1} t + C_{2}  \big] }},
\label{Solution_Toda_Oscillator}
\en
where $C_1$ and $C_2$ are arbitrary constants. \esref{Solution_Toda_Oscillator} and \re{lzsppp} show that $ l_\perp\to0$ and $l_z\to\mathrm{const}$ as $t\to+\infty$.

Then, \eref{Mean-Field_2}, where now $W=0$ and $\omega_A=-\omega_B=\delta/2=0$, implies that the time derivatives
  $\dot{s}^{\tau}_{x}, \dot{s}^{\tau}_{y}$ and $\dot{s}^{\tau}_{z}$ vanish since $l_\perp\to0$ at large times.
  Thus, the asymptotic solution for an arbitrary initial condition $\big(s^{A}_{x0}, s^{A}_{y0}, s^{A}_{z0}, s^{B}_{x0}, s^{B}_{y0}, s^{B}_{z0}\big)$ is
\begs
\begin{align} 
 s^{A}_{x}(\infty) &= - s^{B}_{x}(\infty) = s_{x\infty}, \\  
 s^{A}_{y}(\infty) &= - s^{B}_{y}(\infty) = s_{y\infty}, \\
 s^{\t}_{z}(\infty) &= \pm\sqrt{|\bm s^{\t}_{0}|^{2} -  s_{x\infty}^{2} -  s_{y\infty}^{2}},\quad \tau=A, B.\label{signs}  
\end{align}
\label{Origin_Asymtotic}
\ens%
In the last equation we took into account that the spin length is conserved in the absence of pumping.  The choice of signs in \eref{signs}  depends on the initial condition, but should be such that $l_{z}\le0$. Indeed, a linear stability analysis with the Jacobian~\re{6_by_6_Jacobian} reveals that  fixed points \re{Origin_Asymtotic} are unstable when $l_{z}> 0$  and stable when $l_{z}<0$. These fixed points   are  distinct from the TSS and generally do not retain the $\mathbb{Z}_{2}$ symmetry.

One describes the resultant dynamics as follows. Start with two arbitrary pointing spin vectors. Since $\delta = 0$, $\omega_\t=0$ in \eref{W=0} and the spins do not precess. Due to the Landau-Lifshitz-Gilbert damping, both  spins are pushed towards the   $z$-axis. Eventually, they align in such a way that  $l_{x}=l_{y}=0$ and $l_z\le0$. At that point all the time derivatives  vanish and the spins get stuck.

\section{Power Spectrum of Radiated Electric Field}\label{Expt}

In this appendix, we  show that the emission spectrum of atomic ensembles coupled through a heavily damped cavity mode  is proportional to $|l_{-}(f)|^{2}$, where $l_{-}(f)$ is   defined in \eref{l_Minus_f}. Our derivation combines the approaches of \Rsref{Carmichael_1, Carmichael_3}. For simplicity, we examine a ring cavity at $z = 0$ coupled to a 1D reservoir of length $\mathsf{L}$, which is a collection of standing waves with nodes at $\pm \mathsf{L}/2$.  The   Hamiltonian is,
\begin{equation}
\hat H = \hat H_{S} + \hat H_{R} + \hat H_{SR},
\label{Expt_Full_H}
\end{equation}%
where,  
\begin{equation}
\hat H_{S}  =\omega_{0}\hat{a}^{\dagger}\hat{a},\quad \hat H_{R} = \sum_{j} \omega_{j} \hat{r}_{j}^{\dagger}\hat{r}_{j},\quad \hat H_{SR}  =\sum_{j} \big(\kappa_{j}^{*}\hat{a}\hat{r}_{j}^{\dagger} + \kappa_{j}\hat{a}^{\dagger}\hat{r}_{j}\big),
\label{Expt_Explicit_H}
\end{equation}%
and $\hat{a}(\hat{a}^{\dagger})$ and $\hat{r}_{j}(\hat{r}_{j}^{\dagger})$  are the annihilation (creation) operators  for the cavity and the reservoir modes, respectively.  The periodic boundary condition  allows us to replace the sum over $\omega_{j}$ with an integration over $\omega$ with the density of states
 $g(\omega) = \mathsf{L}/2\pi c$,
where $c$ is the speed of light.  Assuming the reservoir is in the state of thermal equilibrium at temperature $T_0$  and adopting the Born-Markov approximation, we obtain the following master equation:
\begin{equation}
\dot{\rho} = -\imath\omega_{0}\big[\hat{a}^{\dagger}\hat{a}, \rho\big] + \underbrace{\kappa(1+\bar{n})\mathcal{L}[\hat{a}]\rho}_\textrm{Emission} + \underbrace{\kappa\bar{n}\mathcal{L}[\hat{a}^{\dagger}]\rho}_\textrm{Absorption},
\label{Expt_Master_Eq}
\end{equation}%
where $\bar{n} \equiv \frac{\exp{(-\omega_{0}/k_{B}T_0)}}{1 - \exp{(-\omega_{0}/k_{B}T_0)}}$, $\kappa \equiv 2\pi g(\omega_{0})|\kappa(\omega_{0})|^{2} = \Xi c/2\mathsf{L}$ and $\Xi<1$ is the transmission coefficient of the mirror separating the cavity from its surrounding.  At temperatures close to 0 K, we neglect all   terms proportional to $\bar{n}$ in \eref{Expt_Master_Eq}. As a result, we are left  only with the spontaneous emission of the cavity mode as shown in \eref{Full_Master}. We   also neglect the Lamb shift.

The measurable output electric field  in terms of the reservoir operators  reads,
\begin{equation}
\hat{\textbf{E}}(z,t) = \hat{\textbf{E}}^{(+)}(z,t) + \hat{\textbf{E}}^{(-)}(z,t),
\label{Expt_Output_E_1}
\end{equation}%
where, 
\begin{equation}
\hat{\textbf{E}}^{(+)}(z,t) \equiv \imath \hat{e}_{0}\sum_{k}\sqrt{\frac{ \omega_{k}}{2\varepsilon_{0}\mathsf{A}\mathsf{L}}} \hat{r}_{k}(t)e^{\imath \omega_{k} z/c  + \zeta(z)},\quad
\hat{\textbf{E}}^{(-)}(z,t) \equiv \hat{\textbf{E}}^{(+)}(z,t)^{\dagger}.
\label{Expt_Output_E_2}
\end{equation}%
 The phase shift  $\zeta(z)$ due to the mirror separating the cavity from the reservoir is $\zeta(z) = \zeta_{R}$ for $z>0$ and zero otherwise.  The unit vector $\hat{e}_{0}$  is perpendicular to the $z$-axis  and $A$ is the cross-sectional area of the cavity mode. 
 
 To relate the evolution of the reservoir modes $\hat{r}_{k}(t)$  to that of the cavity mode $\hat a(t)$,  consider the equation of motion for $\hat{r}_{k}(t)$,
 \begin{equation}
\dot{\hat{r}}_{k}(t) = -\imath \omega_{k}\hat{r}_{k}(t) - \imath \kappa_{k}^{*}\hat{a}(t).
\label{Expt_r_k_Heisenberg}
\end{equation}%
The solution of the above equation is 
\beg
\hat{r}_{k}(t) = \hat{r}_{k}(0)e^{-\imath\omega_{k}t}-\imath\kappa_{k}^{*} \int_{0}^{t}dt' \hat a(t) e^{\imath\omega_{k}  (t'-t)}.  
\label{Expt_r_k_Solun}
\en
\eref{Expt_Output_E_2} becomes
\begin{equation}
\hat{\textbf{E}}^{(+)}(z,t) = \hat{\textbf{E}}^{(+)}_{f}(z,t) + \hat{\textbf{E}}^{(+)}_{s}(z,t).
\label{Expt_Output_Free_Plus_Source}
\end{equation}%
Here $\hat{\textbf{E}}^{(+)}_{f}(z,t)$ is the freely evolving part of the electric field involving only $\hat{r}_{k}$ and $\hat{\textbf{E}}^{(+)}_{s}(z,t)$ originating from the second term on the right hand side of \eref{Expt_r_k_Solun} describes the effect of the cavity. We are interested in the   autocorrelation function,
\begin{equation}
\mathrm{Auto}(\tau) = \int_{-\infty}^{+\infty}dt\;\left\langle\hat{\textbf{E}}^{(-)}(z,t + z/c)\hat{\textbf{E}}^{(+)}(z,t + z/c + \tau)\right\rangle,
\label{Expt_AutoCorrelation}
\end{equation}%
 where the angular brackets denote taking a trace over the system and the reservoir degrees of freedom.   At low temperatures the state of the reservoir is very close to the vacuum electromagnetic field, $\bar{n}(\omega_{j}, T_0) \approx 0$. Therefore,  the contribution of the freely evolving part of the electric field is negligible, i.e.,
\begin{equation}
\left\langle\hat{\textbf{E}}^{(-)}(z,t + z/c)\hat{\textbf{E}}^{(+)}(z,t + z/c + \tau)\right\rangle \approx \left\langle\hat{\textbf{E}}^{(-)}_{s}(z,t + z/c)\hat{\textbf{E}}^{(+)}_{s}(z,t + z/c + \tau)\right\rangle.
\label{Expt_AutoCorrelation_Source}
\end{equation}%
Following the steps outlined in Sect. 1.4 of Ref.~\onlinecite{Carmichael_3}, we obtain, 
\begin{equation}
\hat{\textbf{E}}^{(+)}_{s}(z,t) = 
\begin{cases}
\hat{e}_{0}\sqrt{\frac{ \omega_{0}}{2\varepsilon_{0}\mathsf{A}c}}\sqrt{2\kappa}e^{\imath \zeta_{R}}\hat{a}(t - z/c), & \text{if } ct > z > 0\\
0, & \text{if } z < 0
\end{cases}
\label{Expt_E_in_Terms_of_Source_Operators}
\end{equation}%
According to \eref{Ad_Elimination} in the bad cavity limit,
\begin{equation}
\mean{\hat{a}(t)} \propto l_{-}(t).
\label{Expt_Bad_Cavity}
\end{equation}%
Outside of the cavity, using \esref{Expt_E_in_Terms_of_Source_Operators},  (\ref{Expt_Bad_Cavity}) and   the mean-field approximation  in \esref{Expt_AutoCorrelation} and  (\ref{Expt_AutoCorrelation_Source}), we  derive,
\begin{equation}
\mathrm{Auto}(\tau) \propto \int_{-\infty}^{+\infty}dt\;l_{+}(t)l_{-}(t + \tau).
\label{Expt_AutoCorrelation_Final}
\end{equation}%
Finally, the Fourier transform of  the autocorrelation function (the power spectrum) is, 
\begin{equation}
 \mathrm{Auto}(f) \propto |l_{-}(f)|^{2},
\label{Expt_Power_Spectrum}
\end{equation} %
 where we  used the Wiener-Khintchine theorem.



\begin{thebibliography}{999}
\bibitem{Resonant_Hopf}  Since a pair of complex conjugate characteristic values acquire  positive real parts, this is a double-Hopf bifurcation. When $\omega_A + \omega_B = 0$, all four of these characteristic values have same imaginary parts $\sqrt{\delta^2 - 1}/2$ at criticality -- making it a $1:1$ resonant Hopf bifurcation. This brings out the innate codimension 2 (in terms of $\delta$ and $W$) nature of the bifurcation.
\bibitem{Elliptic} Note that in \eref{W=1_Limit} $s_{y}\propto\sin{(\omega t)}$, whereas here we find $s_{y}\propto \cos{bt}$. This is not a contradiction, because while solving for \esref{Symm_One_Spin_Eqn_s_y} and (\ref{W=del=1_s_y}) we have not fixed the origin of time. By suitably choosing the origin in one of the solutions, one can match the phases of the  two functions exactly.
\end{thebibliography}
\end{document}